\documentclass[10pt]{iopart}
\usepackage{latexsym}
\usepackage{amsfonts}
\usepackage{amsbsy}
\usepackage[mathscr]{euscript}
\usepackage{accents}
\usepackage{hyperref}
\hypersetup{colorlinks=true,linkcolor=blue,urlcolor=blue,citecolor=blue}
\usepackage{cite}

\usepackage{comment}
\usepackage[cal=boondox]{mathalfa}

\usepackage{tensor}

\usepackage{color}

\begin{document}

\title[$BF$ gravity]
{$BF$ gravity}

\author{Mariano Celada$^1$, Diego Gonz\'alez$^1$ and Merced Montesinos$^{1,2}$}

\address{$^1$ Departamento de F\'{\i}sica, Cinvestav, Instituto Polit\'ecnico Nacional 2508,
San Pedro Zacatenco, 07360, Gustavo A. Madero, Ciudad de M\'exico, M\'exico.}
\address{$^2$ Departamento de Matem\'aticas, Instituto de Ciencias, Benem\'erita Universidad Aut\'onoma de Puebla, Ciudad 
Universitaria, 72572, Puebla, Puebla, M\'exico.}

\eads{\mailto{mcelada@fis.cinvestav.mx}, \mailto{dgonzalez@fis.cinvestav.mx} and \mailto{merced@fis.cinvestav.mx}
}

\begin{abstract} 
$BF$ gravity comprises all the formulations of gravity that are based on deformations of $BF$ theory. Such deformations consist of either constraints or potential terms added to the topological $BF$ action that turn some of the gauge degrees of freedom into physical ones, particularly giving rise to general relativity. The $BF$ formulations have provided new and deep insights into many classical and quantum aspects of the gravitational field, setting the foundations for the approach to quantum gravity known as spinfoam models. In this review, we present a self-contained and unified treatment of the $BF$ formulations of $D$-dimensional general relativity and other related models, focusing on the classical aspects of them and including some new results.
\end{abstract}

\section{Introduction}

Consider a principal bundle $P$ on an arbitrary $D$-dimensional manifold $\mathcal{M}$ ($D\geq2$) with structure group a Lie group $G$. A $BF$ theory is a diffeomorphism-invariant gauge theory whose action principle has the form
\begin{eqnarray}
S[A,B]=\int_\mathcal{M}\Tr(B\wedge F[A]),
\end{eqnarray}
where $B$ is a ($D-2$)-form taking values in the Lie algebra $\mathfrak{g}$ of $G$, $A$ is a connection on $P$, $F:=dA+A\wedge A$ is the curvature of this connection, and the trace is taken with respect to the nondegenerate Killing-Cartan metric of $\mathfrak{g}$. The particular form of the action is the reason why this kind of models are called $BF$ theories, which started to be known as such after they were properly treated as field theories by themselves in~\cite{horowitz1989,blau1989new,Blau1991130} (the Abelian case was first considered in~\cite{Schwarzlmp2}). These models have been studied exhaustively in the literature~\cite{horowitz1989,blau1989new,Blau1991130,birmingham1991,cuesta2007,Montesinos2008,caicedo1995brst,Baez2000,cattaneo2000}, since they define topological field theories and provide an arena for exploring different quantization approaches.

\paragraph{A brief history of $BF$ gravity.} 

General relativity in four dimensions, the geometrical theory of the gravitational field devised by Einstein, can be reformulated as a modified $BF$ theory. This is somewhat surprising since, in contrast to $BF$ theories, general relativity possesses propagating degrees of freedom. As a result, the $BF$ formulations of general relativity have drawn the attention of the community in recent decades, establishing the foundations of the covariant approach to quantum gravity known as spinfoam models~\cite{alexandrov2012spin,perez2013-16-3,rovelli2014covariant}, where a path integral quantization of the gravitational field is intended. The first $BF$-type formulation of general relativity dates back to the work of Plebanski in 1977 (although at that time the $BF$ terminology had not been invented yet). It was shown by Plebanski that the $B$ field (the self-dual complex 2-form)  can be used as the fundamental variable to describe the gravitational field, rather than the metric tensor. The idea behind the Plebanski formulation is to supplement the $BF$ theory with a constraint--the simplicity constraint--on the $B$ field. The solution of this constraint then brings in the tetrad field into the formalism, collapsing Plebanski action to the self-dual formulation of general relativity~\cite{Samuel1987,Jacobson198739,Jacobsoncqg5}. In this way, such a constraint breaks the topological character of the $BF$ theory, and the complete theory acquires local dynamics. As a bonus of this approach, the Lagrange multiplier used to impose the constraint gets identified {\it a posteriori} with the self-dual part of the Weyl curvature tensor.

Nine years afterwards, Ashtekar~\cite{ashtekar1986-57-2244,ashtekar1987-36-1587} was able to simplify the Hamiltonian formulation of general relativity by introducing a new set of complex canonical variables. Notably, in terms of these variables the resulting constraints resemble those of Yang-Mills theory.  In 1990, the Plebanski formulation was rediscovered by  Capovilla \textit{et al}~\cite{capo1991841}, who showed that by performing the canonical analysis of Plebanski's action, the  Hamiltonian formulation of general relativity in terms of the Ashtekar variables emerges. The Plebanski formulation has also served as departure point to reach other interesting formulations of general relativity based on a gauge connection.  Among these formulations we find the one due to Capovilla, Dell, and Jacobson (CDJ)~\cite{capo198963,capo1991859,beng1990261,peld19918,capo199207}, which can be regarded as an almost pure gauge connection formulation since the variables involved in it are an $SO(3,\mathbb{C})$-connection and a scalar density. More recently, in 2011, and also based on Plebanski's work, a pure connection formulation of general relativity was given by Krasnov in~\cite{kras2011106}. An alternative derivation of this formulation was reported by the authors of the present review in~\cite{cgm-pure-2015}, where the link between the Plebanski, CDJ, and pure connection formulations was clarified as well. Besides, later on Krasnov realized that Plebanski's action can be generalized into a large class of ``non-metric gravity theories''~\cite{kras200606} that requires neither new fields, new dimensions, nor a different gauge group. 

Given that Plebanski's action employs complex variables, to arrive at real general relativity one needs to impose algebraic reality conditions on the $B$ field~\cite{capo1991859}.  At the classical level, these conditions do not represent a drawback. In fact, it is possible to include them in the action principle~\cite{pleb1977118,robi1995-36-7,Lewandowski0264}. However, at the quantum level, the reality conditions are difficult to deal with, which led people to look elsewhere for a suitable real formulation of general relativity to circumvent the difficulties posed by them. In the late 1990s, it was realized that general relativity can be described in the framework of real $BF$ theories~\cite{Reisencqg147,PietriFrei}. In these formulations the idea of introducing constraints on the $B$ fields is also adopted, which justifies considering them as nonchiral versions of the Plebanski formulation. Shortly afterwards, the inclusion of the Immirzi parameter~\cite{Immirzicqg1410} in this framework was achieved in~\cite{CMPR2001}, by considering a generalization of the constraint on the $B$ fields. An equivalent $BF$ action using a different mechanism to incorporate the Immirzi parameter is also possible~\cite{EPRnuc798,MercSIGMA72011}. The canonical analysis of these formulations leads to the same phase space of the Ashtekar-Barbero formalism~\cite{perez2013-16-3,CelMontesRom}, which, in the 1990s, established the foundations of loop quantum gravity~\cite{rovelli2004quantum,thiemann2007modern,Rovellicag2815}. Although from the classical point of view the Immirzi parameter has a topological character (it is introduced into the action as a coupling constant of a topological term~\cite{LiuPRD81}; see also~\cite{Sengupta2009,Sengupta2012}), it plays a nontrivial role  at the quantum level and becomes meaningful~\cite{AshtLewacqg14,RoveSmolinNucP442,RovPRLt.77.3288,RovThiePRD57}. However, its quantum relevance seems to be a matter of choice of the connection variables used to build the quantum theory~\cite{Alexprd652,AlexVassprd644,AlexLivprd674,NouiSIGMA72011, Nouiprd84044002}. 

Gravity in dimensions other than four can also be recast as $BF$ theories.  Lower-dimensional gravity allows us to address some fundamental issues that in four dimensions are technically more demanding. In two dimensions, it is well-known that  gravity cannot be described by general relativity, since in this case the theory turns out to be meaningless. However, there are alternative models with a rich and interesting structure. In the 1980s, the first and simplest model was proposed by Jackiw~\cite{Jackiw1984} and Teitelboim~\cite{Teitelboim198341,Teitelboim1984}. Not long after that, the model was formulated as a $BF$ theory~\cite{fukuyama1985259,isler1989-63,chamseddine198975}. The idea of this approach is to combine the frame and spin-connection into the gauge connection. On the other side, in three dimensions, Palatini's action coincides with the pure $BF$ action, and by including a cosmological constant, it is possible to obtain a pure connection formulation for three-dimensional Einstein's theory~\cite{PeldanNPB395239,peld1994115}. Furthermore, Palatini's action corresponds to a gauged version of a Holst-like action principle for three dimensions that has been used to explore the role of the Immirzi parameter in the quantum theory~\cite{AchNouiprd9110}. Moreover, three-dimensional gravity can also be expressed as a constrained $BF$ theory~\cite{GeiNouicqg2913}, which, in turn, arises from a symmetry reduction of its four-dimensional counterpart. On the other hand, general relativity in dimension greater than four also admits a constrained $BF$ formulation, the constraint being a generalization of the one introduced in the four-dimensional case~\cite{Freid_Puz}.

The aim of this review is to provide an introductory, self-contained, and unified approach to the $BF$ formulations of gravity in $D$ dimensions at the classical level. Our review not only includes the latest results and advances in the field of $BF$ gravity, but also presents new results and alternative views of some of the formulations discussed throughout the manuscript. We have made an effort to keep the presentation simple enough for the review  to be of interest to researchers not specialized in the subject.  We have organized the review into three main sections, according to the spacetime dimensionality. Section~\ref{BF-4D} is devoted to describing different $BF$ formulations of general relativity in four dimensions, as well as the generalization of the Plebanski formulation and the $BF$ formulation of the Husain-Kucha\v{r} model. We begin by giving a detailed description of the Plebanski formulation in the $SO(3,\mathbb{C})$ formalism, which also includes a systematic derivation of the CDJ and pure connection formulations. Subsequently, taking as starting point the pure connection formulation, we show that it is possible to arrive at a Plebanski-like action that has some additional interesting features as compared to Plebanski's action~\cite{cgm-PlebLike-2016}. After this, we present the formulation of the Husain-Kucha\v{r} model as a constrained $BF$ theory~\cite{aip_paper}. Then, we discuss the non-metric gravity theories and give an alternative expression for the gauge connection. We conclude this section with a complete analysis of the formulation of real general relativity with the Immirzi parameter as a constrained $BF$ theory~\cite{CelMontesRom,CelMontcqg2920}, followed by an outline of the $BF$ formulation of MacDowell-Mansouri gravity~\cite{StarodubtsevArxiv}. In section~\ref{BF-lower}, we focus on lower dimensions and review the $BF$ formulations of the Jackiw-Teitelboim (JT) model and three-dimensional Einstein's theory, showing how the latter leads to a pure connection formulation for three-dimensional gravity. In addition, we discuss a Plebanski-like formulation for general relativity in three dimensions~\cite{GeiNouicqg2913}. Finally, in section~\ref{BF-higher} we analyze  the $BF$ formulation of Palatini's action in higher dimensions.

\paragraph{Yang-Mills theory as a deformed $BF$ theory.}
Before going into the main subject of this review, which is casting general relativity in the framework of $BF$ theories, we discuss the case of Yang-Mills theory. As in the general relativity case, the $BF$ formulation of Yang-Mills theory (or first-order formalism for gauge theories) also appears before the $BF$ terminology was coined~\cite{Kholy1973so,Halpern.16.1798}. 

We focus on Yang-Mills theory on a four-dimensional curved  manifold $\mathcal{M}$ with fixed metric tensor $g_{\mu\nu}$ of signature ($\sigma$,+1,+1,+1), where $\sigma=-1$ ($\sigma=+1$) for the Lorentzian (Riemannian) case. We assume that the Lie algebra $\mathfrak{g}$ of the Yang-Mills gauge group $G$ has dimension $N$.  The action principle for Yang-Mills theory in the $BF$ formalism is~\cite{fucito1997bf}
\begin{eqnarray}
S[A,B]=\int_{\mathcal{M}}\left[B_I\wedge F^I[A]+e^2B_I\wedge *B^I\right],\label{BFYM_1}
\end{eqnarray}
where $B^I$ are $\mathfrak{g}$-valued 2-forms, $F^I:=dA^I+(1/2)\tensor{f}{^I_{JK}}A^J\wedge A^K$ is the field strength of the Yang-Mills connection $A^I$, $*$ is the spacetime Hodge dual operator\footnote{Let $\Lambda^p(\mathcal{M})$ be space of $p$-forms on a $D$-dimensional spacetime $(\mathcal{M},g)$. The Hodge dual operator $\ast$ is the linear map from $\Lambda^p(\mathcal{M})$ to $\Lambda^{D-p}(\mathcal{M})$ given by
\begin{center}
${\ast(dx^{\mu_1}\wedge\dots \wedge dx^{\mu_p})=\frac{1}{(D-p)!} \sqrt{|\det(g_{\mu\nu})|} \underaccent{\tilde}{\eta}{}_{\nu_1\dots\nu_p\nu_{p+1}\dots{\nu_D}} g^{\nu_1\mu_1}\dots g^{\nu_p\mu_p} dx^{\nu_{p+1}} \wedge\dots\wedge dx^{\nu_D}}$.
\end{center}}, and $e$ is a nonvanishing coupling constant. The indices $I,J,\dots$, which run from 1 to $N$,  are raised and lowered with the Killing-Cartan metric of $\mathfrak{g}$, and we assume that the structure constants $f_{IJK}$ are totally antisymmetric. The equations of motion emerging from (\ref{BFYM_1}) are
\begin{eqnarray}
\delta A_I:\hspace{3mm} &DB^I:=dB^I+\tensor{f}{^I_{JK}}A^J\wedge B^K=0,\label{YM-eq1}\\
\delta B_I:\hspace{3mm}&F^I+2e^2 *\!B^I=0.\label{YM-eq2}
\end{eqnarray}
The last equation can be solved for $B^I$, and replacing the result into equation~(\ref{YM-eq1}), we obtain the Yang-Mills equations, namely, $D*F^I=0$. On the other side, by substitu- ting the solution for $B^I$ from (\ref{YM-eq2}) back into (\ref{BFYM_1}), we obtain the Yang-Mills action
\begin{eqnarray}
S[A]=-\frac{\sigma}{4e^2}\int_{\mathcal{M}}*F_I\wedge F^I.\label{Y-M action}
\end{eqnarray} 
Consequently, the action (\ref{BFYM_1}) and Yang-Mills theory are classically equivalent (the equivalence also holds at the quantum level~\cite{martellini1997feynman,cattaneo1998four}). Notice that for $e=0$ the action (\ref{BFYM_1}) is the topological $BF$ theory. Besides the invariance of the latter under local $G$-transformations, the pure $BF$ action also has the topological symmetry $\delta A^I=0$, $\delta B^I=D\xi^I$, where $\xi^I$ is a 1-form. These two symmetries are responsible for depriving $BF$ theory of local dynamics. Therefore, for a nonvanishing $e$ the second term on the right-hand side of (\ref{BFYM_1}) breaks the topological symmetry and generates the degrees of freedom of Yang-Mills theory. In this sense, Yang-Mills theory is a deformation of topological $BF$ theory.

The canonical analysis of the action (\ref{BFYM_1}) leads to the following. The phase space is parametrized by the canonical pairs built up from the spatial components of the connection $A_{aI}$ and their conjugate canonical momenta\footnote{Our convention is that we write spacetime indices to the left of internal ones. However, that is not the case for the curvature, whose internal indices are written to the left of the spacetime ones to agree with the usual notation of the Riemann tensor.} $\tilde{E}^{aI}:=(1/2)\tilde{\eta}^{abc}\tensor{B}{_{bc}^I}$, where $\tilde{\eta}^{abc}:=\tilde{\eta}^{tabc}$, with $\tilde{\eta}^{\mu\nu\lambda\rho}$ the Levi-Civita tensor density ($\tilde{\eta}^{t123}=+1$). After integrating out $\tensor{B}{_{ta}^I}$ in the action (\ref{BFYM_1}), the Hamiltonian density takes the form $\tilde{\mathcal{H}}=-A_{tI}\tilde{\mathcal{G}}^I-(N/\sqrt{h})h_{ab}\left[e^2\tilde{E}^{aI}\tensor{\tilde{E}}{^b_I}-(\sigma/4e^2)\tilde{B}^{aI}\tensor{\tilde{B}}{^b_I}\right]+N^a\tilde{E}^{bI}F_{Iab}$, where $\tilde{\mathcal{G}}^I:=D_a \tilde{E}^{aI}=\partial_a \tilde{E}^{aI}+\tensor{f}{^I_{JK}}\tensor{A}{_a^J}\tilde{E}^{aK}\approx 0$ is the only constraint of the theory that is first class as well, $\tilde{B}^{aI}:=(1/2)\tilde{\eta}^{abc}\tensor{F}{^I_{bc}}$, $N$ and $N^a$ are respectively the lapse function and the shift vector associated to the ADM form of the background metric $g_{\mu\nu}$~\cite{Arnowitt2008}, and $h_{ab}:=g_{ab}$ is the spatial part of the metric whose determinant is denoted by $h$. As a consequence of this, the action (\ref{BFYM_1}) propagates $2N$ degrees of freedom, the same as Yang-Mills theory.

\section{$BF$ formulations of four-dimensional gravity}\label{BF-4D}

\subsection{Plebanski formulation of complex general relativity}\label{sec:plebanskiformulation}

\subsubsection{Action principle.}

In 1977 Plebanski~\cite{pleb1977118} wrote general relativity as an $SL(2,\mathbb{C})$-invariant $BF$ theory supplemented with a constraint on the $B$ field using the spinorial formalism. Let $\mathcal{M}$ be a four-dimensional manifold. The first-order Plebanski's action of complex general relativity is, equivalently,  given by
\begin{eqnarray}
	\fl S[A, \Sigma,\Psi,\rho]=\int_\mathcal{M} \left[ \Sigma_i \wedge F^i[A] -\frac{1}{2} \left(\Psi_{ij}+\frac{\Lambda}{3} \delta_{ij}  \right)  \Sigma^i \wedge \Sigma^j -\rho {\rm Tr}\Psi  \right],   \label{plebanAct}
\end{eqnarray}
where $\Sigma^i$ are $\mathfrak{so}(3,\mathbb{C})$-valued 2-forms that play the role of the $B$ fields, $F^i=d A^i + (1/2) {\varepsilon}^i{}_{jk}  A^j  \wedge A^k$ is the curvature of the $SO(3,\mathbb{C})$-connection $A^i$, $\Psi$ is a symmetric complex $3\times 3$  matrix, $\rho$ is a nonvanishing complex-valued 4-form, and $\Lambda$ is the cosmological constant. The indices $i,j=1,2,3$ are raised and lowered with the three-dimensional Euclidean metric $\delta_{ij}$, and ${\varepsilon}_{ijk}$ is the Levi-Civita symbol  (${\varepsilon}_{123}=+1$). If the connection $A^i$ is dimensionless as a 1-form and $[\Lambda]$ stands for the dimension of the cosmological constant, then $\Sigma^i$, $\Psi$, and $\rho$ have respectively dimension $[\Lambda]^{-1}$, $[\Lambda]$, and $[\Lambda]^{-2}$.

The equations of motion that follow from the action (\ref{plebanAct}) are
\begin{eqnarray}
	\delta \Psi_{ij}&:& \Sigma^i \wedge \Sigma^j + 2 \rho \delta^{ij}=0,  \label{pleban1}\\
	\delta A^i&:&          D \Sigma^i:=d \Sigma^i  +  \varepsilon^i{}_{jk} A^j \wedge \Sigma^k =0,  \label{pleban2}\\
	\delta \Sigma^i&:&   F^i=\Psi^i{}_j\Sigma^j+\frac{\Lambda}{3} \Sigma^i, \label{pleban3}\\
	\delta \rho&:& {\rm Tr}\Psi=0,\label{pleban4}
\end{eqnarray}
where $D$ is the $\mathfrak{so}(3,\mathbb{C})$-covariant derivative defined by $A^i$. Let us note that, as a consequence of the Bianchi identities, $DF^i=0$, and equations~(\ref{pleban2}) and (\ref{pleban3}), we get
\begin{eqnarray}
	D \Psi^i{}_j \wedge \Sigma^j=0.\label{ple:DPsi}
\end{eqnarray}

The gauge symmetries of the action principle (\ref{plebanAct}) can be obtained from identities involving the equations of motion. In fact, from the identity
\begin{eqnarray}
\fl  - D \Sigma^i \wedge D \varepsilon_i + \left ( F^i - \Psi^i\,_l \Sigma^l -\frac{\Lambda}{3}   \Sigma^i \right ) \wedge \left (- \varepsilon_{ijk} \varepsilon^j\Sigma^k \right ) \nonumber\\
 - \frac12 \left ( \Sigma^i \wedge \Sigma^j + 2 \rho \delta^{ij} \right ) \left ( \varepsilon^k{}_{li} \varepsilon^l \Psi_{kj}  + \varepsilon^k{}_{lj} \varepsilon^l \Psi_{ki}\right ) = d \left ( \varepsilon_i D \Sigma^i \right ),
\end{eqnarray}
where $\varepsilon^i$ is the gauge parameter, we get the infinitesimal $SO(3,\mathbb{C})$ gauge transformation of the fields
\begin{eqnarray}
\fl \delta A^i = D \varepsilon^i, \quad \delta \Sigma^i = - \varepsilon^i{}_{jk}  \varepsilon^j \Sigma^k, \quad \delta \Psi_{ij} &=& \varepsilon^k{}_{li} \varepsilon^l \Psi_{kj}  + \varepsilon^k{}_{lj} \varepsilon^l \Psi_{ki} , \quad \delta \rho=0.
\end{eqnarray}
Similarly, the action principle (\ref{plebanAct}) is also diffeomorphism-invariant. The infinitesimal diffeomorphism transformation of the fields 
\begin{eqnarray}
\delta A^i = {\mathcal L}_v A^i, \quad \delta \Sigma = {\mathcal L}_v \Sigma^i, \quad \delta \Psi_{ij} = {\mathcal L} _v \Psi_{ij}, \quad \delta \rho = {\mathcal L}_v \rho, 
\end{eqnarray}
generated by the vector field $v$ is associated to the identity
\begin{eqnarray}
 \fl - D \Sigma_i \wedge {\mathcal L}_v A^i + \left ( F_i - \Psi_{ij} \Sigma^j - \frac{\Lambda}{3} \Sigma_i \right ) \wedge {\mathcal L}_v \Sigma^i - {\rm Tr} \Psi {\mathcal L}_v \rho \nonumber\\
- \frac12 \left ( \Sigma^i \wedge \Sigma^j + 2 \rho \delta^{ij} \right ) {\mathcal L}_v \Psi_{ij}  = - d \left ( \Sigma_i \wedge {\mathcal L}_v A^i - v \cdot L \right ),
\end{eqnarray}
where $v \cdot L$ stands for the contraction of $v$ with the Lagrangian 4-form $L$ and ${\mathcal L}_v$ is the Lie derivative along the vector field $v$.

\subsubsection{From Plebanski's to Einstein's equations.}\label{Pleb-to-Eins}

It is instructive to see how the field equations of the Plebanski formulation, together with suitable conditions and hypotheses, get related to Einstein's equations of general relativity.  The point of departure is to determine the geometrical meaning of the 2-forms~$\Sigma^i$. 

From equation (\ref{pleban1}) we can obtain $\rho=-(1/6)\Sigma^k \wedge \Sigma_k$, which substituted back into (\ref{pleban1}) yields
\begin{eqnarray}
\Sigma^i \wedge \Sigma^j - \frac13 \delta^{ij} \Sigma^k \wedge \Sigma_k=0. \label{simplicityconstraint}
\end{eqnarray}
This relation is known as the \textit{simplicity constraint} and it implies, for nonvanishing\footnote{The case of vanishing $\Sigma^k \wedge \Sigma_k$ is addressed in~\cite{REISENBERGER1995643}.} $\Sigma^k \wedge \Sigma_k$, the solutions
\begin{eqnarray}
\Sigma^i  = \rmi e^0 \wedge e^i - \frac{1}{2} \varepsilon^i{}_{jk} e^j \wedge e^k,  \label{pleb-sigma} 
\end{eqnarray}
where $e^I=\{e^0,e^i\}$ ($I=0,1,2,3$) is a set of four linearly independent and nondegenerate (complex) 1-forms, and $\rmi$ denotes the imaginary unit. Notice that $\Sigma^i$ is defined up to a rescaling $\Sigma^i\to\Omega \Sigma^i$, for $\Omega\in\mathbb{C}$, and $SO(3,\mathbb{C})$ rotations $\Sigma^i\to O^i{}_j \Sigma^j$, where $O\in SO(3,\mathbb{C})$. To recover general relativity in the Lorentzian signature, the simplicity constraint must be supplemented with the following \textit{reality conditions}~\cite{pleb1977118}
\begin{eqnarray}
	\Sigma^{i} \wedge {\overline \Sigma}{}^{j} = 0, \label{reality1}\\
	\Sigma_{i} \wedge \Sigma^{i} + {\overline \Sigma}{}_{i} \wedge {\overline \Sigma}{}^{i} =0, \label{reality2}
\end{eqnarray}
where ${\overline \Sigma}{}^{i}$ is the complex conjugate of $\Sigma^{i}$. With these conditions, the set of 1-forms $e^I$ in (\ref{pleb-sigma}) can be chosen to be real, which implies that ${\overline \Sigma}{}^{i}$ is also solution of (\ref{simplicityconstraint}). In the case of Euclidean signature, all the variables of the action (\ref{plebanAct}) are $\mathfrak{so}(3,\mathbb{R})$-valued, and hence no reality conditions are needed. This means that the solution for $\Sigma^i$, which has to be real, is given by (\ref{pleb-sigma}) with $\rmi$ replaced by $1$ (or $-1$) and real 1-forms $e^I$. From now on, we will focus on the Lorentzian case.

To better understand the role of the reality conditions (\ref{reality1}) and (\ref{reality2}), we briefly discuss them in the following lines. First, let us recall that on a four-dimensional manifold $\mathcal{M}$, the six-dimensional space $\Lambda^2(\mathcal{M})$ of 2-forms can be decomposed into two three-dimensional complex subspaces, ${}^+\Lambda^2(\mathcal{M})$ and ${}^-\Lambda^2(\mathcal{M})$, so that elements belonging to different subspaces are wedge orthogonal; that is, ${}^+\Lambda^2(\mathcal{M})\wedge{}^-\Lambda^2(\mathcal{M})=0$. These subspaces ${}^+\Lambda^2(\mathcal{M})$ and ${}^-\Lambda^2(\mathcal{M})$ are, respectively, the spaces of self-dual and anti-self-dual 2-forms.  According to this, equation~(\ref{reality1}) implies that the 2-forms $\Sigma^{i}$ and ${\overline \Sigma}{}^{i}$ belong to different subspaces of $\Lambda^2(\mathcal{M})$. In other words, if $\Sigma^{i}\in {}^\pm\Lambda^2(\mathcal{M}$), then ${\overline \Sigma}{}^{i}\in {}^\mp\Lambda^2(\mathcal{M}$).  However, if we take into account the fact that $\Sigma^{i}$ and ${\overline \Sigma}{}^{ i}$ are complex conjugate to each other, then equation~(\ref{reality1}) entails that the Urbantke metric in terms of $\Sigma^i$~\cite{urba198425} (see also~\cite{capo1991841}),
\begin{eqnarray}
	\tilde{g}_{\mu\nu}:=\varepsilon_{ijk} {\widetilde \eta}^{\alpha\beta\gamma\delta} \Sigma_{\mu\alpha}{}^i \Sigma_{\beta\gamma}{}^j \Sigma_{\delta\nu}{}^k, \label{urbametric}
\end{eqnarray}
is conformally related to a real Lorentzian metric\footnote{Two metrics, $g_{\mu\nu}$ and $h_{\mu\nu}$, are conformally related if $g_{\mu\nu}=\xi h_{\mu\nu}$, where $\xi$ is the conformal factor.}~\cite{capo1991841,beng199512}. Here ${\widetilde \eta}^{\alpha\beta\gamma\delta}$ is a totally antisymmetric density of weight~1 (with ${\widetilde\eta}^{0123}=+1$) and we have written $\Sigma^i=(1/2)\Sigma_{\mu\nu}{}^i dx^\mu \wedge dx^\nu$. Recall that the Urbantke metric is the metric with respect to which $\Sigma^{i}$ is (anti-)self-dual. In turn, the second condition, equation~(\ref{reality2}), indicates that the volume 4-form  $\rmi \Sigma_{i} \wedge \Sigma^i$ is real, implying then that the resulting conformal factor is purely imaginary. Concretely, using (\ref{pleb-sigma}), the Urbantke metric becomes $\tilde{g}_{\mu\nu}=12\rmi e \eta_{IJ} \tensor{e}{_{\mu}^I} \tensor{e}{_{\nu}^J}$, where $e:=\det(\tensor{e}{_{\mu}^I})$ and $(\eta_{IJ})= \mbox{diag} (-1,1,1,1)$ is the  Minkowski metric. In the case of ${\overline \Sigma}{}^i$, this metric takes the form $\tilde{g}_{\mu\nu}=-12\rmi e \eta_{IJ} \tensor{e}{_{\mu}^I} \tensor{e}{_{\nu}^J}$. Then, both $\Sigma^i$ and ${\overline \Sigma}{}^i$ lead to a metric conformally related to $g_{\mu\nu}:=\eta_{IJ} \tensor{e}{_{\mu}^I} \tensor{e}{_{\nu}^J}$, but with (purely imaginary) conformal factors that differ by a sign. Since the Hodge dual operator of 2-forms is independent of the conformal factors of the metrics\footnote{The converse statement is also true~\cite{dray1989-6-30}, that is to say, two metrics defining the same Hodge dual operator of 2-forms are conformally related.}, we can consider without loss of generality, that the Urbantke metric defined by  $\Sigma^i$ (or ${\overline \Sigma}{}^{i}$) is simply $g_{\mu\nu}$. Then, by using $g_{\mu\nu}$ to define the Hodge dual operator $\ast$, it can be verified that 
\begin{eqnarray}
	\ast \Sigma^{i}  =  \rmi \Sigma^{i}, \qquad     \ast {\overline \Sigma}{}^{i}  = - \rmi {\overline \Sigma}{}^{i}, \label{pleb:dual}
\end{eqnarray}
which states that $\Sigma^{i}$ is self-dual and ${\overline \Sigma}{}^{i}$ is anti-self-dual, respectively. Actually, the 2-forms $\Sigma^{i}$ and ${\overline \Sigma}{}^{i}$ provide an orthogonal basis for the three-dimensional spaces ${}^+\Lambda^2(\mathcal{M})$ and ${}^-\Lambda^2(\mathcal{M})$, respectively.

The next step is to obtain the geometrical meaning of the internal connection $A^i$. This is carried out by solving the equation of motion~(\ref{pleban2}), which states that $A^i$ can be uniquely determined in terms of $\Sigma^i$. Indeed, equation~(\ref{pleban2}) is a linear system of 12  equations and 12 unknowns $A_J{}^i$ ($A^i:=A_J{}^i e^J$) that  is nondegenerate for $\Sigma^i$ given by (\ref{pleb-sigma}). Solving~(\ref{pleban2}) for  $A^i$ we have~\cite{gmv201204}
\begin{eqnarray}
	A^i =  \frac{\rmi}{12} \left(  -\delta^i{}_j \eta_{LN} + \varepsilon^i{}_{jk} \Sigma_{LN}{}^{k} \right) Q_{IJK}{}^{j}  \varepsilon^{IJKL} e^N, \label{plebconn}
\end{eqnarray}
where $d\Sigma^i=(1/3!) Q_{IJK}{}^i e^I \wedge e^J \wedge e^K$ and $\varepsilon^{0ijk}=-\varepsilon^{ijk}$ ($\varepsilon^{IJKL}$ is totally antisymmetric). It is perhaps remarkable that $A^i$ does not require additional reality conditions, since it is constructed from $\Sigma^i$, which already satisfies the reality conditions. Similarly, since both $\rho$ and $\Psi$ can be expressed, on-shell, in terms of $\Sigma^i$, then we do not need to impose new reality conditions on them. If we take into account that  (\ref{pleb-sigma}) satisfies the following relation
\begin{eqnarray}
	\Sigma_{IK}{}^{i}\Sigma^{K}{}_J{}^j = -\delta^{ij} \eta_{IJ} + \varepsilon^{ij}{}_{k} \Sigma_{IJ}{}^{k}, \label{pleb:quater}
\end{eqnarray}
where $\Sigma^i=(1/2)\Sigma_{IJ}{}^{i}e^I\wedge e^J$ and the indices $I,J,\dots$ are raised and lowered with $\eta_{IJ}$, then we can rewrite~(\ref{plebconn}) in a more compact form 
\begin{eqnarray}
	A^i =  \frac{\rmi}{12}  Q_{IJK}{}^{j}  \varepsilon^{IJKL} \Sigma_{LM}{}^{i}\Sigma^{M}{}_N{}_{j}  e^N. \label{plebconn-simply}
\end{eqnarray}

Now, we have to relate the internal connection $A^i$ to a spacetime connection $\omega^I{}_J$. However, this requires further assumptions. Recall that a spacetime connection is defined by specifying its torsion and the spacetime metric, and the Plebanski formulation, being a genuine gauge theory, does not involve these concepts. Therefore, to link both connections, we make the following assumptions: $\omega^I{}_J$ is torsion-free, and the spacetime metric is given by the Urbantke metric. Specifically, the  spacetime connection is defined by
\begin{eqnarray}
	d e^I + \omega^I{}_J \wedge e^J = 0, \label{torsion} \\
	d \eta_{IJ} - \omega^K{}_I \eta_{KJ}- \omega^K{}_J \eta_{IK} = 0. \label{nometricidad}
\end{eqnarray}
Using (\ref{torsion}) and (\ref{nometricidad}), it can be verified that the internal connection (\ref{plebconn-simply}), in terms  of $\omega^I{}_J$, becomes
\begin{eqnarray}
	A^i =\rmi \omega^i{}_0 + \frac12 \varepsilon^{ij}{}_k \omega^k{}_j, \label{plebconnection} 
\end{eqnarray}
showing that $A^i$ is the self-dual part of the spin-connection $\omega^I{}_J$. 

To proceed further, we need to compute the curvature  of the connection (\ref{plebconnection}), which yields to 
\begin{eqnarray}
	F^i =  \rmi R^i{}_0+\frac{1}{2}\varepsilon^{ij}{}_{k} R^k{}_j, \label{plebcurvature}
\end{eqnarray}
where $R^I{}_J=d\omega^I{}_J+\omega^I{}_K\wedge\omega^K{}_J$ is the curvature of $\omega^I{}_J$. This equation means that $F^i$ is nothing but the self-dual part of $R^I{}_J$. For our purposes, it is convenient to rewrite the right-hand side of (\ref{plebcurvature}) as
\begin{eqnarray}
\fl F^i=\frac{1}{2} \left[ \left( F^i{}_j + E^i{}_j \right) - \rmi \left( Q^i{}_j +  H^i{}_j \right) \right] \Sigma^j + \frac{1}{2} \left[\left( F^i{}_j - E^i{}_j \right) + \rmi \left( Q^i{}_j -  H^i{}_j \right)\right] {\overline \Sigma}{}^j,  \label{plebcurvbasis}
\end{eqnarray}
where $F$, $E$, $Q$, and $H$ are real $3\times3$ matrices  associated with the 36 components of the curvature\footnote{Recall that for the curvature the internal indices are written to the left of the spacetime ones  to agree with the usual convention for the Riemann tensor.}, $R_{IJ}=(1/2) R_{IJKL} e^K\wedge e^L$, via the following relations: $E^i{}_j=R^{0i}{}_{0j}$, $H^i{}_j=(1/2) \varepsilon_j{}^{kl} R^{0i}{}_{kl}$, $Q^i{}_j=-(1/2) \varepsilon^i{}_{kl} R^{kl}{}_{0j}$, and $F^i{}_j=(1/4) \varepsilon^i{}_{kl} \varepsilon_j{}^{mn}  R^{kl}{}_{mn}$.  Let us note that because of the first Bianchi identities, $R^I{}_J \wedge e^J =0$, these matrices have the properties $E= E^T$, $F= F^T$, $H= Q^T$, and ${\rm Tr}H= 0$. Substituting (\ref{plebcurvbasis}) into the left-hand side of equations~(\ref{pleban3}), we obtain
\begin{eqnarray}
	E= F =E^T, \label{relationEF}\\
	H= Q= H^T, \label{relationQH}\\
	E-\rmi H = \Psi+\frac{\Lambda}{3} I. \label{relationEH}
\end{eqnarray}
Furthermore, by taking the trace of equation~(\ref{relationEH}) and using equation~(\ref{pleban4}), we obtain
\begin{eqnarray}
	{\rm Tr}E=\Lambda,\qquad {\rm Tr}H=0. \label{Pleb:TrEH}
\end{eqnarray}

Remarkably, equations~(\ref{relationEF}), (\ref{relationQH}), and (\ref{Pleb:TrEH}) are exactly the set of equations  resulting from writing Einstein's equations in terms of $E$, $F$, $H$, and  $Q$. It can also be verified that these equations are insensitive to the specific sector of 2-forms, $\Sigma^i$ or ${\overline \Sigma}{}^i$, that we use from the very beginning. This signifies that, to recover general relativity from Plebanski's equations, it is enough to know either the  self-dual or anti-self dual sector only~\cite{pleb1977118}. 

Alternatively, by substituting the solution (\ref{pleb-sigma}) in  Plebanski's action (\ref{plebanAct}), we can obtain the self-dual Palatini action~\cite{Samuel1987,Jacobson198739,Jacobsoncqg5}. Indeed, using (\ref{pleb-sigma}) (bear in mind (\ref{pleban1})) with $e^I$ real (which means that $\Sigma^i$ satisfies the reality conditions), equation~(\ref{plebanAct}) reduces to
\begin{equation}
\fl S[A,e]=\int_{\mathcal{M}}\left[\frac{1}{2}\Sigma_{IJi}e^I\wedge e^J\wedge F^i[A]+\frac{\mathrm{i}\Lambda}{24}\varepsilon_{IJKL}e^I\wedge e^J\wedge e^K\wedge e^L\right].\label{PlebPalt1}	
\end{equation}
Now, let us introduce a real internal Lorentz connection $\Gamma^{IJ}$ such that $A_i:=(1/2)\Sigma_{IJi}\Gamma^{IJ}$, that is, $A^i$ is the self-dual part of $\Gamma^{IJ}$. Then, $F^i$ is the self-dual part of the curvature $R^{IJ}$ of $\Gamma^{IJ}$, and using $\tensor{\Sigma}{_{IJ}^i}\tensor{\Sigma}{^{KL}_i}=\delta_I^K\delta_J^L-\delta_I^L\delta_J^K-\mathrm{i}\tensor{\varepsilon}{_{IJ}^{KL}}$, (\ref{PlebPalt1}) reads
\begin{equation}
\fl S[e,{}^{\scriptscriptstyle (+)}\Gamma]=\int_{\mathcal{M}}\left\{e_I\wedge e_J\wedge R^{IJ}[{}^{\scriptscriptstyle (+)}\Gamma]+\frac{\mathrm{i}\Lambda}{24}\varepsilon_{IJKL}e^I\wedge e^J\wedge e^K\wedge e^L\right\},\label{PlebPalt2}
\end{equation}
where ${}^{\scriptscriptstyle (+)}\Gamma^{IJ}:={}^{\scriptscriptstyle (+)}\mathcal{P}\Gamma^{IJ}$, with ${}^{\scriptscriptstyle (+)}\mathcal{P}:=(1-\mathrm{i}*)/2$ being the projector on the self-dual component of the Lorentz algebra. The action (\ref{PlebPalt2}) is, up to a global factor, the self-dual Palatini action with a cosmological constant. Notice that the equation of motion for $\Gamma^{IJ}$ leads to the torsion-free condition and hence it becomes the spin-connection $\omega^{IJ}$. This result, together with the equation of motion corresponding to $e^I$, imply Einstein's equations.

\subsubsection{Weyl curvature tensor.}

Plebanski's equations also provide information about the Weyl curvature tensor. First, let us recall that the Weyl tensor $C_{IJKL}$ is constructed in such a way that it only contains the trace-free part of the Riemann tensor; that is
\begin{eqnarray}
	C_{IJKL}&=&R_{IJKL}-S_{IJKL} + U_{IJKL},\label{weyltensor}
\end{eqnarray}
with
\begin{eqnarray}
	S_{IJKL}&:=&\frac{1}{2} \left(\eta_{IK} \mathcal{R}_{JL}-\eta_{JK} \mathcal{R}_{IL}+\eta_{JL} \mathcal{R}_{IK}-\eta_{IL} \mathcal{R}_{JK} \right), \label{weyl-defS}\\
	U_{IJKL}&:=&\frac{1}{6} \mathcal{R} \left(\eta_{IK}\eta_{JL}- \eta_{IL}\eta_{JK}  \right),\label{weyl-defU}
\end{eqnarray}
where $\mathcal{R}_{JL}:=\eta^{IK}R_{IJKL}$ is the Ricci tensor and $\mathcal{R}:= \eta^{IJ}\mathcal{R}_{IJ}$ is the  scalar curvature. It turns out that the scalar field $\Psi^{ij}$ is the self-dual part of $C_{IJKL}$, namely $ \Sigma^{IJi}\Sigma^{KLj} C_{IJKL}$. This statement is shown in what follows. Taking into account equations~(\ref{pleban3}), (\ref{pleb:quater}) and (\ref{plebcurvature}), and the symmetry of the Ricci tensor by virtue of the first Bianchi identities, the  self-dual part of each term on the right-hand side of  (\ref{weyltensor}) yields
\begin{eqnarray}
	\Sigma^{IJi} \Sigma^{KLj} R_{IJKL}&=& 8\left(\Psi^{ij}+\frac{\Lambda}{3} \delta^{ij} \right), \label{pleb:PPRiem2}\\
	\Sigma^{IJi} \Sigma^{KLj} S_{IJKL}&=& 2\delta^{ij} \mathcal{R}, \label{pleb:PPS2}\\
	\Sigma^{IJi} \Sigma^{KLj} U_{IJKL}&=& \frac{4}{3} \delta^{ij} \mathcal{R}. \label{pleb:PPU2}
\end{eqnarray}
Collecting these results together and using $\mathcal{R}=4\Lambda$, which comes from  Einstein's equations,  we finally find
\begin{eqnarray}
	\Sigma^{IJi}\Sigma^{KLj} C_{IJKL}= 8\Psi^{ij}, \label{selfdualWeyl}
\end{eqnarray}
showing that $\Psi^{ij}$ is, on-shell, the self-dual part of the Weyl curvature tensor.

\subsubsection{Canonical analysis.}\label{Pleb-HamilAna}

We now turn to the canonical analysis of Plebanski's action~(\ref{plebanAct}). We consider that the four-dimensional manifold $\mathcal{M}$ has a topology $\mathbb{R}\times\Omega$,  where $\Omega$ is a spatial three-dimensional manifold without a boundary. We then choose a global time function $t$ such that the hyper-surfaces $t=$ constant, which have the topology of $\Omega$, foliate spacetime $\mathcal{M}$. Let $x^a$ ($a=1,2,3$) be the coordinates on $\Omega$; the action~(\ref{plebanAct}) takes the form
\begin{eqnarray}
\fl S[A, \Sigma,\Psi,\rho]=\int_{\mathbb{R}}dt\int_{\Omega}d^3x \left[\tilde{\Pi}^{ai} \dot{A}_{ai} +A_{ti} \tilde{\mathcal{G}}^i +  \Sigma_{tai} \tilde{E}^{ai} - \Psi_{ij} \left( \Sigma_{ta}{}^i\tilde{\Pi}^{aj} + \tilde{\rho} \delta^{ij} \right)\right],\label{PlebH:Hamil1}
\end{eqnarray}
where we have defined $\tilde{\rho}d^4x:=\rho$, $\tilde{\Pi}^{ai}:=(1/2)\tilde{\eta}^{abc}\Sigma_{bc}{}^i$, $\tilde{E}^{ai}=\tilde{B}^{ai}-(\Lambda/3)\tilde{\Pi}^{ai}$ with $\tilde{B}^{ai}:=(1/2)\tilde{\eta}^{abc}F^i{}_{bc}$, and $\tilde{\mathcal{G}}^i:=\mathcal{D}_a\tilde{\Pi}^{ai}$. Here, $\mathcal{D}_a$ is the $SO(3,\mathbb{C})$-covariant derivative corresponding to the spatial connection $A_a{}^i$, a dot over a variable stands for its time derivative, and $\tilde{\eta}^{abc}$ ($\underaccent{\tilde}{\eta}_{abc}$)  is a totally antisymmetric tensor density of weigh 1~($-1$),  so that $\tilde{\eta}^{123}=1$~($\underaccent{\tilde}{\eta}_{123}=1$).

Following~\cite{capo1991841} (see also~\cite{aip_paper}), the analysis can be simplified considerably by using the equation of motion for $\Psi_{ij} $, namely
\begin{eqnarray}
\delta \Psi_{ij}:  \Sigma_{ta}{}^i\tilde{\Pi}^{aj}+\Sigma_{ta}{}^j\tilde{\Pi}^{ai} + 2\tilde{\rho} \delta^{ij}=0, \label{PlebH:constPsi}
\end{eqnarray}
to express the Lagrange multiplier $\Sigma_{tai}$ in terms of $\tilde{\Pi}^{ai}$ and $\tilde{\rho}$, and then get rid of this variable from the action. Indeed, equation~(\ref{PlebH:constPsi}) constitutes a system of 6 equations for the 9 unknowns $\Sigma_{ta}{}^i$, whose general solution is
\begin{eqnarray}
\Sigma_{tai}=\left(-\tilde{\rho} \delta_{ij} + \varepsilon_{ijk} \tilde{N}^k  \right) \underaccent{\tilde}{\Pi}_a{}^j, \label{PlebH:constPsisol}
\end{eqnarray}
where $\tilde{N}^k$ is an arbitrary internal 3-vector and we have assumed that $\tilde{\Pi}^{aj}$ is invertible, with inverse given by $\underaccent{\tilde}{\Pi}_{aj}$; that is, $\underaccent{\tilde}{\Pi}_{ai}\tilde{\Pi}^{aj}=\delta^j_i$ and $\underaccent{\tilde}{\Pi}_{ai}\tilde{\Pi}^{bi}=\delta^b_a$ (the case of degenerate $\tilde{\Pi}^{ai}$ is treated in~\cite{REISENBERGER1995643}). By substituting (\ref{PlebH:constPsisol}) into (\ref{PlebH:Hamil1}), and after some algebra, the Hamiltonian form of Plebanski's action turns out to be
\begin{eqnarray}
\fl S[A_{a},\tilde{\Pi}^{a},A_{t},N^a,\underaccent{\tilde}{N}]=\int_{\mathbb{R}}dt\int_{\Omega}d^3x\left(\tilde{\Pi}^{ai}\dot{A}_{ai}+A_{ti}\tilde{\mathcal{G}}^i+N^a\tilde{\mathcal{V}}_a+\underaccent{\tilde}{N}\tilde{\tilde{\mathcal{H}}}\right),\label{PlebH:hamil}
\end{eqnarray}
where $A_{ti}$, $N^a:=-(\det\tilde{\Pi})^{-1}\tilde{\Pi}^a{}_i \tilde{N}^i$, and $\underaccent{\tilde}{N}:=-3(\det\tilde{\Pi})^{-1}\tilde{\rho}$ appear linearly in (\ref{PlebH:hamil}), and thus play the role of Lagrange multipliers imposing the constraints
	\begin{eqnarray}
	\tilde{\mathcal{G}}^i&=&\mathcal{D}_a\tilde{\Pi}^{ai}\approx 0,\label{PlebH:gauss}\\
	\tilde{\mathcal{V}}_a&:=&\tilde{\Pi}^{bi}F_{iba}\approx 0,\label{PlebH:Vect}\\
	\tilde{\tilde{\mathcal{H}}}&:=&\frac{1}{6}\underaccent{\tilde}{\eta}_{abc}\varepsilon_{ijk}\tilde{\Pi}^{ai}\tilde{\Pi}^{bj}\tilde{B}^{ck}-\frac{\Lambda}{18} \underaccent{\tilde}{\eta}_{abc}\varepsilon_{ijk}\tilde{\Pi}^{ai}\tilde{\Pi}^{bj}\tilde{\Pi}^{ck}\approx 0.\label{PlebH:scalar}
	\end{eqnarray}
The constraint algebra reads
\begin{eqnarray}
\fl	\{\tilde{\mathcal{G}}^i(x),\tilde{\mathcal{G}}^j(y)\}&=&\tensor{\varepsilon}{^{ij}_k}\tilde{\mathcal{G}}^k\delta^3(x,y),\label{GGPleb}\\ 
\fl	\{\tilde{\mathcal{G}}^i(x),\tilde{\mathcal{V}}_a(y)\}&=&0,\label{GVPleb}\\
\fl	\{\tilde{\mathcal{G}}^i(x),\tilde{\tilde{\mathcal{H}}}(y)\}&=&0,\label{GHPleb}\\
\fl	\{\tilde{\mathcal{V}}_a(x),\tilde{\mathcal{V}}_b(y)\}&=&\left[\tilde{\mathcal{V}}_a(y)\frac{\partial}{\partial y^b}-\tilde{\mathcal{V}}_b(x)\frac{\partial}{\partial x^a}-F_{iab}\tilde{\mathcal{G}}^i\right]\delta^3(x,y),\label{VVPleb}\\
\fl	\{\tilde{\mathcal{V}}_a(x),\tilde{\tilde{\mathcal{H}}}(y)\}&=&\left[\tilde{\tilde{\mathcal{H}}}(y)\frac{\partial}{\partial y^a}-\tilde{\tilde{\mathcal{H}}}(x)\frac{\partial}{\partial x^a}-\frac{1}{3}\underaccent{\tilde}{\eta}_{abc}\varepsilon_{ijk}\tilde{\Pi}^{bj}\left(\tilde{B}^{ck}-\frac{\Lambda}{2}\tilde{\Pi}^{ck}\right)\tilde{\mathcal{G}}^i\right]\delta^3(x,y),\label{VHPleb}\nonumber\\
\fl \\
\fl	\{\tilde{\tilde{\mathcal{H}}}(x),\tilde{\tilde{\mathcal{H}}}(y)\}&=&\frac{1}{9}\left[\tilde{\tilde{h}}^{ab}(x)\tilde{\mathcal{V}}_a(x)\frac{\partial}{\partial x^b}-\tilde{\tilde{h}}^{ab}(y)\tilde{\mathcal{V}}_a(y)\frac{\partial}{\partial y^b}\right]\delta^3(x,y),\label{HHPleb}
\end{eqnarray}
where $\tilde{\tilde{h}}^{ab}:=\tilde{\Pi}^{ai}\tensor{\tilde{\Pi}}{^b_i}$. Since the algebra closes, the evolution of the constraints is trivial and no new constraints arise. This means that $\tilde{\mathcal{G}}^i$, $\tilde{\mathcal{V}}_a$, and $\tilde{\tilde{\mathcal{H}}}$, which are known as the Gauss, vector, and scalar constraints, respectively, are first class. In brief, we ended up with a phase space parametrized by the nine canonical pairs $(A_{ai},\tilde{\Pi}^{ai})$ subject to seven first-class constraints, implying that the theory propagates two (complex) physical degrees of freedom, the same as complex general relativity (these degrees of freedom become real once the reality conditions are taken into account).  In addition, the aforementioned constraints generate the gauge symmetries of the theory; namely, $\tilde{\mathcal{G}}^i$ generates local $SO(3,\mathbb{C})$ rotations, whereas $\tilde{\mathcal{V}}_a$ and $\tilde{\tilde{\mathcal{H}}}$ generate correspondingly spatial and time diffeomorphisms. It is worth pointing out that the true generator of spatial diffeomorphisms is given by  $\tilde{\mathcal{D}}_a:=\tilde{\mathcal{V}}_a+A_{ai} \tilde{\mathcal{G}}^i$, which is called diffeomorphism constraint.

The constraints (\ref{PlebH:gauss})-(\ref{PlebH:scalar}) are the same as those of the Ashtekar formulation of general relativity, which were originally found through a complex canonical transformation on the constraints of the ADM formulation~\cite{ashtekar1986-57-2244,ashtekar1987-36-1587}. Consequently, the Ashtekar formulation also results from the canonical analysis of the Plebanski formulation.

\subsubsection{Matter fields.} \label{pleb-matter}

Our discussion will now be focused on the coupling of matter to Plebanski's action (\ref{plebanAct}), which has been carried out in~\cite{capo1991841} in the spinorial approach. We shall see that the matter fields can be coupled to gravity through the 2-forms $\Sigma^i$, showing that, instead of the spacetime metric, we can actually take them as the fundamental gravitational variables. In this section, we consider the coupling of the scalar and the Yang-Mills field in the $SO(3,\mathbb{C})$ formalism.

We begin with the coupling of a real scalar field $\phi$ to the action (\ref{plebanAct}),  which we now rename as $S_{\rm Ple}[A, \Sigma,\Psi,\rho]$. The action is given by
\begin{eqnarray}
\fl S[A, \Sigma,\Psi,\rho,\pi,\phi]=& S_{\rm Ple}[A, \Sigma,\Psi,\rho] \nonumber\\
&+\int_{\mathcal{M}}\left[ \rmi \Sigma_i \wedge \Sigma^i (a\pi^{\mu}\partial_{\mu}\phi + b V(\phi))+\alpha \rmi \tilde{g}_{\mu\nu}\pi^{\mu}\pi^{\nu}d^4 x\right], \label{PlebM:scalaraction}
\end{eqnarray}
where $\pi^{\mu}$ are auxiliary fields, $V(\phi)$ is the scalar field potential, $\tilde{g}_{\mu\nu}$ is the Urbantke metric defined in (\ref{urbametric}), and $a$,  $b$, and $\alpha$ are constants. Notice that the coupling terms added are quadratic and cubic in $\Sigma^i$, and hence the action is polynomial in this variable. Now, since the equations of motion for $\Psi$, $A$, and $\rho$ are not modified, the solution for $\Sigma^i$ in terms of the frame $e^I$ is still given by equation~(\ref{pleb-sigma}) and $A^i$ is the self-dual part of the spin-connection compatible the metric $g_{\mu\nu}:=\eta_{IJ} \tensor{e}{_{\mu}^I} \tensor{e}{_{\nu}^J}$. A direct way to verify that (\ref{PlebM:scalaraction}) describes general relativity coupled to a scalar field is by expressing it in terms of the tetrad field. The equation of motion for $\pi^{\mu}$ is
\begin{eqnarray}
\delta \pi^\mu:  a \Sigma_i \wedge \Sigma^i \partial_{\mu}\phi+2 \alpha \tilde{g}_{\mu\nu}\pi^{\nu}d^4 x=0, \label{PlebM:scalar}
\end{eqnarray}
whose solution for $\pi^{\mu}$ reads
\begin{eqnarray}
\pi^\mu= \frac{a}{4 \alpha} g^{\mu\nu} \partial_{\nu}\phi, \label{PlebM:pi}
\end{eqnarray}
where we have used equation~(\ref{pleb-sigma}), which leads to $\tilde{g}_{\mu\nu}=
12\rmi e g_{\mu\nu}$ with $e:=\det(\tensor{e}{_{\mu}^I})$. Then, substituting equations~(\ref{pleb-sigma}) and (\ref{PlebM:pi}) into (\ref{PlebM:scalaraction}) yields
\begin{eqnarray}
\fl S[\phi,e,{}^{\scriptscriptstyle (+)}\Gamma]=S[e,{}^{\scriptscriptstyle (+)}\Gamma] + \frac{3}{2}\int_{\mathcal{M}}d^4x\sqrt{-g}\left(\frac{a^2}{2\alpha} g^{\mu\nu}\partial_{\mu}\phi \, \partial_{\nu}\phi + 4 b V(\phi)\right),
\end{eqnarray}
where ${}^{\scriptscriptstyle (+)}\Gamma^{IJ}=(1/2)\Sigma^{IJ}{}_i A^i$ and $S[e,{}^{\scriptscriptstyle (+)}\Gamma]$ is given by equation~(\ref{PlebPalt2}). Therefore, we recover the self-dual Palatini action coupled to a scalar field with potential $V(\phi)$.

Now we move to the coupling of the Yang-Mills field, which can be accomplished by considering the action
\begin{eqnarray}
\fl S[A, \Sigma,\Psi,\rho,\mathcal{A},\phi]= S_{\rm Ple}[A, \Sigma,\Psi,\rho]
+\int_{\mathcal{M}}\left[\beta_1 \mathcal{F}^a[\mathcal{A}]\wedge \Sigma^i \phi_{ai} - \frac{\beta_2}{2}  \Sigma^i \wedge \Sigma^j \phi^a{}_{i} \phi_{aj} \right], \nonumber\\ \label{PlebM:YMcoupling}
\end{eqnarray}
where $\mathcal{F}[\mathcal{A}]:=d\mathcal{A}+\mathcal{A}\wedge\mathcal{A}$ is the field strength  of the Yang-Mills connection 1-form $\mathcal{A}$, $\tensor{\phi}{^a_i}$ are auxiliary fields taking values in Yang-Mills group's Lie algebra, and $\beta_1$ and $\beta_2$ are parameters. The internal index $a$ is raised and lowered with the nondegenerate inner product of the group Lie algebra.  Note that again the action is polynomial in $\Sigma^i$ and that the equations of motion for $\Psi$, $A$, and $\rho$ are still given by equations~(\ref{pleban1}), (\ref{pleban2}), and (\ref{pleban4}), respectively. To verify the equivalence of (\ref{PlebM:YMcoupling}) and general relativity we shall follow a procedure analogous to that of the scalar field case. The equation of motion for $\tensor{\phi}{^a_i}$ is 
\begin{eqnarray}
\delta \tensor{\phi}{^a_i}:  \beta_1 \mathcal{F}_a\wedge \Sigma^i - \beta_2 \Sigma^i \wedge \Sigma^j  \phi_{aj}=0, \label{PlebM:YMphii}
\end{eqnarray}
which we want to solve for $\tensor{\phi}{^a_i}$. Expanding the Yang-Mills field strength in the basis of 2-forms, namely $\mathcal{F}^a={}^{\scriptscriptstyle (+)}\mathcal{F}^a{}_i \Sigma^i + {}^{\scriptscriptstyle (-)}\mathcal{F}^a{}_i {\overline \Sigma}{}^{i}$, the solution of this equation can be written as
\begin{eqnarray}
\tensor{\phi}{^a_i} = \frac{\beta_1}{\beta_2} {}^{\scriptscriptstyle (+)}\mathcal{F}^a{}_i.
\end{eqnarray}
Putting this and~(\ref{pleb-sigma}) back into the action (\ref{PlebM:YMcoupling}), and using ${}^{\scriptscriptstyle (+)}\mathcal{F}^a{}_i \Sigma^i={}^{\scriptscriptstyle (+)}\mathcal{P} \mathcal{F}^a$, we find
\begin{eqnarray}
S[\mathcal{A},e,{}^{\scriptscriptstyle (+)}\Gamma]=S[e,{}^{\scriptscriptstyle (+)}\Gamma] + \frac{(\beta_1)^2}{4\beta_2} \int_{\mathcal{M}}  \left( \mathcal{F}^a\wedge \mathcal{F}_a - \rmi \mathcal{F}^a\wedge \ast \mathcal{F}_a \right) .
\end{eqnarray}
where ${}^{\scriptscriptstyle (+)}\Gamma^{IJ}$ and $S[e,{}^{\scriptscriptstyle (+)}\Gamma]$ are the same  as above. Since the Pontrjagin term does not modify the dynamics, we conclude that this is the self-dual Palatini action coupled to Yang-Mills theory.

\paragraph{\bf Remark.} The coupling of spin 1/2 and 3/2 fermions to Plebanski's action has been performed in~\cite{capo1991841} using the spinorial approach. In the case of the $SO(3,\mathbb{C})$ formalism, the coupling has not been attained yet.

\subsubsection{Gauge connection formulations from Plebanski formulation.}\label{ple-to-kras} 

By eliminating auxiliary fields in Plebanski's action, it is possible to derive interesting alternative formulations of general relativity. Such is the case of the CDJ formulation~\cite{capo198963,capo1991859}, the pure connection formulation~\cite{kras2011106}, and the gauge connection formulations of~\cite{gm201591024021}. The aim of this section is to review how these formulations emerge from the Plebanski formulation. Since the common variables involved in these formulations are the gauge connection and a nondynamical field (the action of~\cite{kras2011106} does not have this variable), we shall begin by following a general procedure, based on that described in~\cite{cgm-pure-2015}, which consists in eliminating $\Sigma^i$ and $\Psi^{ij}$ from (\ref{plebanAct}) to get an action $S[A,\rho]$ that will be analyzed separately in each case.  Of course, the pure connection formulation requires to eliminate $\rho$, thus leaving the $SO(3,\mathbb{C})$-connection $A^i$ as the only dynamical variable in the final action. This approach to reach the pure connection action turns out to be more straightforward than the one followed in the original derivation~\cite{kras2011106}, which only addresses the Euclidean case. It is also worth mentioning that in the final step of the latter approach,  the real matrix $\Psi$  is integrated out by using a basis where it is diagonal; such a procedure, however, is not always possible in the Lorentzian case, since $\Psi$ is a complex matrix and might not be diagonalizable. Although we deal with the Lorentzian signature, our analysis applies also to the Euclidean case.

We start by assuming that the matrix $X:=\Psi+(\Lambda/3) I$ is nonsingular, which is usual in this type of formulations~\cite{capo1991859,capo199207,kras2011106}. Using this assumption,  equation~(\ref{pleban3}) implies $\Sigma^i=(X^{-1})^i{}_jF^j$, which substituted back into (\ref{plebanAct}) results in 
\begin{eqnarray}
	S[A,\Psi,\rho]&=&\int_\mathcal{M} \left[ \frac{1}{2} (X^{-1})_{ij}F^i\wedge F^j-\rho {\rm Tr}\Psi  \right]. \label{Act1-KRAS}
\end{eqnarray}
A remarkable fact of this action is that now $\Psi$ is an auxiliary field, in contrast to Plebanski's action where this field enters linearly. This fact will be used to eliminate this variable. The variation of~(\ref{Act1-KRAS}) with respect to $\Psi$ gives 
\begin{eqnarray}
\delta \Psi_{ij}: \tilde{M}+2\tilde{\rho}X^2=0, \label{motionPsi-KRAS}
\end{eqnarray}
where we have defined $\tilde{M}^{ij}d^4x:=F^i\wedge F^j$. Solving this matrix equation for $\Psi$, we obtain~\cite{cgm-pure-2015}
\begin{eqnarray}
\Psi=\frac{\rmi \epsilon}{\sqrt{2}} \tilde{\rho}^{-1/2} \tilde{M}^{1/2}-\frac{\Lambda}{3} I,  \label{Xsol-kras}
\end{eqnarray}
where $\tilde{M}^{1/2}$ is a symmetric square root of $\tilde{M}$; that is, $\tilde{M}^{1/2} \tilde{M}^{1/2}=\tilde{M}$, and $\epsilon=\pm1$ accounts for the branch of this square root. A few remarks are in order regarding equation~(\ref{Xsol-kras}). First, the existence of $\tilde{M}^{1/2}$ is guaranteed by the fact that $\tilde{M}$ is invertible~\cite{horn1994topics}, which in turn follows from equation~(\ref{motionPsi-KRAS}) and the fact that $X$ is nonsingular. Second, $\tilde{M}^{1/2}$ is not unique in general. In the particular case when $\tilde{M}$ is positive definite, $\tilde{M}^{1/2}$ is unique (see~\cite{zhangmatrixtheory}, theorem 3.5). We point out, however, that it is not necessary to restrict ourselves to such a particular case, since (i)~we might be excluding square roots satisfying (\ref{motionPsi-KRAS}) and (ii)~the procedure followed does not depend on any particular~$\tilde{M}^{1/2}$. Moreover, equation~(\ref{Xsol-kras}) involves choosing a $\tilde{M}^{1/2}$, which, because of the constraint imposed by $\rho$, is not completely arbitrary. In fact, if we take the trace of~(\ref{Xsol-kras}) and use $ {\rm Tr}\Psi=0$, which results from the variation of (\ref{Act1-KRAS}) with respect to $\rho$, we obtain 
\begin{eqnarray}
{\rm Tr}\tilde{M}^{1/2}= -  \sqrt{2} \rmi \epsilon \Lambda  \tilde{\rho}^{1/2}.\label{cond-1-M-mu}
\end{eqnarray}
Then, among all possible symmetric square roots $\tilde{M}^{1/2}$,  we have to consider only that
(or those) with trace given by (\ref{cond-1-M-mu}). It is worth noting that even in the case of Euclidean signature, where $\tilde{M}$ is a real symmetric matrix, the square root $\tilde{M}^{1/2}$ may not be unique. 

Continuing with the procedure, we now substitute~(\ref{Xsol-kras}) back into (\ref{Act1-KRAS}) to obtain~\cite{cgm-pure-2015}
\begin{eqnarray}
	S[A,\rho]=-\int_\mathcal{M} d^4x \left( \sqrt{2} \rmi \epsilon\tilde{\rho}^{1/2} {\rm Tr}\tilde{M}^{1/2}-\Lambda\tilde{\rho} \right) . \label{Amu-action-kras}
\end{eqnarray}
Notice that (\ref{Xsol-kras}) and (\ref{Amu-action-kras}) are {\it results}, not hypotheses as in~\cite{capo1991859}. At this point, it is important to keep in mind that both cases of  the cosmological constant, $\Lambda=0$ and 
$\Lambda\neq0$, are supported by the action (\ref{Amu-action-kras}).  

Finally, by making the variation of (\ref{Amu-action-kras}) with respect to $\rho$ we obtain
\begin{eqnarray}
\delta \rho: \frac{\rmi\epsilon}{\sqrt{2}} \tilde{\rho}^{-1/2} {\rm Tr}\tilde{M}^{1/2}-\Lambda=0. \label{eqmotion-mu-kras}
\end{eqnarray}
This equation will be used in what follows.

\paragraph{CDJ formulation.}
In the case $\Lambda=0$,  equation~(\ref{eqmotion-mu-kras}) cannot be solved for $\rho$, and hence this variable cannot be integrated out in~(\ref{Amu-action-kras}) to obtain a pure connection formulation. Instead, it can be shown that equation~(\ref{Amu-action-kras}) leads to the CDJ action~\cite{capo1991859}. 
To do this, we first compute $\tilde{M}^{1/2}$ by using  the characteristic equation for a $3\times3$ matrix $Z$:
\begin{eqnarray}
Z^3-({\rm Tr}Z)Z^2+\frac{1}{2}\left[\left({\rm Tr}Z\right)^2-{\rm Tr}Z^2 \right]Z-\det Z=0. \label{characteristicequation}
\end{eqnarray}
If this equation is multiplied by $Z$, and then we set $Z^2=\tilde{M}$ with ${\rm Tr}\tilde{M}^{1/2}=0$, which comes from equation~(\ref{eqmotion-mu-kras}) with $\Lambda=0$, then we have
\begin{eqnarray}
Z=\frac{1}{(\det\tilde{M})^{1/2}}\left[ \tilde{M}^{2} - \frac{1}{2} ({\rm Tr}\tilde{M}) \tilde{M} \right]. \label{cdj-charac}
\end{eqnarray}
The CDJ action emerges when (\ref{cdj-charac}) is substituted into~(\ref{Amu-action-kras}), leading to
\begin{eqnarray}
S[A,\underaccent{\tilde}{\eta}]= \int_\mathcal{M} d^4x \, \underaccent{\tilde}{\eta} \left[ {\rm Tr}\tilde{M}^{2} - \frac{1}{2} ({\rm Tr}\tilde{M})^2 \right]  , \label{ActionCDJ}
\end{eqnarray}
where we have introduced the new variable $\underaccent{\tilde}{\eta}:=-\sqrt{2} \rmi \epsilon (\tilde{\rho}/\det\tilde{M})^{1/2}$. 

An alternative procedure to arrive at the CDJ action (\ref{ActionCDJ}), starting from Plebanski's action, is carried out in ~\cite{capo199207}, where is also achieved the case with a nonvanishing cosmological constant. Moreover, it is also possible to derive the CDJ formulation, for both cases of the cosmological constant, taking as departure point the Ashtekar formalism~\cite{peld19918}. The coupling of a scalar field to the action~ (\ref{ActionCDJ}) was studied in~\cite{peld199062} by using a different but equivalent form of the Urbantke metric in terms of the curvature. An interesting generalization of~(\ref{ActionCDJ}), keeping both the $SO(3,\mathbb{C})$-connection $A^i$ and the scalar density $\underaccent{\tilde}{\eta}$ as basic variables, was proposed in~\cite{beng1991254} (see also~\cite{capo1992233}), where it is also  shown that such a generalization still describes two degrees of freedom per space point.

\paragraph{Pure connection formulation.}
In the case $\Lambda\neq0$, equation~(\ref{eqmotion-mu-kras}) implies
\begin{eqnarray}
\tilde{\rho} = - \frac{1}{2 \Lambda^2}  ( {\rm Tr}\tilde{M}^{1/2})^2, \label{Amu-rho-kras}
\end{eqnarray}
and so we can get rid of $\rho$. It should be stressed that $\rho$ is {\it not} an auxiliary field neither for Plebanski's action (\ref{plebanAct}) nor for the action (\ref{Act1-KRAS}). Finally, by substituting (\ref{Amu-rho-kras}) into (\ref{Amu-action-kras}) we get~\cite{cgm-pure-2015}
\begin{eqnarray}
	S[A]=\frac{1}{2 \Lambda} \int_\mathcal{M} d^4x ({\rm Tr}\tilde{M}^{1/2})^2, \label{ActionKrasnovprl}
\end{eqnarray}
which is the pure connection  action of general relativity with a nonvanishing cosmological constant (compare with the approach of~\cite{kras2011106}). Let us point out that, by construction, the action (\ref{ActionKrasnovprl}) and Plebanski's action~(\ref{plebanAct}) are equivalent, which holds provided that $X$ (or $\tilde{M}$) is invertible and  $\Lambda\neq0$. Furthermore, the canonical analysis of (\ref{ActionKrasnovprl}) leads to Ashtekar's Hamiltonian formulation of general relativity~\cite{cgm-pure-2015}. It may be added that (\ref{ActionKrasnovprl})  is a particular member of the class of theories proposed in~\cite{kras201081}, which depends only on an  $SO(3,\mathbb{C})$-connection and propagates two degrees of freedom per space point.

\paragraph{Alternative gauge connection formulations.}
Let us return to the action (\ref{Amu-action-kras}) and rewrite it as
\begin{eqnarray}
S[A,\rho]=-\int_\mathcal{M} \rho\left( 2 {\rm Tr}\Psi[A,\rho] + \Lambda\right), \label{gcf:1}
\end{eqnarray}
where the matrix $\Psi[A,\rho]$  is understood as a function of $A^i$ and $\rho$ via $\tilde{M}+2\tilde{\rho}X^2=0$ with $X:=\Psi+(\Lambda/3) I$. Since $\Psi[A,\rho]$ is a scalar function, it is natural to extend the action (\ref{gcf:1}) to
\begin{eqnarray}
S[A,\rho]=\int \rho f[{\rm Tr}\Psi[A,\rho]],\label{Action}
\end{eqnarray}
where $f$ is an arbitrary holomorphic function that depends on ${\rm Tr}\Psi$.  Notice that $f$ must have the same dimension of the cosmological constant. Strikingly, when $f$ satisfies the following two properties:
\begin{enumerate}
	\item The only zero of $f- \frac{1}{2} \left( {\rm Tr}\Psi + \Lambda \right) f'$, as a function of ${\rm Tr}\Psi$, is ${\rm Tr}\Psi=0$;
	\item $f'_0:=\left. f' \right|_{{\rm Tr}\Psi=0} \neq 0$;
\end{enumerate}
where  $f'=d f / d {\rm Tr}\Psi$,  the action~(\ref{Action})  describes general relativity~\cite{gm201591024021}. In particular, these properties are enjoyed by $f =\alpha_1 (2 {\rm Tr}\Psi+\Lambda)+ \alpha_2 ({\rm Tr}\Psi+\Lambda)^2$ if the constants $\alpha_1$ and $\alpha_2$ are such that $\alpha_1\neq0$ and
$\alpha_1+\alpha_2\Lambda\neq0$ with $\Lambda=0$ or $\Lambda\neq0$. In the case $\alpha_2=0$, we recover (\ref{gcf:1}).  Another function with interesting features is $f =  -  \Lambda \beta \exp\left(2{\rm Tr}\Psi /\Lambda \right) + \alpha ({\rm Tr}\Psi+\Lambda)^2$ where the constants $\alpha$ and $\beta$ must satisfy $\beta\neq0$ and  $\beta\neq\Lambda \alpha$ with $\Lambda\neq0$.

\subsubsection{From the pure connection formulation to the Plebanski formulation.}\label{kras-to-ple}

In this section, in contrast to what was done in the section~\ref{ple-to-kras}, we show how to arrive at Plebanski's action taking as starting point the pure connection formulation~(\ref{ActionKrasnovprl}). The idea is to add one by one the fields $\rho$, $\Psi$, and $\Sigma^i$ present in the Plebanski formulation to the pure connection action by means of classically equivalent actions. An advantage of introducing auxiliary fields is that the resulting actions might describe general relativity for both cases of the cosmological constant, despite the fact that the action~(\ref{ActionKrasnovprl}) only works for the case of a nonvanishing cosmological constant. However, the equivalence between the actions involving the auxiliary fields and the action~(\ref{ActionKrasnovprl}) holds only if the cosmological constant is nonzero.

Let us begin by considering the action
\begin{eqnarray}
S[A,\rho,\mu]=-\int_\mathcal{M} d^4x \left[ \Lambda\tilde{\mu} + 2 \tilde{\rho}^{1/2} \left( \frac{\rmi\epsilon}{\sqrt{2}} {\rm Tr}\tilde{M}^{1/2} - \Lambda  \tilde{\mu}^{1/2} \right)\right], \label{KraPle:act1}
\end{eqnarray}
where $\tilde{\rho} d^4x:=\rho$ and $\tilde{\mu} d^4x:=\mu$ are nonvanishing 4-forms, and $\epsilon=\pm1$ stands for the branch of $\tilde{M}^{1/2}$. Notice that the equation of motion corresponding to $\rho$ leads to $\tilde{\mu}=(-1/2\Lambda^2) ({\rm Tr}\tilde{M}^{1/2})^2$,  which substituted into the action gives back (\ref{ActionKrasnovprl}). This shows that  the actions (\ref{ActionKrasnovprl}) and (\ref{KraPle:act1}) are equivalent, provided that $\Lambda\neq0$. On the other hand, the variation of~(\ref{KraPle:act1}) with respect to $\mu$ leads to $\tilde{\mu}^{1/2}=\tilde{\rho}^{1/2}$, and after substituting this into~(\ref{KraPle:act1}), we precisely obtain the action $S[A,\rho]$ of equation~(\ref{Amu-action-kras}). As was pointed out in section~\ref{ple-to-kras}, the equivalence between (\ref{Amu-action-kras}) and (\ref{ActionKrasnovprl}) holds when $\Lambda\neq0$, but  in contrast to (\ref{ActionKrasnovprl}), the action (\ref{Amu-action-kras}) supports the case of a nonvanishing cosmological constant due to the presence of $\rho$. 

We continue our procedure by noting that the action~(\ref{Amu-action-kras}) is equivalent to
\begin{eqnarray}
\fl S[A,\rho,X,Y]=\int_\mathcal{M} d^4x \left[\Lambda\tilde{\rho}-2\tilde{\rho}  {\rm Tr}Y + \frac{1}{2} (X^{-1})_{ij} \left( \tilde{M}^{ij}+2\tilde{\rho}Y^i{}_k Y^{jk}  \right)\right], \label{KraPle:act2}
\end{eqnarray}
where we have introduced the symmetric and nonsingular $3\times 3$ matrices~$X$ and~$Y$. In fact, the equation of motion for $X$ implies $Y=(\rmi\epsilon/\sqrt{2}) \tilde{\rho}^{-1/2} \tilde{M}^{1/2}$, which replaced in~(\ref{KraPle:act2}) leads to~(\ref{Amu-action-kras}). Now, the variation of (\ref{KraPle:act2}) with respect to $Y$ gives $Y=X$. When this is substituted into the action and $X$ is redefined as $X:=\Psi+(\Lambda/3) I$, the action (\ref{Act1-KRAS}) follows.

In the final step towards Plebanski's action, we introduce the 2-form fields. To this end, from (\ref{Act1-KRAS}) we construct the following action 
\begin{eqnarray}
\fl S[A,\rho,\Psi,\Sigma,B]=\int d^4x \left[ \frac{1}{2} X_{ij} B^i\wedge B^j -\rho {\rm Tr}\Psi  + \Sigma_i \wedge\left( F^i[A] - X^i{}_j B^j \right) \right],\label{KraPle:act3}
\end{eqnarray}
where $\Sigma^i$ and $B^i$ are 2-forms, and we have used the definition of  $X$ in terms of $\Psi$. The constraint resulting from the variation of ~(\ref{KraPle:act3}) with respect to the Lagrange multiplier $\Sigma^i$ implies $B^i=(X^{-1})^i{}_j F^j$, and by replacing this into~(\ref{KraPle:act3}),~(\ref{Act1-KRAS}) is obtained, showing then the equivalence between both actions. On the other hand, notice that Plebanski's action emerges from~(\ref{KraPle:act3}) when  $B^i$ is integrated out. Indeed, the variation with respect to $B^i$ leads to $B^i=\Sigma^i$, which in turn substituted into~(\ref{KraPle:act3}) gives precisely Plebanski's action~(\ref{plebanAct}),  thus concluding our procedure. It is important to emphasize that~(\ref{plebanAct}) works well for both $\Lambda=0$ and $\Lambda\neq0$, and that its equivalence with the pure connection action~(\ref{ActionKrasnovprl}) holds whenever $X$ is invertible and  $\Lambda\neq0$.

\subsection{Plebanski-like formulation of general relativity and anti-self-dual gravity}\label{sec_Plebanski-like}

The action of complex general relativity as an $SO(3,\mathbb{C})$-invariant $BF$ theory is not unique. Recently, in~\cite{Krasn2015} it has been proposed a $BF$-type action for general relativity, which only involves the $SO(3,\mathbb{C})$-connection and the $B$ fields as fundamental variables, reinforcing the analogy between general relativity and gauge theories. However,  it is not possible to derive this formulation directly from Plebanski's action. The reason is that, as we have seen in section~\ref{ple-to-kras}, in the Plebanski formulation it is not possible to get rid of the Lagrange multipliers $\Psi$ and $\rho$, without first eliminating the $B$ fields. This was the motivation of~\cite{cgm-PlebLike-2016} for introducing a Plebanski-like action---an action using the same variables of Plebanski's action but with a different functional form---from which be possible to derive, through the elimination of some auxiliary fields, the $BF$-type action of~\cite{Krasn2015}.  Taking a different point of view, in this  section we  show, by following an analogous procedure to the one outlined  in section~\ref{kras-to-ple},  how the $BF$-type action of~\cite{Krasn2015} and the Plebanski-like formulation of~\cite{cgm-PlebLike-2016} emerge from the pure connection action (\ref{ActionKrasnovprl}) plus a topological term.

Let us pose the following action
\begin{eqnarray}
S[A]=\frac{1}{2 \Lambda} \int_\mathcal{M} d^4x ({\rm Tr}\tilde{M}^{1/2})^2+\frac{1}{2 \lambda} \int_\mathcal{M} F_i\wedge F^i, \label{PleLike-KrasPontr}
\end{eqnarray}
where $\lambda$ is a nonvanishing constant with the same dimension as $\Lambda$. Clearly, this action describes general relativity with a nonvanishing cosmological constant since the topological term does not modify the classical dynamics. However, the presence of this term is precisely the key point in what follows. We begin with the observation that equation~(\ref{PleLike-KrasPontr}) is equivalent to
\begin{eqnarray}
\fl S[A,B,Q]= \int_\mathcal{M}  \left[ \frac{1}{2\Lambda} ({\rm Tr}\tilde{L}^{1/2})^2 d^4x + \frac{1}{2\lambda} Q_i \wedge Q^i + B_i \wedge (F^i - Q^i)   \right], \label{PleLike-act1}
\end{eqnarray}
where $B^i$ and $Q^i$ are 2-forms, and $\tilde{L}$ is a nonsingular matrix defined by $\tilde{L}^{ij}d^4x:=Q^i\wedge Q^j$. Indeed, the equation of motion for the Lagrange multiplier $B^i$, namely $Q^i=F^i$, combined with (\ref{PleLike-act1}) leads to (\ref{PleLike-KrasPontr}). Now we want to integrate out $Q^i$ in~(\ref{PleLike-act1}). The equation of motion for $Q^i$ is given by
\begin{eqnarray}
\delta Q^i: B^i=\frac{1}{\Lambda} {\rm Tr}\tilde{L}^{1/2} (\tilde{L}^{-1/2})^i{}_jQ^j+\frac{1}{\lambda} Q^i, \label{PleLike-act1-Q}
\end{eqnarray}
and it can be solved for $Q^i$ provided that $3\lambda+\Lambda\neq0$. We find
\begin{eqnarray}
Q^i=\lambda B^i -\alpha {\rm Tr}\tilde{N}^{1/2}  (\tilde{N}^{-1/2})^i{}_j  B^j,  \label{PleLike-act1-Qsol}
\end{eqnarray}
where $\alpha:=\lambda^2/(3\lambda+\Lambda)$ and  $\tilde{N}^{ij}d^4x:=B^i\wedge B^j$. By replacing (\ref{PleLike-act1-Qsol}) back into (\ref{PleLike-act1}), we obtain
\begin{eqnarray}
S[A,B]= \int_\mathcal{M}  \left[ B_i \wedge F^i[A]  + \frac{\alpha}{2}  ({\rm Tr}\tilde{N}^{1/2})^2 d^4x   -\frac{\lambda}{2} B_i \wedge B^i  \right]. \label{PleLike-act2}
\end{eqnarray}
This is the action proposed in~\cite{Krasn2015} to describe general relativity with a nonvanishing cosmological constant. Solving $B^i$ from its own equation of motion gives $B^i=(1/\Lambda){\rm Tr}\tilde{M}^{1/2}(\tilde{M}^{1/2})^i{}_jF^j+(1/\lambda)F^i$, and substituting this back into the action~(\ref{PleLike-act2}) we recover~(\ref{PleLike-KrasPontr}), as expected. Furthermore, in~\cite{Krasn2015} it is shown that the equations of motion arising from (\ref{PleLike-act2})  imply Plebanski's equations of general relativity. Notice that (\ref{PleLike-act2}) is of the form
\begin{eqnarray}
S [A,B] = \int_{\mathcal{M}} \left [ B_i \wedge F^i + V(B) \right ],
\end{eqnarray}
with the potential $V(B)= (\alpha/2) ({\rm Tr}\tilde{N}^{1/2})^2 d^4x   - (\lambda/2) B_i \wedge B^i$.

Next, by introducing the 4-forms $\rho=\tilde{\rho} d^4x$ and $\mu=\tilde{\mu} d^4x$, the action~(\ref{PleLike-act2}) is equivalent to
\begin{eqnarray}
\fl S[A,B,\rho, \mu]= \int_\mathcal{M}  \left[ B_i \wedge F^i[A] + \frac{\alpha}{2} \mu-\frac{\lambda}{2} B_i \wedge B^i + \tilde{\rho}^{1/2} \left(\varepsilon{\rm Tr}\tilde{N}^{1/2} - \tilde{\mu}^{1/2}\right)d^4x  \right], \label{PleLike-act3}
\end{eqnarray}
where $\varepsilon=\pm1$ stands for the branch of $\tilde{N}^{1/2}$. In fact, the constraint imposed by $\rho$, that is $ \tilde{\mu}^{1/2}=\varepsilon{\rm Tr}\tilde{N}^{1/2}$, together with (\ref{PleLike-act3}), imply (\ref{PleLike-act2}), showing the equivalence. Now, we proceed to integrate out $\mu$ in the action. The equation of motion for $\mu$ implies $\tilde{\mu}^{1/2}=\tilde{\rho}^{1/2}/\alpha$, which replaced in (\ref{PleLike-act3}) gives the action
\begin{eqnarray}
 \fl S[A,B,\rho]= \int_\mathcal{M}  \left[ B_i \wedge F^i[A] +\varepsilon\tilde{\rho}^{1/2}{\rm Tr}\tilde{N}^{1/2} d^4x -\frac{\lambda}{2} B_i \wedge B^i -\frac{3\lambda+\Lambda}{2\lambda^2} \rho \right]. \label{PleLike-act4}
\end{eqnarray}
Although the procedure involves $\Lambda\neq0$ and $3\lambda+\Lambda\neq0$, this action does not require these conditions to describe general relativity and hence works well for {\it both} a vanishing and a nonvanishing cosmological constant.  A direct way to check it is to show that (\ref{PleLike-act4}) is equivalent to (\ref{Amu-action-kras}), which allows for both cases of the cosmological constant. To do this, we only have to integrate out $B^i$ from (\ref{PleLike-act4}). The equation of motion for $B^i$  leads to $B^i=(\varepsilon/\lambda)\tilde{\rho}^{1/2}(\tilde{M}^{-1/2})^i{}_jF^j+(1/\lambda)F^i$, which substituted in (\ref{PleLike-act4}) gives
\begin{eqnarray}
S[A,\rho']= \int_\mathcal{M} d^4x \left(\sqrt{2} \varepsilon \tilde{\rho}'^{1/2}{\rm Tr}\tilde{M}^{1/2} - \Lambda \tilde{\rho}' \right) + \frac{1}{2 \lambda} \int_\mathcal{M} F_i\wedge F^i,
\end{eqnarray}
where we have made the change of variable $\rho':=\rho/2\lambda^2$. This expression corresponds to the action (\ref{Amu-action-kras}), modulo constant factors and a topological term. Therefore, (\ref{PleLike-act4}) supports both cases of the cosmological constant. In the particular case when $\Lambda\neq0$ and $3\lambda+\Lambda\neq0$, we can recover (\ref{PleLike-act2}) from (\ref{PleLike-act4}). Given this fact, (\ref{PleLike-act4}) can be regarded as a generalization of (\ref{PleLike-act2}). Let us point out that the relationship between the actions (\ref{PleLike-act2}) and (\ref{PleLike-act4}) is analogous to that between the actions (\ref{ActionKrasnovprl}) and (\ref{Amu-action-kras}), in the sense that only after the introduction of the 4-form $\rho$, the actions are able to describe general relativity with $\Lambda=0$.

In the final step, we complete our approach by introducing a symmetric $3\times 3$ matrix in~(\ref{PleLike-act4}). To accomplish this, we note that~(\ref{PleLike-act4}) is equivalent to
\begin{eqnarray}
\fl S[A,B,\rho,\psi,\phi]= \int_\mathcal{M}  && \left[ B_i \wedge F^i[A] +\rho{\rm Tr} \phi - \frac{\lambda}{2} B_i \wedge B^i -\frac{3\lambda+\Lambda}{2\lambda^2} \rho \right. \nonumber\\
&& \ \ + \left.  \frac{1}{2}\psi_{ij}\left(\tilde{N}^{ij}-\tilde{\rho} \phi^i{}_k \phi^{jk} \right) d^4x \right], \label{PleLike-act5} 
\end{eqnarray}
where $\psi$ and $\phi$ are nonsingular symmetric $3\times 3$ matrices. The constraint imposed by $\psi$ leads to $\phi=\varepsilon\tilde{\rho}^{-1/2}\tilde{N}^{1/2}$, and replacing this back into (\ref{PleLike-act5}) leads to~(\ref{PleLike-act4}).  On the other hand, the equation of motion for $\phi$, namely, $\phi\psi+\psi\phi=2 I$, is a particular case of the Lyapunov matrix equation and  has a unique solution for $\phi$ if and only if the eigenvalues $\lambda_1$, $\lambda_2$, $\lambda_3$ of $\psi$ satisfy $\lambda_i + \lambda_j \neq 0$ ($i,j=1,2,3$) (see, for instance,~\cite{lancaster1985theory}, page 414). The unique solution turns out to be $\phi=\psi^{-1}$, and after substituting it into (\ref{PleLike-act5}), we finally obtain
\begin{eqnarray}
\fl S[A,B,\rho,\psi]= \int_\mathcal{M}  \left[ B_i \wedge F^i[A] +\frac{1}{2}\left(\psi_{ij}-\lambda \delta_{ij} \right) B^i \wedge B^j +\frac{1}{2} \rho \left( {\rm Tr} \psi^{-1} -\frac{3\lambda+\Lambda}{\lambda^2}  \right)\right]. \nonumber\\ \label{PleLike-act6}
\end{eqnarray}
By construction, this action describes general relativity with either a vanishing or a nonvanishing cosmological constant, provided that $\lambda\neq0$ (no additional conditions are needed on $\psi$). In fact, solving $\psi$ from its equation of motion gives $\psi=\varepsilon\tilde{\rho}^{1/2}\tilde{N}^{-1/2}$, and substituting this result back into (\ref{PleLike-act6}) we recover (\ref{PleLike-act4}), which describes general relativity for both cases of the cosmological constant. 

Having verified the equivalence of (\ref{PleLike-act6}) with general relativity, some remarks are worth mentioning. First, the main difference between Plebanski's action (\ref{plebanAct}) and (\ref{PleLike-act6}) is the way that the symmetric matrix enters into the respective actions;  while in~(\ref{plebanAct}) the matrix $\Psi$ is a Lagrange multiplier, in~(\ref{PleLike-act6}) the matrix $\psi$ is an auxiliary field. More precisely, the term with ${\rm Tr}\Psi $ that appears in~(\ref{plebanAct}) is replaced in~(\ref{PleLike-act6}) by a term involving ${\rm Tr} \psi^{-1}$. Consequently, in contrast to the Plebanski formulation, in the action principle (\ref{PleLike-act6}) the matrix field can be integrated out  from the very beginning, leading to (\ref{PleLike-act4}) and to (\ref{PleLike-act2}) after integrating out $\rho$, which is only possible for $\Lambda \neq 0$ and $3\lambda+\Lambda\neq0$. Second, as expected, the geometrical meaning of the variables changes. Indeed, the resulting equations of motion
\begin{eqnarray}
\delta \psi_{ij}&:& B^i\wedge B^j-\rho(\Psi^{-1})^{ik}\tensor{(\Psi^{-1})}{^j_k}=0,  \label{PleLike-1}\\
\delta A^i&:&          DB^i:=dB^i+\tensor{\varepsilon}{^i_{jk}}A^j\wedge B^k=0,  \label{PleLike-2}\\
\delta B^i&:&   F^i+(\tensor{\Psi}{^i_j}-\lambda\delta^i_j)B^j=0, \label{PleLike-3}\\
\delta \rho&:& {\rm Tr} \psi^{-1} -\frac{3\lambda\!+\!\Lambda}{\lambda^2}=0,\label{PleLike-4}
\end{eqnarray}
make it clear that the $B$ fields do not satisfy the simplicity constraint and that the matrix $\psi$ is no longer the self-dual part of the Weyl tensor. However, since these equations are equivalent to Plebanski's equations~(\ref{pleban1})-(\ref{pleban4}) (see~\cite{cgm-PlebLike-2016}), we can identify the 2-forms and the matrix field with the mentioned properties.  Third, the canonical analysis of this formulation shows that the Hamiltonian form of the action (\ref{PleLike-act6}) is given by~(\ref{PlebH:hamil}), where the Gauss and vector constraints remain unchanged, but the scalar constraint takes the form
\begin{eqnarray}
\fl \tilde{\tilde{\mathcal{H}}}:=\frac{3\lambda+\Lambda}{2\lambda^2}  BBB-3\left(1+\frac{\Lambda}{2\lambda}\right)\Pi B B
+\frac{3}{2}(\lambda+\Lambda)\Pi\Pi B-\frac{1}{2}\lambda\Lambda\ \Pi\Pi\Pi\approx 0,\label{scalar}
\end{eqnarray}   
where $\tilde{\Pi}^{ai}:=(1/2)\tilde{\eta}^{abc}\tensor{B}{_{bc}^i}$, 
$\tilde{B}^{ai}:=(1/2)\tilde{\eta}^{abc}\tensor{F}{^i_{bc}}$, and we have used the shorthand $\Pi\Pi B:=(1/6)\underaccent{\tilde}{\eta}_{abc}\varepsilon_{ijk}\tilde{\Pi}^{ai}\tilde{\Pi}^{bj}\tilde{B}^{ck}$, etc. The terms $BBB$ and $\Pi B B$ of this constraint are the result of the presence of the  topological term implicit in~(\ref{PleLike-act6}). Nevertheless, the contribution of this topological term can be canceled. In fact, by performing the canonical transformation $(A_{ai},\tilde{\Pi}^{ai})\rightarrow(A_{ai},\tilde{\Pi}^{ai}+\theta \tilde{B}^{ai})$~\cite{ashtekar1989-04-06-1493,beng1991254}, the Gauss, vector, and scalar constraints take the same form as the ones of the Ashtekar formulation, showing again that the action~(\ref{PleLike-act6}) describes general relativity. 

Finally, the action (\ref{PleLike-act6}) is a particular case of the Plebanski-like action~\cite{cgm-PlebLike-2016}
\begin{eqnarray}
\fl S[A,B,\rho,\psi]= \int_\mathcal{M}  \left[ B_i \wedge F^i[A] +\frac{1}{2}\left(\psi_{ij}-\lambda \delta_{ij} \right) B^i \wedge B^j + \rho \left( \beta {\rm Tr} \psi^{-1} -\gamma  \right)\right], \label{PleLike}
\end{eqnarray}
where $\lambda$,  $\beta$, and $\gamma$ are arbitrary constants with appropriate units. The advantage of this action is that it allows us to describe not only general relativity, but also anti-self-dual gravity when a particular choice of the constants is taken. Explicitly, if $\lambda=0$ and $\beta\neq 0$ in (\ref{PleLike}), then the resulting action describes conformally anti-self-dual gravity. Actually, the equations of motion of this case imply $F^i\wedge F^j \sim \delta^{ij}$ which is the instanton equation~\cite{Torreprd41,Capo-7-1-001}, and by eliminating $B^i$ and $\rho$ from the action and making the change $X:=-\Psi^{-1}$, we can arrive at the action of~\cite{torre1990topological} used to describe the moduli space of anti-self-dual gravitational instantons.

\subsection{Husain-Kucha\v{r} model as a constrained $BF$ theory}\label{HusainKuchar}

\paragraph{$BF$ formulation.} An interesting background-independent model closely related to general relativity is the one proposed by Husain and Kucha\v{r}~\cite{HusainKucharmodel}. The appealing feature of this model is that it lacks the scalar constraint; that is, its phase space variables, which are the same of general relativity, are subject to the Gauss and vector constraints only. From the quantum viewpoint,  the absence of the scalar constraint could be advantageous, since we would not have to deal with the complications entailed by it. This, in turn, could shed some light on the role of the same constraint in quantum gravity. There are two approaches to describe the Husain-Kucha\v{r} model in the framework of  $BF$ theories: one that consists of two interacting unconstrained $BF$ theories~\cite{BarberoVillasenor2001}, and a second that employs a single constrained $BF$ theory~\cite{aip_paper}. In this section we focus on the latter approach in the $SO(3,\mathbb{C})$ formalism. 

The constrained $BF$ action to describe the Husain-Kucha\v{r} model can be obtained from Plebanski's action~(\ref{plebanAct}) by removing the constraint on ${\rm Tr}\Psi$~\cite{capo1991841}. The action principle, which uses the same base manifold and structure group as the Plebanski formulation, reads
\begin{eqnarray}
	S[A, \Sigma,\Psi]=\int_\mathcal{M} \left[ \Sigma_i \wedge F^i[A] -\frac{1}{2} \Psi_{ij}  \Sigma^i \wedge \Sigma^j \right], \label{HK-action}
\end{eqnarray}
where $\Psi$ is a symmetric $3\times 3$  matrix; then it has six independent components. Here we have not considered the term with the cosmological constant of the action~(\ref{plebanAct}), since its effect can be reversed just by making the change of variables $X:=\Psi+(\Lambda/3) I$. Alternatively, the action~(\ref{HK-action}) follows from the Plebanski-like action by eliminating the condition on ${\rm Tr} \psi^{-1}$~\cite{cgm-PlebLike-2016}. More precisely, to arrive at (\ref{HK-action}) starting from (\ref{PleLike}), we only have to set $\beta=0$, consider the change of variables $X:=\psi-\lambda I$, and either take $\gamma=0$ or use the fact that the variation with respect to $\rho$ leads to $\gamma=0$.

The equations of motion arising from the action~(\ref{HK-action}) are
\begin{eqnarray}
	\delta \Psi_{ij}&:& \Sigma^i \wedge \Sigma^j =0,  \label{HK-eq1}\\
	\delta A^i&:&          D \Sigma^i:=d \Sigma^i  +  \varepsilon^i{}_{jk} A^j \wedge \Sigma^k =0,  \label{HK-eq2}\\
	\delta \Sigma^i&:&   F^i=\Psi^i{}_j\Sigma^j. \label{HK-eq3}
\end{eqnarray}
Thereby, as a consequence of removing the condition on $\Psi$ we have that, instead of the five equations given in the simplicity constraint of general relativity  (see equation~({\ref{simplicityconstraint}})), the 2-forms $\Sigma^i$ satisfy the six equations given in~(\ref{HK-eq1}). These equations have the solution (modulo a rescaling of $e^i$)~\cite{aip_paper}
\begin{eqnarray}
	\Sigma^i  = \varepsilon^i{}_{jk} e^j \wedge e^k,  \label{HK-sigma} 
\end{eqnarray}
where $e^i$ are three 1-forms. Thus, the 18 variables $\Sigma_{\mu\nu}{}^i$ have been paremetrized in terms of the 12 free functions $e_\mu{}^i$. Having this solution, it is straightforward to establish the equivalence of the $BF$ action (\ref{HK-action}) with the Husain-Kucha\v{r} model. In fact, by substituting (\ref{HK-sigma}) back into (\ref{HK-action}), we obtain 
\begin{eqnarray}
	S[e,A]=\int_{\mathcal{M}}   \varepsilon_{ijk} e^j \wedge e^k  \wedge F^i[A],\label{HK-model}
\end{eqnarray}
which is precisely the action principle for the Husain-Kucha\v{r} model~\cite{HusainKucharmodel}.

\paragraph{Canonical analysis.}
Using the notation of section~\ref{Pleb-HamilAna}, the (3+1)-decomposition of the action~(\ref{HK-action}) reads
\begin{eqnarray}
	\fl S[A, \Sigma,\Psi]=\int_{\mathbb{R}}dt\int_{\Omega}d^3x \left[\tilde{\Pi}^{ai} \dot{A}_{ai} +A_{ti} \tilde{\mathcal{G}}^i +  \Sigma_{tai} \tilde{B}^{ai} - \Psi_{ij} \Sigma_{ta}{}^i\tilde{\Pi}^{aj} \right].\label{HK-Hamil1}
\end{eqnarray}
As in that section, we can get rid of $\Psi_{ij}$ by using the constraint that it imposes. The variation of (\ref{HK-Hamil1}) with respect to $\Psi_{ij}$ gives the six constraints
\begin{eqnarray}
	\delta \Psi_{ij}:  \Sigma_{ta}{}^i\tilde{\Pi}^{aj}+\Sigma_{ta}{}^j\tilde{\Pi}^{ai}=0, \label{HK-constPsi}
\end{eqnarray}
which can be solved for nine variables $\Sigma^i{}_{ta}$ as
\begin{eqnarray}
	\Sigma_{tai}=\varepsilon_{ijk} \tilde{N}^k\underaccent{\tilde}{\Pi}_a{}^j, \label{HK-constsol}
\end{eqnarray}
where $\tilde{N}^k$ is an arbitrary internal 3-vector and $\underaccent{\tilde}{\Pi}_{ai}$ is the inverse of $\tilde{\Pi}^{ai}$. If $\tilde{\Pi}^{ai}$ is not assumed to be invertible, we can get other sectors, as in Plebanski's case. Substituting~(\ref{HK-constsol}) into~(\ref{HK-Hamil1}), the Hamiltonian form of $BF$ action~(\ref{HK-action}) turns out to be
\begin{eqnarray}
	S[A_{a},\tilde{\Pi}^{a},A_{t},N^a]=\int_{\mathbb{R}}dt\int_{\Omega}d^3x\left(\tilde{\Pi}^{ai}\dot{A}_{ai}+A_{ti}\tilde{\mathcal{G}}^i+N^a\tilde{\mathcal{V}}_a\right),\label{HK-hamil}
\end{eqnarray}
where $A_{ti}$ and $N^a:=-(\det\tilde{\Pi})^{-1}\tilde{\Pi}^a{}_i \tilde{N}^i$ are Lagrange multipliers imposing the Gauss and vector constraints, $\tilde{\mathcal{G}}^i=\mathcal{D}_a\tilde{\Pi}^{ai}\approx 0$ and $\tilde{\mathcal{V}}_a=\tilde{\Pi}^{bi}F_{iba}\approx 0$, of the Husain-Kucha\v{r} model. Note that the scalar constraint is missing in  (\ref{HK-hamil}) because the lapse function does not appear in (\ref{HK-constsol}) (compare with equation~(\ref{PlebH:constPsisol})), which was already noticed in the spinorial approach~\cite{capo1991841}.  Since we have nine configuration variables $A_{a}{}^i$ and six first-class constraints, $\tilde{\mathcal{G}}^i$ and $\tilde{\mathcal{V}}_a$,  the action (\ref{HK-action}) locally propagates three degrees of freedom, as it should be for a formulation for the Husain-Kucha\v{r} model. Finally, it is worth mentioning  that the spinfoam quantization of the Husain-Kucha\v{r} model is outlined in~\cite{Bojowald2009}.

\subsection{Non-metric gravity theories}\label{nonmetric}

The main subject of this section is to review the so-called ``non-metric gravity'' theories proposed in~\cite{kras200606}, which are defined by $BF$-type actions  resulting from a modification of Plebanski's action without adding any new fields. Interestingly, this class of modified gravity theories still propagates two complex degrees of freedom~\cite{beng20072222,kras2008100}. However, as we shall see, the geometrical meaning of the fundamental fields changes with respect to the original ones of the Plebanski formulation~\cite{frei20080812,kras2008-25-2,cmv201027}.

\subsubsection{Action principle.}

This class of modified gravity theories is obtained from the Plebanski formulation by promoting the cosmological constant~$\Lambda$ to an arbitrary function of the independent invariants that can be constructed from the trace-free part of~$\Psi$. To make this more precise, it is useful to first rewrite Plebanski's action~(\ref{plebanAct}) in the following  alternative form
\begin{eqnarray}
S[A, \Sigma,\phi]=\int_\mathcal{M} \left[ \Sigma_i \wedge F^i[A] -\frac{1}{2} \left(\phi_{ij}+\frac{\Lambda}{3} \delta_{ij}  \right)  \Sigma^i \wedge \Sigma^j  \right],   \label{plebanAct-no-rho}
\end{eqnarray}
where $\phi$ is a symmetric traceless $3\times 3$ matrix. This action follows from equation~(\ref{plebanAct}) by eliminating $\rho$ and making the change of variable $\phi:=\Psi-(1/3) {\rm Tr}\Psi I$. The $BF$-type action defining the ``non-metric gravity'' theories is obtained by promoting~(\ref{plebanAct-no-rho}) to~\cite{kras200606} 
\begin{eqnarray}
S[A,B,\phi]=\int_\mathcal{M} \left\{ B_i \wedge F^i[A] -\frac{1}{2}\left[ \phi_{ij} + \left(\Lambda + \Phi \right)\delta_{ij} \right] B^i \wedge   B^j\right\}, \label{krasnovAct}
\end{eqnarray}
where $\Lambda$ is three times the cosmological constant that appears in (\ref{plebanAct-no-rho}), and 
\begin{eqnarray}
\Phi=\Phi ({\rm Tr}\phi^2, {\rm Tr}\phi^3),
\end{eqnarray}
is a complex function that depends on the independent scalar invariants ${\rm Tr}\phi^2$ and ${\rm Tr}\phi^3$. As before, the indices $i,j,k$ are raised and lowered with the three-dimensional Euclidean metric. Notice that $\Phi$ does not depend on the full set of invariants of $\phi$, namely ${\rm Tr}\phi$, ${\rm Tr}\phi^2$, and ${\rm Tr}\phi^3$. The reason is quite simple: the scalar field $\phi$ has been defined so that it is traceless. On the other hand, since $\phi$ is a tensor density of weight zero, we can regard $\Phi$ as an arbitrary (holomorphic) function of its arguments. This means that (\ref{krasnovAct}) actually describes an infinite class of theories of gravitation with two complex degrees of freedom.  Clearly,  we would need to specify how $\Phi$ depends on $\phi$, in order to fix the theory we are considering. In particular, when $\Phi$ is constant, the Plebanski formulation of complex general relativity emerges. 

 The variation of the action (\ref{krasnovAct}) with respect to the independent fields gives
	\begin{eqnarray}
	\delta \phi_{ij} &: & B^i \wedge   B^j -  \frac{1}{3} \Delta^{ij} B^k \wedge   B_k =0, \label{kras-simpl}\\
	\delta A^i&: & D B^i:=d B^i  +  {\varepsilon}^i{}_{jk} A^j \wedge B^k =0,  \label{krasnov2}\\
	\delta B^i&: & F^i = \phi^i{}_j  B^j +\left( \Lambda + \Phi \right)  B^i, \label{krasnov3}
	\end{eqnarray}
where
\begin{eqnarray}\label{defDelta}
\Delta^{ij}:=\delta^{ij} - 3\left(  \Theta^{ij} -\frac{1}{3} \delta^{ij}   {\rm Tr} \Theta\right), \qquad {\rm Tr}\Delta=3,
\end{eqnarray}
with
\begin{eqnarray}
\Theta^{ij}:= 2\, \phi^{ij} \frac{\partial \Phi}{\partial{\rm Tr}\phi^2} + 3\, {\phi}^i{}_k{\phi}^{kj} \frac{\partial \Phi}{\partial{\rm Tr}\phi^3}. \label{krasmod-theta}
\end{eqnarray}
In view of (\ref{kras-simpl}), it is clear that one of the consequences of the presence of the function $\Phi$ is that now the simplicity constraint of the Plebanski formulation is relaxed to admit a complex symmetric matrix $\Delta$. This immediately means that the $B$ fields of the modified theories are no longer  the Plebanski 2-forms. In fact, note that if $B^i$ is a solution of (\ref{kras-simpl}), then its complex conjugate is not, which contrasts with the properties of the Plebanski 2-forms. Moreover, since the geometrical structure of $B^i$ is modified by $\Phi$ (more precisely by (\ref{krasmod-theta})), then, by virtue of (\ref{krasnov2}), the meaning of $A^i$ is modified; that is, $A^i$ is no longer the self-dual part of the spin-connection. Consequently, and because of (\ref{krasnov3}), the meaning of $\phi^{ij}$ is also changed. Of course, when $\Phi$ is constant, we have $\Theta^{ij}\to0$ (implying $\Delta^{ij} \to \delta^{ij}$), and then equations~(\ref{kras-simpl}), (\ref{krasnov2}), and (\ref{krasnov3}) reduce to  Plebanski's equations of general relativity with a redefined cosmological constant. It is worth mentioning that, as a consequence of the modification, we have now
\begin{eqnarray}
D\phi^i{}_j \wedge B^j + \Theta^j{}_k D\phi^k{}_j \wedge B^i  =0,\label{krasmod-Dphi}
\end{eqnarray}
instead of equation~(\ref{ple:DPsi}), where we have used the Bianchi identities, $DF^i=0$, and equations~(\ref{krasnov2}) and (\ref{krasnov3}).

\subsubsection{Field equations.}

The aim here is to review how the introduction of the function $\Phi$ modifies the geometrical structure of the fields involved in equation~(\ref{krasnovAct}) with respect to those of the Plebanski formulation. 

\paragraph{Solving the constraints on the $B$'s.} As the first step, we can consider, without loss of generality, that the 2-forms $B^i$ of these modified theories have the form~\cite{frei20080812,kras2008-25-2,speziale2009-26,kras2009-79}
\begin{eqnarray}
B^i_{\scriptscriptstyle(\epsilon)}=b_{\scriptscriptstyle(\epsilon)}^i{}_j \Sigma_{\scriptscriptstyle(\epsilon)}^j,\label{krasmod-solB}
\end{eqnarray}
where $b_{\scriptscriptstyle(\epsilon)}$ is a complex $3\times3$ matrix and 
\begin{eqnarray}
\Sigma_{\scriptscriptstyle(\epsilon)}^i  = \epsilon\rmi e^0 \wedge e^i - \frac{1}{2} \varepsilon^i{}_{jk} e^j \wedge e^k,  \label{krasmod-sigma} 
\end{eqnarray}
with $\epsilon=\pm1$ and $e^I=\{e^0,e^i\}$ ($I=0,1,2,3$) being four linearly independent (complex) 1-forms. Let us note that the matrix $b_{\scriptscriptstyle(\epsilon)}$ is defined up to a rescaling $b_{\scriptscriptstyle(\epsilon)}^i{}_j\to \Omega^{-1} b_{\scriptscriptstyle(\epsilon)}^i{}_j$ when $\Sigma_{\scriptscriptstyle(\epsilon)}^i\to\Omega \Sigma_{\scriptscriptstyle(\epsilon)}^i$,  and  $SO(3,\mathbb{C})$ rotations $b_{\scriptscriptstyle(\epsilon)}^i{}_j \to  O^i{}_k O_j{}^l b_{\scriptscriptstyle(\epsilon)}^k{}_l $, where $O\in SO(3,\mathbb{C})$. It follows that $b_{\scriptscriptstyle(\epsilon)}^i{}_j$ are $9-1-3=5$ independent components. Now, since the 2-form $\Sigma_{\scriptscriptstyle(\epsilon)}^i$ satisfies the five equations~(\ref{simplicityconstraint}), its 18 components are reduced to 13. Therefore, the parametrization (\ref{krasmod-solB}) preserves the total number ($18=5+13$) of components in $B_{\scriptscriptstyle(\epsilon)}^i$. 
 
Substituting~(\ref{krasmod-solB}) in~(\ref{kras-simpl}) and using the simplicity constraint (\ref{simplicityconstraint}), we obtain $b_{\scriptscriptstyle(\epsilon)}^i{}_k b_{\scriptscriptstyle(\epsilon)}^{jk}-(1/3) \Delta^{ij} b_{\scriptscriptstyle(\epsilon)}{}_{kl} b_{\scriptscriptstyle(\epsilon)}^{kl}=0$, which (together with ${\rm Tr}\Delta=3$) means that the 2-forms~(\ref{krasmod-solB}) are solution of~(\ref{kras-simpl}) if $b_{\scriptscriptstyle(\epsilon)}$ satisfies
 \begin{eqnarray}
	b_{\scriptscriptstyle(\epsilon)}^i{}_k  b_{\scriptscriptstyle(\epsilon)}^{jk}  = \Delta^{ij}. \label{krasmod-Delta}
\end{eqnarray}
Notice that the matrix $\Delta$ has the same number of components of $b_{\scriptscriptstyle(\epsilon)}$, since it  is symmetric and ${\rm Tr}\Delta=3$. From (\ref{krasmod-Delta}) we also see that since $\Delta$ is complex symmetric, $b_{\scriptscriptstyle(\epsilon)}$ always exists (see~\cite{horn2012matrix}, corollary 4.4.6). However, as we shall see, $b_{\scriptscriptstyle(\epsilon)}$ is not unique. In particular, if $\Delta$ is complex orthogonally diagonalizable; that is, $\Delta=L \lambda L^T$, where  $L\in SO(3,\mathbb{C})$ and $\lambda={\rm diag}(\lambda_1,\lambda_2,\lambda_3)$ with $\lambda_1,\lambda_2,\lambda_3$ being the eigenvalues of $\Delta$, then we can write $\Delta=(L\lambda^{1/2})(L\lambda^{1/2})^T$ where $\lambda^{1/2}={\rm diag}(\sqrt{\lambda_1},\sqrt{\lambda_2},\sqrt{\lambda_3})$, and hence equation~(\ref{krasmod-Delta}) has the solution $b_{\scriptscriptstyle(\epsilon)}=L \lambda^{1/2}$. In the literature~\cite{speziale2009-26,kras200722} this is the common approach to obtain the matrix $b_{\scriptscriptstyle(\epsilon)}$. However, it is important to recall that although a real symmetric matrix is diagonalizable, a complex symmetric matrix may not be diagonalizable. This motivates the following approach to obtain $b_{\scriptscriptstyle(\epsilon)}$, which is based on Takagi's factorization and does not require further assumptions on $\Delta$. Takagi's factorization states that $\Delta$, being a symmetric matrix, can be expressed as $\Delta=Q\xi Q^T$, where $Q$ is a unitary matrix and $\xi={\rm diag}(\xi_1,\xi_2,\xi_3)$ with $\xi_1,\xi_2,\xi_3$ being the  nonnegative square roots of the eigenvalues of the Hermitian matrix $\Delta\overline{\Delta}$ (for more details see~\cite{horn2012matrix}, corollary 4.4.3). Thus, using this factorization, we can write $\Delta=\left(Q\xi^{1/2} \right) \left(Q\xi^{1/2} \right)^T$ where $\xi^{1/2}={\rm diag}(\sqrt{\xi_1},\sqrt{\xi_2},\sqrt{\xi_3})$, and then we conclude that $b=Q\xi^{1/2}$. In the particular case when $b_{\scriptscriptstyle(\epsilon)}\in SO(3,\mathbb{C})$, the 2-forms $B^i$ reduce to those of the Plebanski formulation (modulo an $SO(3,\mathbb{C})$ rotation).

\paragraph{Reality conditions.} Let us note that the Urbantke metric~(\ref{urbametric}) defined by $B^i_{\scriptscriptstyle(\epsilon)}$ is conformally related to a real Lorentzian metric if $\Sigma_{\scriptscriptstyle(\epsilon)}^i$ satisfies the reality conditions of the Plebanski formulation and if $\Delta$ is a nonsingular matrix. Indeed, in this case we have that $\Sigma_{\scriptscriptstyle(+)}^i=\Sigma^i$ and $\Sigma_{\scriptscriptstyle(-)}^i=\overline{\Sigma}{}^i$ are the Plebanski 2-forms---with real 1-forms $e^I$---and that the resulting Urbantke metric constructed from~$B^i_{\scriptscriptstyle(\epsilon)}$ is $\tilde{g}^{\scriptscriptstyle(\epsilon)}_{\mu\nu}=
\epsilon 12\rmi\det(b_{\scriptscriptstyle(\epsilon)})\det(e_\mu{}^I{}) \eta_{IJ} e_\mu{}^I e_\nu{}^J$, where $\det b_{\scriptscriptstyle(\epsilon)}\neq0$. This means that the sector of 2-forms $B_{\scriptscriptstyle(+)}^i$ (respectively $B_{\scriptscriptstyle(-)}^i$) is self-dual (respectively anti-self-dual) with respect to the metric $g_{\mu\nu}:=\eta_{IJ} e_\mu{}^I e_\nu{}^J$. To be more specific, 
\begin{eqnarray}
\ast B_{\scriptscriptstyle(\epsilon)}^i  = \epsilon \rmi B_{\scriptscriptstyle(\epsilon)}^i,  \label{krasmod-dual}
\end{eqnarray}
where $\ast$ is the Hodge dual operator defined by $g_{\mu\nu}$. Now, using equations~(\ref{simplicityconstraint}),~(\ref{reality1}), and~(\ref{reality2}) together with (\ref{krasmod-Delta}), it can be verified that the 2-forms $ B_{\scriptscriptstyle(\epsilon)}^i$ satisfy the relations
\begin{eqnarray}
 B_{\scriptscriptstyle(\epsilon)}^{i} \wedge \overline{B}{}_{\scriptscriptstyle(\epsilon)}^{j} = 0, \label{krasmod-real1}\\
 B_{\scriptscriptstyle(\epsilon)}{}_{i} \wedge B_{\scriptscriptstyle(\epsilon)}^{i} + \overline{B}{}_{\scriptscriptstyle(\epsilon)}{}_{i} \wedge \overline{B}{}_{\scriptscriptstyle(\epsilon)}^{i} =0, \label{krasmod-real2}
\end{eqnarray}
and also that $B_{\scriptscriptstyle(+)}^i$ and $B_{\scriptscriptstyle(-)}^i$ are wedge orthogonal and have opposite volume, namely 
\begin{eqnarray}
B_{\scriptscriptstyle(+)}^{i} \wedge B_{\scriptscriptstyle(-)}^{j} = 0, \label{krasmod-reality1}\\
 B_{\scriptscriptstyle(+)}{}_{i} \wedge B_{\scriptscriptstyle(+)}^{i} + B_{\scriptscriptstyle(-)}{}_{i} \wedge B_{\scriptscriptstyle(-)}^{i} =0. \label{krasmod-reality2}
\end{eqnarray}
First, notice that equations~(\ref{krasmod-real1}) and (\ref{krasmod-real2}) have the same form of Plebanski's reality conditions. However, in contrast to the Plebanski case where $\overline{\Sigma}{}^i$ is also solution of the simplicity constraint, in the modified theories $\overline{B}{}_{\scriptscriptstyle(\epsilon)}^{i}$ is no longer a solution of (\ref{kras-simpl}). In the case of equations~(\ref{krasmod-reality1}) and~(\ref{krasmod-reality2}), we have that  $B_{\scriptscriptstyle(+)}^{i}$ and $B_{\scriptscriptstyle(-)}^{i}$, which are solutions of (\ref{kras-simpl}), are not complex conjugate to each other, in contrast to what happens with $\Sigma^i$ and $\overline{\Sigma}{}^i$. To set the reality conditions directly on the 2-forms $B_{\scriptscriptstyle(\epsilon)}^i$, we only have to impose equations~(\ref{krasmod-real1}) and (\ref{krasmod-real2})~\cite{kras2008-25-2}, which, in turn, imply that $\Sigma_{\scriptscriptstyle(\epsilon)}^i$ satisfies equations~(\ref{reality1}) and~(\ref{reality2}).

  \paragraph{The solution of the connection $A$.} The solution of equation~(\ref{krasnov2}) for the gauge connection in terms of the parametrization (\ref{krasmod-solB}) has been already given in~\cite{frei20080812} (see also~\cite{kras201081,speziale2010-82}). Here, we provide an independent and straightforward derivation of the connection. From now on, we only consider the sector $\epsilon=+1$ and write $B^i=B_{\scriptscriptstyle(+)}^{i}=b^i{}_j \Sigma^j$ (the case $\epsilon=-1$ is completely analogous). Let us begin by decomposing the gauge connection $A^i$ as 
 \begin{eqnarray}
 A^i=A_{\rm P}^i+K^i, \label{krasmod-conex0}
 \end{eqnarray}
where $A_{\rm P}^i$ is the gauge connection of the Plebanski formulation and $K^i$ is an unknown 1-form. Then, substituting (\ref{krasmod-solB}) and (\ref{krasmod-conex0}) into (\ref{krasnov2}), we obtain
 \begin{eqnarray}
 \mathscr{D} b^i{}_j \wedge \Sigma^j+\varepsilon^i{}_{jk} K^j \wedge b^k{}_l \Sigma^l =0, \label{krasmod-conex1}
 \end{eqnarray}
 where $\mathscr{D}b^{ij}:=db^{ij}+\varepsilon^i{}_{lk}A_{\rm P}^l b^{kj}+\varepsilon^j{}_{lk} A_{\rm P}^l b^{ik}$, and we have used $\mathscr{D}\Sigma^i:=d \Sigma^i  +  \varepsilon^i{}_{jk} A_{\rm P}^j \wedge \Sigma^k =0$. 
 
 The problem now is solving equation~(\ref{krasmod-conex1}) for $K^i$. To this end, we define $k^i$ via $K^i=:b^i{}_j k^j$  and rewrite this equation as 
  \begin{eqnarray}
 (\det b)^{-1} b_k{}^i  \mathscr{D}b^k{}_j\wedge \Sigma^j+\varepsilon^i{}_{jk} k^j \wedge \Sigma^k =0, \label{krasmod-conex2}
 \end{eqnarray}
This equation can be solved for $k^i$ by following the same strategy used to solve equation~(\ref{pleban2}) for the gauge connection. In fact, equation~(\ref{krasmod-conex2}) is a nonsingular system of 12 equations for 12 unknowns in $k^i$, whose unique solution is (compare with equation~(\ref{plebconn-simply}))
\begin{eqnarray}
 k^i =  \frac{\rmi}{4}  (\det b)^{-1} b_{lj}  \mathscr{D}_I b^l{}_k \Sigma_{JK}{}^k  \varepsilon^{IJKL} \Sigma_{LM}{}^i\tensor{\Sigma}{^M_N^j}  e^N, \label{krasmod-conex3}
\end{eqnarray}
where $\mathscr{D}_I $ stands for the components of the covariant derivative in the basis $\{e^I \}$. It remains to simplify this solution by using the self-duality of $\Sigma^i$ and the fact that $\Sigma^i$ satisfies the relation $\tensor{\Sigma}{^I_J^i}\tensor{\Sigma}{^J_K^j} \tensor{\Sigma}{^K_L^k} = -\varepsilon^{ijk} \delta^I_L -\delta^{ij} \tensor{\Sigma}{^I_L^k}-\delta^{jk} \tensor{\Sigma}{^I_L^i}+\delta^{ki} \tensor{\Sigma}{^I_L^j}$. After some algebra, we obtain 
 \begin{eqnarray}
 \fl k^i=\frac{1}{2}(\det b)^{-1} \varepsilon^{ijk} b_{lj} \mathscr{D}b^l{}_k + \frac{1}{4}(\det b)^{-1} ( \delta^{ij} \tensor{\Sigma}{^I_J^k} + \delta^{ik}\tensor{\Sigma}{^I_J^j}-\delta^{jk} \tensor{\Sigma}{^I_J^i}) \mathscr{D}_I (b_{lj} b^l{}_k) e^J.\label{krasmod-conex5} \nonumber\\
 \end{eqnarray}

Using this result together with the definition of $K^i$, we can finally write the expression for the gauge connection as
\begin{eqnarray}
\fl A^i=A_{\rm P}^i-\frac{1}{2} \varepsilon^{ij}{}_k (b^{-1})^l{}_{j} \mathscr{D}b^k{}_l \nonumber\\+ \frac{1}{4} (\det b)^{-1} b^i{}_n ( \delta^{nj} \tensor{\Sigma}{^I_J^k} + \delta^{nk} \tensor{\Sigma}{^I_J^j}-\delta^{jk} \tensor{\Sigma}{^I_J^n}) \mathscr{D}_I (b_{lj} b^l{}_k) e^J. \label{krasmod-conex6}
\end{eqnarray}
 Notice that in the case when $b\in SO(3,\mathbb{C})$, the last term of the right-hand side of this equation vanishes since $b_{lj} b^l{}_k=\delta_{jk}$, and then we are left with  $A^i=b^i{}_jA_{\rm P}^j-(1/2) \varepsilon^{ij}{}_k b_j{}^l db^k{}_{l}$, which is nothing but the transformation law for $A_{\rm P}^i$ under an $SO(3,\mathbb{C})$ rotation, as expected. Having the expression (\ref{krasmod-conex6}) for the connection, the curvature is computed by using $F^i=d A^i + (1/2) {\varepsilon}^i{}_{jk}  A^j  \wedge A^k$. 

\paragraph{\bf Remarks.} (i) This class of modifications has been also studied in the two-component spinor formalism by considering a minimally modified Plebanski's action and a perturbative approach~\cite{cmv201027}. (ii) The coupling of the theories (\ref{krasnovAct}) to Yang-Mills fields can be done directly by means of the $B$ fields ~\cite{kras2008-25-2,speziale2009-26}, in the same fashion as in section~\ref{pleb-matter}, and even though these 2-forms do not satisfy the simplicity constraint of the Plebanski formulation. However, the same strategy of~\cite{capo1991841} for the coupling to the scalar and fermionic fields has not been extended to these modifications.

\subsubsection{Canonical analysis.}

We now review the Hamiltonian description of the class of theories (\ref{krasnovAct})~\cite{beng20072222,kras2008100}. Here we use the notation and terminology of section~\ref{Pleb-HamilAna}. To simplify the analysis, we follow the same approach of~\cite{kras2008100}; that is, after the (3+1)-decomposition of the action we consider a convenient decomposition of the Lagrange multiplier $B_{tai}$ that, despite the arbitrariness of the function $\Phi(\phi)$, allows us to express $\phi_{ij}$ in terms of phase space variables.

The (3+1)-decomposition of the action principle~(\ref{krasnovAct}) gives
\begin{eqnarray}
\fl S[A, B,\phi]=\int_{\mathbb{R}}dt\int_{\Omega}d^3x \left[\tilde{\Pi}^{ai} \dot{A}_{ai} +A_{ti} \tilde{\mathcal{G}}^i +  B_{tai} \tilde{E}^{ai} - \phi_{ij}  B_{ta}{}^i\tilde{\Pi}^{aj} -\Phi B_{tai} \tilde{\Pi}^{ai} \right], \label{krasmod-Hamil1}
\end{eqnarray}
where we have introduced $\tilde{\Pi}^{ai}:=(1/2)\tilde{\eta}^{abc}B_{bc}{}^i$, $\tilde{E}^{ai}=\tilde{B}^{ai}-\Lambda\tilde{\Pi}^{ai}$ with $\tilde{B}^{ai}:=(1/2)\tilde{\eta}^{abc}F^i{}_{bc}$, and $\tilde{\mathcal{G}}^i:=\mathcal{D}_a\tilde{\Pi}^{ai}$, which is the usual expression for the Gauss constraint. We recall that  $\mathcal{D}_a$ is the $SO(3,\mathbb{C})$-covariant derivative corresponding to the spatial connection $A_a{}^i$. Since $\tilde{\Pi}^{ai}$ is assumed to be invertible, by decomposing the term $B_{ta}{}^i\tilde{\Pi}^{aj}$ into its symmetric and antisymmetric parts, we can write the Lagrange multiplier $B_{tai}$ as \begin{eqnarray}
B_{tai}=\left[-\tilde{\rho} (\delta_{ij}+N_{ij}) + \varepsilon_{ijk} \tilde{N}^k  \right] \underaccent{\tilde}{\Pi}_a{}^j, \label{krasmod-constphisol}
\end{eqnarray}
where $N_{ij}$ is a symmetric traceless matrix, $\tilde{\rho}$ and $\tilde{N}^k$ are four arbitrary functions, and $\underaccent{\tilde}{\Pi}_{ai}$ is the inverse of $\tilde{\Pi}^{ai}$. Therefore, the 9 components $B_{tai}$ can be replaced with the 9 components $N_{ij}$, $\tilde{\rho}$, and $\tilde{N}^k$. In terms of these variables, the action~(\ref{krasmod-Hamil1}) reads
\begin{eqnarray}
\fl S[A_{a},\tilde{\Pi}^{a},A_{t},N,\tilde{N},\tilde{\rho},\phi]=& \int_{\mathbb{R}}dt  \int_{\Omega} d^3x \big[ \tilde{\Pi}^{ai} \dot{A}_{ai} +A_{ti} \tilde{\mathcal{G}}^i -  \tilde{\rho} \left(\underaccent{\tilde}{\Pi}_{ai} \tilde{E}^{ai} - 3 \Phi   \right) \nonumber \\ 
&+ \tilde{\rho} N_{ij} \left( \phi^{ij}  - \tilde{B}^{ai} \underaccent{\tilde}{\Pi}_{a}{}^j \right)    + \varepsilon_{ijk}  \tilde{B}^{ai}\underaccent{\tilde}{\Pi}_a{}^j \tilde{N}^k \big]. \label{krasmod-Hamil2}
\end{eqnarray}
Then, the equations of motion that result from the variation of the action with respect to $N_{ij}$, $\tilde{N}^i$, and $\phi_{ij}$ are
\begin{eqnarray}
\delta N_{ij}&:& \phi^{ij}=\left( \tilde{B}^{ai} \underaccent{\tilde}{\Pi}_{a}{}^{j} \right)_{\rm tf},\label{krasmod-N}\\
\delta \tilde{N}^i&:&  \varepsilon_{ijk}  \tilde{B}^{ai}\underaccent{\tilde}{\Pi}_a{}^j = 0, \label{krasmod-Nvect} \\
\delta \phi_{ij}&:& N_{ij}=-3 (\Theta_{ij})_{\rm tf},\label{krasmod-phi}
\end{eqnarray}
where ``${\rm tf}$" means trace-free part and $\Theta^{ij}$ was defined in (\ref{krasmod-theta}). Notice that the right-hand side of equation~(\ref{krasmod-N}) is symmetric as a consequence of equation~(\ref{krasmod-Nvect}). Also, from~(\ref{krasmod-phi}) it is clear that in the case where the modification disappears, we have $N_{ij}=0$ and then equation~(\ref{krasmod-constphisol}) reduces to equation~(\ref{PlebH:constPsisol}) of the Plebanski theory.

Substituting equations~(\ref{krasmod-N}) and (\ref{krasmod-phi}) into the action~(\ref{krasmod-Hamil2}), it takes the Hamiltonian form 
\begin{eqnarray}
\fl S[A_{a},\tilde{\Pi}^{a},A_{t},N^a,\underaccent{\tilde}{N}]=\int_{\mathbb{R}}dt\int_{\Omega}d^3x\left(\tilde{\Pi}^{ai}\dot{A}_{ai}+A_{ti}\tilde{\mathcal{G}}^i+N^a\tilde{\mathcal{V}}_a+\underaccent{\tilde}{N}\tilde{\tilde{\mathcal{H}}}\right),\label{krasmod:hamil}
\end{eqnarray}
where we have made the change of variables $N^a:=-(\det\tilde{\Pi})^{-1}\tilde{\Pi}^a{}_i \tilde{N}^i$ and $\underaccent{\tilde}{N}:=-3(\det\tilde{\Pi})^{-1}\tilde{\rho}$. The phase space variables are $(A_{ai},\tilde{\Pi}^{ai})$, while $A_{ti}$, $N^a$ and $\underaccent{\tilde}{N}$ are Lagrange multipliers imposing the constraints
\begin{eqnarray}
\tilde{\mathcal{G}}^i&=&\mathcal{D}_a\tilde{\Pi}^{ai}\approx 0,\label{krasmod-gauss}\\
\tilde{\mathcal{V}}_a&:=&\tilde{\Pi}^{bi}F_{iba}\approx 0,\label{krasmod-Vect}\\
\tilde{\tilde{\mathcal{H}}}&:=&\frac{1}{6}\underaccent{\tilde}{\eta}_{abc}\varepsilon_{ijk}\tilde{\Pi}^{ai}\tilde{\Pi}^{bj}\tilde{B}^{ck}-\frac{1}{6} \left ( \Lambda + \Phi \right ) \underaccent{\tilde}{\eta}_{abc}\varepsilon_{ijk}\tilde{\Pi}^{ai}\tilde{\Pi}^{bj}\tilde{\Pi}^{ck}\approx 0,\label{krasmod-scalar}
\end{eqnarray}
where $\Phi=\Phi ({\rm Tr}\phi^2, {\rm Tr}\phi^3)$ with the matrix $\phi^{ij}$ being 
\begin{eqnarray}
\phi^{ij}:=\frac{1}{2 \det \tilde{\Pi}} \left(\varepsilon^{klj} F^i{}_{ab} \tilde{\Pi}^{a}{}_k \tilde{\Pi}^{b}{}_l\right)_{\rm tf}. \label{krasmod-def-phi}
\end{eqnarray}
Notice that because of equation~(\ref{krasmod-Vect}), the right-hand side of (\ref{krasmod-def-phi}) is symmetric. This is the same phase space described in~\cite{kras2008100}; equations~(\ref{krasmod-gauss}) and (\ref{krasmod-Vect}) are respectively the Gauss and vector constraints of the Ashtekar formulation of general relativity, namely (\ref{PlebH:gauss}) and (\ref{PlebH:Vect}), whereas (\ref{krasmod-scalar}) is a modified scalar constraint where the cosmological constant is promoted to the function $\Lambda + \Phi(\phi)$ that dependents on the symmetric traceless $\phi^{ij}$ defined by (\ref{krasmod-def-phi}).  In the particular case when $\Phi$ is constant, we recover the usual scalar constraint  of the Ashtekar formalism, modulo a redefinition of the cosmological constant. It can be verified that the constraints (\ref{krasmod-gauss})-(\ref{krasmod-scalar}) are first class~\cite{kras2008100}, implying that the number of (complex) degrees of freedom per space point is two, as in general relativity.

\paragraph{\bf Remark.}  While these modifications propagate two complex degrees of freedom, the canonical analysis of the action that implements the same modifications in real general relativity has locally eight degrees of freedom~\cite{kras2099-26-5} (the real case has been related to certain bimetric theories of gravity~\cite{speziale2010-82,Beke2012}).

\subsection{Real general relativity as a constrained $BF$ theory}\label{sect_Real_F}

\subsubsection{Action principle.}\label{BF4_act}

The Plebanski formulation introduced to describe complex general relativity as a constrained $BF$ theory can be adapted to describe real general relativity. Since in this section we deal with real general relativity, we omit the word real from now on. Let us consider a four-dimensional manifold $\mathcal{M}$.  As internal group we choose $SO(1,3)$ or $SO(4)$, which we denote $SO(\sigma)$. The internal indices are then raised and lowered with the metric $(\eta_{IJ})=\textrm{diag}(\sigma,1,1,1)$, where $\sigma=-1$ ($\sigma=+1$) for the Lorentzian (Euclidean) case. The action principle that describes general relativity as a constrained $BF$ theory (recall that $BF$ theories define topological field theories; see~\cite{MondMontJMP472} for the particular $BF$ term involved in this section) has the form
\begin{eqnarray}\label{actBF4R}
\fl S[A,B,\varphi,\mu]=\int_\mathcal{M} \left[B^{IJ}\wedge F_{IJ}[A]-\varphi_{IJKL}B^{IJ}\wedge B^{KL}-\mu G(\varphi)\right],
\end{eqnarray}
where $B^{IJ}$ is a set of $\mathfrak{so}(\sigma)$-valued 2-forms, $A^{IJ}$ is an $\mathfrak{so}(\sigma)$-valued connection 1-form, $F^{IJ}:=dA^{IJ}+A^I{}_K\wedge A^{KJ}$ is the curvature of $A^{IJ}$, $\varphi_{IJKL}$ is an internal Lagrange multiplier satisfying $\varphi_{IJKL}=\varphi_{[IJ][KL]}=\varphi_{KLIJ}$ (alternatively, we can use a spacetime tensor density $\tilde{\varphi}^{\mu\nu\lambda\sigma}$ with the same index symmetries as $\varphi_{IJKL}$; see~\cite{BuffHennetcqg2122}), and $\mu$ is a 4-form enforcing the constraint $G(\varphi)=0$. This constraint is nothing but a restriction on the trace of $\varphi$, something reminiscent of the condition imposed on the trace of the multiplier $\Psi$ in the Plebanski formulation (see equation (\ref{pleban4})). Since $\mathfrak{so}(\sigma)$ has two Killing forms, namely $\eta_{I[J|}\eta_{K|L]}$ and $\varepsilon_{IJKL}$, there are several ways of taking the trace: either we can use each of the Killing forms separately or we can make a linear combination of them to take it. As a consequence of that, the function $G(\varphi)$ has the following forms~\cite{capo1991859,Reisencqg147,PietriFrei,CMPR2001}
\begin{eqnarray}\label{G_func}
\fl G_1(\varphi)=\tensor{\varphi}{^{IJ}_{IJ}}, \quad G_2(\varphi)=\varepsilon_{IJKL}\varphi^{IJKL}, \quad G_3(\varphi)=a_1\tensor{\varphi}{^{IJ}_{IJ}}+a_2\varepsilon_{IJKL}\varphi^{IJKL}, 
\end{eqnarray}
where $a_1$ and $a_2$ are constants. Let us take the constraint $G_3=0$, for which the action (\ref{actBF4R}) is known as the CMPR action~\cite{CMPR2001}. From (\ref{actBF4R}), the equation of motion corresponding to the Lagrange multiplier $\varphi$ imposes the following constraint on the $B$ field
\begin{eqnarray}\label{Simplic_Re}
	B^{IJ}\wedge B^{KL}+\mu\left(a_1\eta^{I[K|}\eta^{J|L]}+a_2\varepsilon^{IJKL}\right)=0.
\end{eqnarray}
This set of relations is known as {\it simplicity constraint} and has the solution
\begin{eqnarray}\label{SolSimp}
	B^{IJ}=\alpha *(e^I\wedge e^J)+\beta e^I\wedge e^J,
\end{eqnarray}
where $e^I$ is an $SO(\sigma)$-frame assumed to be invertible, $*$ is an internal Hodge dual operator defined by $*\omega_{IJ}=(1/2)\varepsilon_{IJKL}\omega^{KL}$ for any $\mathfrak{so}(\sigma)$-valued object $\omega$, and $\alpha$, $\beta$ are real parameters which are related to $a_1$, $a_2$ via
\begin{eqnarray}\label{relparam}
	\frac{a_2}{a_1}=\frac{\alpha^2+\sigma\beta^2}{4\alpha\beta}.
\end{eqnarray}

It is interesting to note that by substituting (\ref{SolSimp}) back into the action (\ref{actBF4R}), it reduces to the Holst action~\cite{HolstPRD53}
\begin{eqnarray}\label{Holst4d}
	S[e,A]=\int_{\mathcal{M}}\left[\alpha *(e^I\wedge e^J)+\beta e^I\wedge e^J\right]\wedge F_{IJ}[A],
\end{eqnarray}
from which we see that the quotient $\alpha/\beta$ corresponds, modulo a sign, to the Immirzi parameter~\cite{Immirzicqg1410}. This means that by using the function $G_3$ in (\ref{actBF4R}), we are including an intrinsic Immirzi parameter into the $BF$ formulation of general relativity.

Since $G_3$ is a linear combination of $G_1$ and $G_2$, let us now consider both cases separately. The case of $G_1=0$ corresponds to $a_1\neq 0$, $a_2=0$, and, according to (\ref{relparam}), the parameters $\alpha$, $\beta$ turn out to be related via $\beta=\pm\sqrt{-\sigma}\alpha$; this means that real solutions only exist for Lorentzian signature. On the other hand, the case of $G_2=0$ corresponds to $a_1=0$, $a_2\neq 0$, and from (\ref{relparam}) we obtain the possibilities $\alpha\neq 0$ and $\beta=0$, or $\alpha=0$ and $\beta\neq 0$; the former reduces (\ref{Holst4d}) to Palatini's action, while the latter collapses it to the Holst term, which defines a topological field theory~\cite{LiuPRD81}.

\subsubsection{Deriving Einstein's equations.}\label{BFEE_derive}

In this section we use a slightly modified version of the action (\ref{actBF4R}) to show that the equations of motion arising from it lead us to the Einstein field equations. In order to include the Immirzi parameter (which is denoted by $\gamma$) explicitly into the action, we add to the action (\ref{actBF4R}) a $BF$ term involving the internal Hodge dual of the $B$ field~\cite{Montescqg237}  and consider the constraint $G_2=0$ instead of $G_3=0$. The resulting action principle, which is related to the CMPR action via an invertible linear transformation~\cite{MercSIGMA72011}, is given by
\begin{eqnarray}\label{ActBFImm}
 \fl S[A,B,\varphi,\mu]\! =\!\! \int_{\mathcal{M}}\! \left[ \! \left( \! B^{IJ} \! +\frac{1}{\gamma}*B^{IJ}\right)\! \wedge F_{IJ}[A]-\varphi_{IJKL}B^{IJ}\wedge B^{KL}-\mu\varepsilon^{IJKL}\varphi_{IJKL} \right]. \nonumber\\
\end{eqnarray}
By varying the action with respect to the independent variables, we arrive at the following equations of motion
\begin{eqnarray}
	\delta\varphi_{IJKL}&:& B^{IJ}\wedge B^{KL}+\mu\varepsilon^{IJKL}=0,\label{variatphi}\\
	\delta A^{IJ}&:& D\left(B^{IJ}+\frac{1}{\gamma}*B^{IJ}\right)=0,\label{variatA}\\
	\delta B^{IJ}&:& F_{IJ}+\frac{1}{\gamma}*F_{IJ}-2\varphi_{IJKL}B^{KL}=0,\label{variatB}\\
		\delta\mu&:& \varepsilon^{IJKL}\varphi_{IJKL}=0,\label{variatmu}
\end{eqnarray}
where $D$ is the $SO(\sigma)$-covariant derivative. Equation (\ref{variatA}) says that the $B$ field is covariantly constant, while equation (\ref{variatB}) relates the curvature $F^{IJ}$ to both the Lagrange multiplier $\varphi$ and the $B$ field. The remaining equations were already introduced in section~\ref{BF4_act}. As we saw at the end of the previous section, the equation (\ref{variatphi}) has the solutions
\begin{eqnarray}
	B^{IJ}&=\kappa_1*(e^I\wedge e^J),\label{solveq1}\\
	B^{IJ}&=\kappa_2 e^I\wedge e^J,\label{solveq2}
\end{eqnarray} 
where $\kappa_1$, $\kappa_2$ are constants. They both reduce the action (\ref{ActBFImm}) to the Holst action, being $\gamma$ and $\gamma^{-1}$ the Immirzi parameter, respectively. This shows that the action (\ref{ActBFImm}) indeed describes general relativity with the Immirzi parameter. Notice that the frame $e^I$ determines the spacetime metric $g_{\mu\nu}=\eta_{IJ}\tensor{e}{_{\mu}^I}\tensor{e}{_{\nu}^J}$, which coincides up to a constant factor with the Urbantke metric defined by the $B$'s~\cite{PietriFrei,CMPR2001}.

Substituting  (\ref{solveq1}) or (\ref{solveq2}) into equation (\ref{variatA}), it reduces to the condition $De^{I}=0$ in either case provided that $\gamma^2\neq\sigma$, which we assume throughout this section. Therefore, the $SO(\sigma)$-connection $A^{IJ}$ is torsion-free, and, as such, becomes the spin-connection compatible with the frame $e^I$: $A=\omega(e)$. The components of the curvature $F=F[\omega(e)]$ then satisfy the Bianchi identities for vanishing torsion
\begin{eqnarray}\label{Bianchi_1}
	F_{I[JKL]}=0,
\end{eqnarray}
where we have set $F_{IJ}=(1/2)F_{IJKL}e^K\wedge e^L$.

\paragraph{Case $B^{IJ}=\kappa_1*(e^I\wedge e^J)$.}

Equation (\ref{variatB}) can be solved for the components of the curvature, which yields
\begin{eqnarray}\label{compF_1}
	F_{IJKL}=\frac{4\gamma^2\kappa_1}{\gamma^2-\sigma}\left(\varphi^{*}_{IJKL}-\frac{1}{\gamma}{}^{*}\varphi^{*}_{IJKL}\right),
\end{eqnarray}
with ${}^{*}\varphi_{IJKL}:=(1/2)\tensor{\varepsilon}{_{IJ}^{MN}}\varphi_{MNKL}$ and $\varphi^{*}_{IJKL}:=(1/2)\varphi_{IJMN}\tensor{\varepsilon}{^{MN}_{KL}}$ the left and right internal Hodge dual operators, respectively. The Bianchi identities (\ref{Bianchi_1}) then impose the following restrictions on the Lagrange multiplier $\varphi$
\begin{eqnarray}\label{RestBi_1}
\fl	\varphi^{*}_{IJKL}+\varphi^{*}_{IKLJ}+\varphi^{*}_{ILJK}-\frac{1}{\gamma}\left({}^{*}\varphi^{*}_{IJKL}+{}^{*}\varphi^{*}_{IKLJ}+{}^{*}\varphi^{*}_{ILJK}\right)=0.
\end{eqnarray}

Up to here we have not used the equation of motion (\ref{variatmu}) yet. We now compute the Ricci tensor from (\ref{compF_1}). Contracting the first and third indices of (\ref{compF_1}) we obtain, after using (\ref{RestBi_1}),
\begin{eqnarray}\label{Ricci_1}
	\mathcal{R}_{IJ}:=\tensor{F}{^K_I_K_J}=\frac{\kappa_1}{2}\varepsilon_{KLMN}\varphi^{KLMN}\eta_{IJ},
\end{eqnarray}
which is proportional to the constraint (\ref{variatmu}). From this, we conclude that the Ricci tensor vanishes, which is Einstein's equations in vacuum. Notice that the Immirzi parameter has completely dropped out from the equations of motion, as expected due to its topological character.

\paragraph{Case $B^{IJ}=\kappa_2\ e^I\wedge e^J$.}
From (\ref{variatB}), the components of the curvature are now given by
\begin{eqnarray}\label{compF_2}
	F_{IJKL}=\frac{4\gamma^2\kappa_2}{\gamma^2-\sigma}\left(\varphi_{IJKL}-\frac{1}{\gamma}{}^{*}\varphi_{IJKL}\right),
\end{eqnarray}
which implies the relations
\begin{eqnarray}\label{RestBi_2}
\fl	\varphi_{IJKL}+\varphi_{IKLJ}+\varphi_{ILJK}-\frac{1}{\gamma}\left({}^{*}\varphi_{IJKL}+{}^{*}\varphi_{IKLJ}+{}^{*}\varphi_{ILJK}\right)=0, 
\end{eqnarray}
once the Bianchi identities (\ref{Bianchi_1}) are taken into account. Using (\ref{compF_2}) and (\ref{RestBi_2}), the Ricci tensor takes the form
\begin{eqnarray}
	\mathcal{R}_{IJ}=\frac{\sigma\gamma\kappa_2}{2}\varepsilon_{KLMN}\varphi^{KLMN}\eta_{IJ},
\end{eqnarray}
which vanishes as a consequence of (\ref{variatmu}). It follows that Einstein's equations in vacuum are also satisfied. 

In this way, we have shown that the equations of motion arising from the action principle (\ref{ActBFImm}) imply the vacuum Einstein field equations as required. This action principle and the CMPR action have been used as the starting point of some spinfoam models for gravity~\cite{perez2013-16-3,EPRnuc798,Livoriprd654}.

\subsubsection{Canonical analysis.}\label{canon_analy_R}

This section is devoted to the canonical analysis of the action principle (\ref{ActBFImm}), which we perform in a Lorentz-covariant fashion. Since second-class constraints arise during the process, we solve them explicitly while keeping the full gauge symmetry of the theory and show that by gauge fixing the boosts, we recover the Ashtekar-Barbero variables~\cite{Barberoprd5110} used in the canonical approach to quantum gravity. The issue of Lorentz covariance has gained some relevance in the last years due to some recent results from loop quantum gravity. It is well known that in the current approach to loop quantum gravity based on an $SU(2)$-connection, the eigenvalues of the area operator~\cite{AshtLewacqg14,RoveSmolinNucP442} as well as the black hole entropy~\cite{RovPRLt.77.3288,AshBaezcqg105,Meisscqg21,Agulloetcprl100,EngleNuiPRL105} depend explicitly on the Immirzi parameter, which has no role at the classical level because of its topological character, but becomes a quantization ambiguity~\cite{RovThiePRD57}. However, new findings indicate that one can get rid of this parameter from the quantum theory by choosing Lorentz-covariant connection variables~\cite{Alexprd652,AlexVassprd644}. The issue seems to be still open, since, on the other hand, it is also possible to build a Lorentz-covariant connection for which the eigenvalues of the area operator are the same as those of the $SU(2)$ theory, that is, $\gamma$-dependent~\cite{AlexLivprd674,NouiSIGMA72011, Nouiprd84044002} (see also~\cite{Nouiprd9110} for a symmetry-reduced model). Therefore, a Lorentz-covariant canonical analysis of the action (\ref{ActBFImm}) seems required not only as a mere exercise, but in order to provide new insights into the relevance of the Immirzi parameter either in loop quantum gravity or in the spinfoam models for gravity.

\paragraph{\bf Remark.} Historically, the Lorentz-covariant Hamiltonian analysis of $BF$ gravity was first performed on the formulation that employs the Lagrange multiplier with spacetime indices~\cite{BuffHennetcqg2122,Alexandrovcqg2007}. The canonical analysis of the action using the multiplier with internal indices was later carried out in~\cite{kras2099-26-5}, and including the Immirzi parameter  in~\cite{CelMontesRom,CelMontcqg2920}.

\paragraph{{\rm (}$3+1${\rm )}-decomposition of the action.} \label{decomposeact_r}
Let us define the $\gamma$-valued quantities ${}^{\scriptscriptstyle(\gamma)}\omega$ by ${}^{\scriptscriptstyle(\gamma)}\omega{}^{IJ}:=\omega{}^{IJ}+\gamma^{-1}*\omega{}^{IJ}$ for any $\mathfrak{so}(\sigma)$-valued object $\omega$. As in section~\ref{Pleb-HamilAna}, we assume that the spacetime has the topology $\mathbb{R}\times\Omega$,  where $\Omega$ is a spatial three-dimensional manifold without a boundary. The (3+1)-decomposition of the action (\ref{ActBFImm}) takes the form
\begin{eqnarray}
	\fl S[A,B,\phi,\mu]=\int_{\mathbb{R}}dt\int_{\Omega}d^3x&\left[{}^{\scriptscriptstyle(\gamma)}\tilde{\Pi}\tensor{}{^{aIJ}}\dot{A}_{aIJ}+A_{tIJ}D_a{}^{\scriptscriptstyle(\gamma)}\tilde{\Pi}\tensor{}{^{aIJ}}+\frac{1}{2}\tilde{\eta}^{abc}{}^{\scriptscriptstyle(\gamma)}F\tensor{}{_{IJab}}\tensor{B}{_{tc}^{IJ}}\right.\nonumber\\
	&\ -\left(2\tensor{B}{_{ta}^{IJ}}\tilde{\Pi}^{aKL}+\tilde{\mu}\varepsilon^{IJKL}\right)\varphi_{IJKL}\biggr],\label{decompBFR}
\end{eqnarray}
where $\tilde{\Pi}^{aIJ}:=(1/2)\tilde{\eta}^{abc}\tensor{B}{_{bc}^{IJ}}$. Notice that the action (\ref{decompBFR}) indicates that the canonical pairs are either $({}^{\scriptscriptstyle(\gamma)}A_{aIJ},\tilde{\Pi}^{aIJ})$ or $(A_{aIJ},{}^{\scriptscriptstyle(\gamma)}\tilde{\Pi}{}^{aIJ})$. We shall follow the same strategy we used for getting the Hamiltonian formulation of the Plebanski action, which consists in solving the equation of motion corresponding to $\varphi_{IJKL}$ for $B_{taIJ}$, thus reducing the number of variables involved in the analysis. First, the (3+1)-decomposition of equation (\ref{variatphi}) yields
\begin{equation}
	\tensor{B}{_{ta}^{IJ}}\tilde{\Pi}^{aKL}+\tensor{B}{_{ta}^{KL}}\tilde{\Pi}^{aIJ}-\frac{\sigma}{4}\tilde{\mathcal{V}}\varepsilon^{IJKL}=0,\label{decompConst}
\end{equation}
where $\tilde{\mathcal{V}}:=(1/3)\varepsilon_{IJKL}\tensor{B}{_{ta}^{IJ}}\tilde{\Pi}^{aKL}$ is the four-dimensional volume, which we assume nonvanishing anywhere. It can be shown that the solution of this equation is given by~\cite{CelMontcqg2920,kras2099-26-5}
\begin{eqnarray}
	\tensor{B}{_{ta}^{IJ}}=\frac{1}{8}\underaccent{\tilde}{N}h_{ab}\varepsilon^{IJKL}\tensor{\tilde{\Pi}}{^b_{KL}}+\frac{1}{2}\underaccent{\tilde}{\eta}_{abc}\tilde{\Pi}^{bIJ}N^{c}, \label{solvConst}
\end{eqnarray}
together with the constraints
\begin{eqnarray}
	\tilde{\tilde{\Phi}}^{ab}:=-\sigma*\tensor{\tilde{\Pi}}{^a_{IJ}}\tilde{\Pi}^{bIJ}=0,\label{simpliccan}
\end{eqnarray}
where $h_{ab}$ is the inverse of the matrix defined by $h h^{ab}:=(\sigma/2)\tensor{\tilde{\Pi}}{^a_{IJ}}\tilde{\Pi}^{bIJ}$ with $h:=\det(h_{ab})$ (having weight 2), and $\underaccent{\tilde}{N}$ and $N^a$ are an arbitrary scalar density and an arbitrary 3-vector, respectively. This shows that, from the 20 equations (\ref{decompConst}), six of them become the constraints (\ref{simpliccan}) on $\tilde{\Pi}^{aIJ}$, while the  remaining 14 allow us to express the 18 components $\tensor{B}{_{ta}^{IJ}}$ in terms of $\tilde{\Pi}^{aIJ}$ and the four free functions $\underaccent{\tilde}{N}$ and $N^a$.

By substituting equation (\ref{solvConst}) back into the action (\ref{decompBFR}), and taking into account the constraint (\ref{simpliccan}), the action takes the Hamiltonian form
\begin{eqnarray}
	\fl S[A_a,\tilde{\Pi}^a,A_t,\underaccent{\tilde}{N},N^a,\underaccent{\tilde}{\lambda}_{ab}]\nonumber\\
\hspace{-6mm}=\int_{\mathbb{R}}dt\int_{\Omega}d^3x\left[{}^{\scriptscriptstyle(\gamma)}\tilde{\Pi}^{aIJ}\dot{A}_{aIJ}
+A_{tIJ}\tilde{\mathcal{G}}^{IJ}+\underaccent{\tilde}{N}\tilde{\tilde{\mathcal{H}}}+N^a\tilde{\mathcal{H}}_a+\underaccent{\tilde}{\lambda}_{ab}\tilde{\tilde{\Phi}}^{ab}\right],\label{HamiltAct} 
\end{eqnarray}
where $\underaccent{\tilde}{\lambda}_{ab}$ is a Lagrange multiplier enforcing the constraint (\ref{simpliccan}). Therefore, since $A_{tIJ}$, $\underaccent{\tilde}{N}$, and $N^a$ appear linearly in the action, they also play the role of Lagrange multipliers and, together with $\underaccent{\tilde}{\lambda}_{ab}$, impose the primary constraints
\begin{eqnarray}
	\tilde{\mathcal{G}}^{IJ}:= D_a{}^{\scriptscriptstyle(\gamma)}\tilde{\Pi}^{aIJ}\approx 0,\label{gauss4dr}\\
	\tilde{\tilde{\mathcal{H}}}:=\frac{1}{4}\tilde{\eta}^{abc}h_{ad}*{}^{\scriptscriptstyle(\gamma)}\tilde{\Pi}^{dIJ} F_{IJbc}\approx 0,\label{scal4dr}\\
	\tilde{\mathcal{H}}_a:=\frac{1}{2}{}^{\scriptscriptstyle(\gamma)}\tilde{\Pi}^{bIJ}F_{IJba}\approx 0,\label{vect4dr}\\
\tilde{\tilde{\Phi}}^{ab}=\eta\left[{}^{\scriptscriptstyle(\gamma)}\tilde{\tilde{\Phi}}^{ab}+\frac{4\sigma\gamma}{\gamma^{2}+\sigma}{}^{\scriptscriptstyle(\gamma)}(hh^{ab})\right]\approx 0,\label{simplipsv}
\end{eqnarray}
where $\eta:=\gamma^2(\gamma^2+\sigma)/(\gamma^2-\sigma)^2$, and the quantities ${}^{\scriptscriptstyle(\gamma)}\tilde{\tilde{\Phi}}^{ab}$ and ${}^{\scriptscriptstyle(\gamma)}(hh^{ab})$ are $\tilde{\tilde{\Phi}}^{ab}$ and $hh^{ab}$ with $\tilde{\Pi}^{aIJ}$ replaced with ${}^{\scriptscriptstyle(\gamma)}\tilde{\Pi}^{aIJ}$, respectively. Notice that here we took $(A_{aIJ},{}^{\scriptscriptstyle(\gamma)}\tilde{\Pi}^{aIJ})$ as the phase space variables, and that is why the constraints (\ref{gauss4dr})-(\ref{simplipsv}) have been written in terms of them. 

It turns out that the Poisson algebra of the constraints (\ref{gauss4dr})-(\ref{simplipsv}) closes, except for the Poisson bracket between $\tilde{\tilde{\mathcal{H}}}$ and $\tilde{\tilde{\Phi}}^{ab}$. The evolution of $\tilde{\tilde{\Phi}}^{ab}$ then leads to the secondary constraint
\begin{eqnarray}
\fl	\Psi^{ab}:= -2\eta h_{cf}\left(-{}^{\scriptscriptstyle(\gamma)}\tensor{\tilde{\Pi}}{^f_{IJ}}+\frac{2\gamma}{\gamma^{2}+\sigma}*{}^{\scriptscriptstyle(\gamma)}\tensor{\tilde{\Pi}}{^f_{IJ}}\right)\tilde{\eta}^{(a|cd}D_d{}^{\scriptscriptstyle(\gamma)}\tensor{\tilde{\Pi}}{^{|b)IJ}}\approx 0.\label{secund4dr}
\end{eqnarray}
In turn, because the Poisson bracket between this constraint and $\tilde{\tilde{\Phi}}^{ab}$ defines a nonsingular matrix, the evolution of $\Psi^{ab}$ does not generate further constraints and allows us to fix the Lagrange multipliers $\underaccent{\tilde}{\lambda}_{ab}$~\cite{CelMontcqg2920}. 

The classification of the constraints is as follows. The constraints $\tilde{\mathcal{G}}^{IJ}$, $\tilde{\mathcal{H}}_a$, and\footnote{We must shift $\tilde{\tilde{\mathcal{H}}}$ by a term proportional to the constraint $\tilde{\tilde{\Phi}}^{ab}$ coming from the fixing of the Lagrange multiplier $\underaccent{\tilde}{\lambda}_{ab}$ so that the Hamiltonian of the theory be truly first class (see~\cite{peld1994115,CelMontcqg2920}).} $\tilde{\tilde{\mathcal{H}}}$ are first class and generate the gauge symmetries of the theory: $\tilde{\mathcal{G}}^{IJ}$ generates $SO(\sigma)$-local transformations, whereas $\tilde{\mathcal{H}}_a$ and $\tilde{\tilde{\mathcal{H}}}$ generate spatial and time diffeomorphisms, respectively. On the other hand, the constraints $\tilde{\tilde{\Phi}}^{ab}$ and $\Psi^{ab}$ are second class. Since our phase space was initially parametrized by the pairs $(A_{aIJ},{}^{\scriptscriptstyle(\gamma)}\tilde{\Pi}^{aIJ})$, the initial number of variables was 36. However, bearing in mind that the theory possesses 10 first-class and 12 second-class constraints, the reduced phase space is left with only four local degrees of freedom, producing a physical configuration space with the two propagating degrees of freedom of general relativity.

The canonical analysis of the CMPR action of section~\ref{BF4_act}  can also be performed in an analogous way. In that case, one obtains that the kinematical phase space is now described by the pairs $(A_{aIJ},\tilde{\Pi}^{aIJ})$ subject to the constraints
\begin{eqnarray}
	&\tilde{\mathcal{g}}^{IJ}:= D_a\tilde{\Pi}^{aIJ}\approx 0,\label{gaussCMPR}\\
	&\tilde{\tilde{\mathcal{S}}}:=\frac{1}{4}\tilde{\eta}^{abc}h_{ad}*\tilde{\Pi}^{dIJ}F_{IJbc}\approx 0,\label{scalarCMPR}\\
	&\tilde{\mathcal{V}}_a:=\frac{1}{2}\tilde{\Pi}^{bIJ}F_{IJba}\approx 0,\label{vectorCMPR}\\
	&\tilde{\tilde{\varphi}}^{ab}:=\tilde{\tilde{\Phi}}^{ab}+\frac{a_1}{a_2}hh^{ab}\approx 0,\label{PhiCMPR}\\
	&\psi^{ab}:=\frac{1}{2}h_{cf}\left(-\tensor{\tilde{\Pi}}{^f_{IJ}}+\frac{\sigma a_1}{2a_2}*\tensor{\tilde{\Pi}}{^f_{IJ}}\right)\tilde{\eta}^{(a|cd}D_d\tilde{\Pi}^{|b)IJ}\approx 0.\label{PsiCMPR}
\end{eqnarray}
Notice that if we replace $\tilde{\Pi}^{aIJ}$ with ${}^{\scriptscriptstyle(\gamma)}\tilde{\Pi}^{aIJ}$ in the constraints (\ref{gaussCMPR})-(\ref{vectorCMPR}), then they become exactly the same constraints (\ref{gauss4dr})-(\ref{vect4dr}), correspondingly\footnote{The quantities $h_{ab}$ showing up in the constraints (\ref{scal4dr}) and (\ref{scalarCMPR}) are in principle different, since they have the same dependency on $\tilde{\Pi}$, but differ from each other when they are written in terms of the phase space variables. However, we can show that the quantity $hh^{ab}$ defined below equation (\ref{simpliccan}) takes the form $hh^{ab}=c_1{}^{\scriptscriptstyle(\gamma)}(hh^{ab})+c_2\tilde{\tilde{\Phi}}^{ab}$ when it is expressed in terms of the canonical variable ${}^{\scriptscriptstyle(\gamma)}\tilde{\Pi}^{aIJ}$, where $c_1$ and $c_2$ are constants. Since the second term on the right-hand side is proportional to the constraint $\tilde{\tilde{\Phi}}^{ab}$, we can neglect it (alternatively, we can redefine some of the Lagrange multipliers inside the action to absorb the terms proportional to the constraint $\tilde{\tilde{\Phi}}^{ab}$), yielding $hh^{ab}\propto{}^{\scriptscriptstyle(\gamma)}(hh^{ab})$. Thus, the replacement $\tilde{\Pi}^{aIJ}$ with ${}^{\scriptscriptstyle(\gamma)}\tilde{\Pi}^{aIJ}$ in the quantity $hh^{ab}$ for the case of the CMPR action will make it equal (up to a constant factor) to the corresponding one for the action (\ref{ActBFImm}), and the same applies for $h_{ab}$.}. On the other side, the constraints (\ref{PhiCMPR}) and (\ref{PsiCMPR}) coincide with (\ref{simplipsv}) and (\ref{secund4dr}), respectively, if the same replacement is made on them and besides we identify $a_1/a_2$ with $4\sigma\gamma/(\gamma^{2}+\sigma)$; this identification is precisely the one stated in equation (\ref{relparam}) for $\gamma=\sigma\alpha/\beta$. We conclude that the canonical variable $\tilde{\Pi}^{aIJ}$ used in the canonical analysis of the CMPR action is the $\gamma$-valued variable ${}^{\scriptscriptstyle(\gamma)}\tilde{\Pi}^{aIJ}$ used in the canonical analysis of the action (\ref{ActBFImm}) for which $\gamma$ satisfies the relation $a_1/a_2=4\sigma\gamma/(\gamma^{2}+\sigma)$, showing that the Hamiltonian form of both action principles is the same.

\paragraph{Solution of the second-class constraints.}\label{solve2ndclass}
When dealing with second-class systems, there are two {\it equivalent} approaches that can be followed: either we can modify the symplectic structure by introducing the {\it Dirac bracket}, or we can {\it solve the second-class constraints explicitly}. Here, we follow the latter, which in turn allows us to eliminate some of the variables involved in the formalism.

In order to solve the second-class constraints, we use the canonical pairs $({}^{\scriptscriptstyle(\gamma)}A_{aIJ},\tilde{\Pi}^{aIJ})$, which turn out to be simpler to work with. We denote by ``0'' the internal time component, while we use the Latin letters starting at ``$i$'' to label the internal spatial components. The solution of the constraint (\ref{simpliccan}) or (\ref{simplipsv}) is given by
\begin{eqnarray}
	\tilde{\Pi}^{a0i}=&\tilde{E}^{ai},  \label{solPhia}\\
	\tilde{\Pi}^{aij}=&2\tilde{E}^{a[i}\chi^{j]}, \label{solPhib}
\end{eqnarray}
where, given that $hh^{ab}=\eta_{ij}\tilde{E}^{ai}\tilde{E}^{bj}$ with\footnote[1]{The quantity $h_{ab}$ is interpreted as the spatial metric; see~\cite{CelMontesRom}.} $\eta_{ij}: =\left(1+\sigma\chi_{k} \chi^{k} \right)\delta_{ij}-\sigma\chi_{i}\chi_{j}$, $\tilde{E}^{ai}$ is a nonorthogonal densitized frame, and $\chi^{i}$ is an arbitrary internal 3-vector, which are our new variables; notice that the internal spatial indices are raised and lowered with the Euclidean metric. Now, we have to rewrite all the expressions of the previous section in terms of the $\tilde{E}^{ai}$ and  $\chi^{i}$. Using (\ref{solPhia}) and (\ref{solPhib}), the symplectic structure collapses to
\begin{equation}\label{reducedpsv}
	\int_{\Omega}d^{3}x\left(\tilde{\Pi}^{aIJ}\frac{\partial}{\partial t}{}^{\scriptscriptstyle(\gamma)}A_{aIJ}\right) = 2\int_{\Omega}d^{3}x\left( \tilde{E}^{ai}\dot{A}_{ai} +  \tilde{\zeta}_i \dot{\chi}^i\right),
\end{equation}
with the definitions
\begin{eqnarray}
	A_{ai} & := & {}^{\scriptscriptstyle(\gamma)}A_{a0i} + {}^{\scriptscriptstyle(\gamma)}A_{aij}\chi^{j}, \label{Aai}\\
	\tilde{\zeta}_i & := & {}^{\scriptscriptstyle(\gamma)}A_{aij}\tilde{E}^{aj}.\label{zeta}
\end{eqnarray}
Accordingly, the reduced phase space will be parametrized by the canonical pairs $(A_{ai},\tilde{E}^{ai})$ and $(\chi_i,\tilde{\zeta}^i)$. The idea is then to use the constraint (\ref{secund4dr}) to restrict some of the components ${}^{\scriptscriptstyle(\gamma)}\tensor{A}{_a^{IJ}}$. Equation (\ref{zeta}) can be regarded as a set of inhomogeneous equations for the unknowns ${}^{\scriptscriptstyle(\gamma)}A_{aij}$ whose solution takes the form
\begin{equation}\label{wij}
	{}^{\scriptscriptstyle(\gamma)}A_{aij}= \frac{1}{2}\varepsilon_{ijk}\underaccent{\tilde}{E}_{al}\tilde{M}^{kl} -\underaccent{\tilde}{E}_{a[i}\tilde{\zeta}_{j]},
\end{equation}
where $\underaccent{\tilde}{E}_{ai}$ is the inverse of $\tilde{E}^{ai}$ and $\tilde{M}^{ij}$ is an arbitrary symmetric matrix. We then use the constraint (\ref{secund4dr}) to fix the six components of this matrix. Substituting (\ref{wij}) and ${}^{\scriptscriptstyle(\gamma)}A_{a0i}$ from (\ref{Aai}) into (\ref{secund4dr}), the solution for $\tilde{M}^{ij}$ is~\cite{CelMontesRom,BarrosESa}
\begin{eqnarray}\label{Mij}
\fl	\tilde{M}^{ij} =  \frac{1}{1+\sigma \chi_{m}\chi^{m}}\left[ \left(\tensor{\tilde{f}}{^k_k}+\sigma \tilde{f}_{kl}\chi^k\chi^l \right)\delta^{ij} + \left(\sigma \tensor{\tilde{f}}{^k_k} -\tilde{f}_{kl}\chi^k\chi^l \right)\chi^i\chi^j \right. \nonumber \\
	\hspace{7mm} -\tilde{f}^{ij} -\tilde{f}^{ji} - \left. \sigma\left(\tilde{f}^{ik}\chi^j + \tilde{f}^{jk}\chi^i + \tilde{f}^{ki}\chi^j +\tilde{f}^{kj}\chi^i \right)\chi_k \right],
\end{eqnarray}
where $\tilde{f}_{ij}$ is given by
\begin{eqnarray}
\fl	\tilde{f}_{ij}=  -\varepsilon_{ikl}\tilde{E}^{ak}\left[\left(1-\sigma\gamma^{-2}\right) \underaccent{\tilde}{E}_{bj} \partial_a \tilde{E}^{bl} + \sigma\chi^l A_{aj} \right] +\frac{\sigma}{\gamma}\left(\tilde{E}^{ak}A_{ak}\delta_{ij}-A_{ai}\tensor{\tilde{E}}{^a_j}+\tilde{\zeta}_i\chi_j \right).\label{fij} \nonumber\\
\end{eqnarray}
Notice that only the symmetric part of $\tilde{f}_{ij}$ enters in (\ref{Mij}). Thereby, we have succeeded at solving the second-class constraints. All that is left to do is to rewrite the first-class constraints (\ref{gauss4dr})-(\ref{vect4dr}) in terms of the new phase space variables. The Gauss constraint $\tilde{\mathcal{G}}^{IJ}$ is broken down into (local) rotation and boost generators
\begin{eqnarray}
	\fl \tilde{\mathcal{G}}^i_{\mbox{{\scriptsize boost}}} :=  \tilde{\mathcal{G}}^{0i}  =  \partial_a\left(\tilde{E}^{ai}+\frac{\sigma}{\gamma}\tensor{\varepsilon}{^i_{jk}}\tilde{E}^{aj} \chi^k  \right) +2\sigma A_{aj} \tilde{E}^{a[j}\chi^{i]}+\sigma\tilde{\zeta}_j\chi^j\chi^i+\tilde{\zeta}^i, \label{gaussboost}\\
	\fl \tilde{\mathcal{G}}^i_{\mbox{{\scriptsize rot}}} :=  \frac{1}{2}\tensor{\varepsilon}{^i_{jk}}\tilde{\mathcal{G}}^{jk} =  \partial_a\left(\tensor{\varepsilon}{^i_{jk}}\tilde{E}^{aj}\chi^k  + \frac{1}{\gamma}\tilde{E}^{ai} \right) - \tensor{\varepsilon}{^i_{jk}}\left(A_a{}^j\tilde{E}^{ak} -\tilde{\zeta}^j\chi^k\right), \label{gaussrot}
\end{eqnarray}
while the vector and scalar constraints, $\tilde{\mathcal{H}}_a$ and $\tilde{\tilde{\mathcal{H}}}$, respectively, read
\begin{eqnarray}
	\fl\tilde{\mathcal{H}}_a  =  2\tilde{E}^{bi}\partial_{[b}A_{a]i}-\tilde{\zeta}_i\partial_a \chi^i +\frac{\gamma^2}{\gamma^2-\sigma}\biggl[2\sigma \tilde{E}^{b[i}\chi^{j]}A_{ai}A_{bj}  \nonumber\\
	\left. - A_{ai}\left(\tilde{\zeta}^i +\sigma \tilde{\zeta}_j\chi^j\chi^i\right) + \frac{\sigma}{\gamma}\varepsilon_{ijk}\left( \tilde{E}^{bi}A_b{}^j+\tilde{\zeta}^i\chi^j \right)A_a{}^k  \right],\label{vectsolve2nd}\\
	\fl\tilde{\tilde{\mathcal{H}}} =  -\sigma \tilde{E}^{ai}\chi_i \tilde{\mathcal{H}}_a + \left(1+\sigma\chi_{k} \chi^{k} \right) \left( \tilde{E}^{ai}\partial_a\tilde{\zeta}_i + \frac{1}{2}\tilde{\zeta}_i \tilde{E}^{ai}\tilde{E}^{bj}\partial_a\underaccent{\tilde}{E}_{bj} \right) \nonumber \\
	\hspace{-17mm}-\frac{\sigma\gamma^2}{\gamma^2-\sigma}\left\lbrace \left(1+\sigma\chi_{l} \chi^{l} \right) \left[ \tilde{E}^{ai}\tilde{E}^{bj}A_{a[i|}A_{b|j]} + \tilde{\zeta}_i\chi^i A_{aj}\tilde{E}^{aj} + \frac{3}{4}\left(\tilde{\zeta}_i \chi^i\right)^2 + \frac{3}{4}\sigma\tilde{\zeta}_i \tilde{\zeta}^i  \right.\right. \nonumber \\
	\hspace{-17mm} - \frac{1}{\gamma}\varepsilon_{ijk}\tilde{E}^{ai} A_{a}{}^j \tilde{\zeta}^k \Bigr] -\frac{\sigma}{4}\left(\tilde{f}^i{}_i\right)^2 +\frac{\sigma}{2}\tilde{f}^{ij}f_{(ij)}  - \frac{1}{2}\tilde{f}_{ij}\chi^i\chi^j \left(\tilde{f}^k{}_k-\frac{\sigma}{2} \tilde{f}_{kl}\chi^k\chi^l\right)  \nonumber \\
	\hspace{-17mm}\left.  + \frac{1}{4}\left(\tilde{f}_{ij}+\tilde{f}_{ji} \right)\left(\tilde{f}^{ik}+\tilde{f}^{ki} \right)\chi^j\chi_k  \right\rbrace. \label{scalsolve2nd}
\end{eqnarray}
In conclusion, we ended up with a phase space whose points are labeled by the 12 canonical pairs $(A_{ai},\tilde{E}^{ai})$ and $(\chi_i,\tilde{\zeta}^i)$, subject to the 10 first-class constraints $\tilde{\mathcal{G}}^i_{\mbox{{\scriptsize boost}}}$, $\tilde{\mathcal{G}}^i_{\mbox{{\scriptsize rot}}}$, $\tilde{\mathcal{H}}_a$, and $\tilde{\tilde{\mathcal{H}}}$. Therefore, after solving the second-class constraints, the theory still propagates two degrees of freedom. We remark that although we split the Gauss constraint into boost and rotation generators, the theory still possesses full local $SO(\sigma)$ invariance; this is analogous to the fact that we can write electrodynamics in terms of electric and magnetic fields without breaking the Lorentz invariance of the theory.

\paragraph{Derivation of the Ashtekar-Barbero variables.}

In order to arrive at the Ashtekar-Barbero variables, we must break the $SO(\sigma)$ local symmetry down to its compact subgroup $SO(3)$. This is accomplished by choosing the time gauge, which fixes the boost transformations. It turns out to be easier to apply the time gauge before solving the second-class constraints than directly on the constraints obtained in the previous section. 

In the time gauge, we impose the extra conditions $\chi^i=0$, which form a second-class pair with the boost part of the Gauss constraint and imply that the solution of (\ref{simpliccan}) now reads
\begin{eqnarray}
	\tilde{\Pi}^{a0i}=\tilde{E}^{ai},  \label{solPhiafix}\\
	\tilde{\Pi}^{aij}=0, \label{solPhibfix}
\end{eqnarray}
where now $\tilde{E}^{ai}$ is an orthonormal (densitized) frame with respect to $h^{ab}$. With this at hand,  (\ref{Aai}) and (\ref{wij}) become
\begin{eqnarray}
	 {}^{\scriptscriptstyle(\gamma)}A_{a0i} = A_{ai}, \\
	 {}^{\scriptscriptstyle(\gamma)}A_{aij} = \varepsilon_{ijk}\left[\left(1 - \sigma\gamma^{-2}\right)\Gamma_{a}{}^k +\sigma \gamma^{-1}A_a{}^k\right],
\end{eqnarray}
where $\Gamma_{ai}:=(1/2)\varepsilon_{ijk}\tensor{A}{_a^{jk}}$ is the rotational part of $A_{aij}$. Using these expressions, the constraints (\ref{gauss4dr}) and (\ref{secund4dr}) read
\begin{eqnarray}
	\tilde{\mathcal{G}}^i_{\mbox{{\scriptsize boost}}}  =  \partial_a \tilde{E}^{ai}+\varepsilon^{ijk}  \left[\left(1-\sigma\gamma^{-2}\right)\Gamma_{ak}+\sigma\gamma^{-1}A_{ak}\right]\tilde{E}^{a}{}_{j}\approx 0, \label{fix1}\\
	\tilde{\mathcal{G}}^i_{\mbox{{\scriptsize rot}}}  =  \gamma^{-1}\partial_a\tilde{E}^{ai}-\varepsilon^{ijk}A_{aj}\tilde{E}^a{}_k\approx 0, \label{fix2}\\
	\Psi^{ab} = -4\sigma \epsilon_{ijk}\tilde{E}^{ci}\tilde{E}^{(a|k}\left(\partial_c\tilde{E}^{|b)j}-\varepsilon^{jlm}\Gamma_{cl}\tilde{E}^{|b)}{}_m\right)\approx 0. \label{fix3}
\end{eqnarray}
Now, the constraints (\ref{fix1}) and (\ref{fix3}) must be simultaneously solved\footnote{In fact, in the canonical analysis of the Holst action, the analog of these constraints are the components of the spatial part of the torsion~\cite{Geillergrg459}, which further indicates that they must be solved at the same time. However, it is worth recalling that the concept of torsion is not initially involved in the gauge formulation of general relativity as a constrained $BF$ theory.}, which allows us to fix $\Gamma_{ai}$. By eliminating the variable $A_{ai}$ between (\ref{fix1}) and (\ref{fix2}), and combining the result with (\ref{fix3}), the solution for $\Gamma_{ai}$ is
\begin{equation}
	\Gamma_{ai}=\epsilon_{ijk}\left( \partial_{[b}\underaccent{\tilde}{E}_{a]}{}^j+\underaccent{\tilde}{E}_{a}{}^{[l|}\tilde{E}^{c|j]}\partial_b\underaccent{\tilde}{E}_{cl}\right)\tilde{E}^{bk}.
\end{equation}
As a result, this connection turns out to be the the torsion-free spin-connection compatible with the densitized triad $\tilde{E}^{ai}$. The remaining constraints can now be written in terms of the new phase space variables ($A_{ai},\tilde{E}^{ai}$), and their final expressions are
\begin{eqnarray}
	\tilde{\mathcal{G}}^i & = & \gamma^{-1}\mathfrak{D}_{a}\tilde{E}^{ai},\label{constg}\\
	\tilde{\mathcal{H}}_a & = & \tilde{E}^{bi}F_{iba} + \left( 1 -\sigma \gamma^{-2} \right)\left(\gamma A_{ai} - \Gamma_{ai}\right)\tilde{\mathcal{G}}^i, \label{consth}\\
	\tilde{\tilde{\mathcal{H}}} & = & \frac{\sigma}{2\gamma} \varepsilon_{ijk} \tilde{E}^{aj} \tilde{E}^{bk} \left[ \tensor{F}{^i_{ab}} + \left( \sigma\gamma -  \gamma^{-1}\right) \tensor{R}{^i_{ab}} \right],\label{consthh}
\end{eqnarray}
where $\mathfrak{D}_{a}\tilde{E}^{ai}:=\partial_a\tilde{E}^{ai}-\gamma\varepsilon^{ijk}A_{aj}\tilde{E}^a{}_k$, and $F_{iab}$ and $R_{iab}$ are the curvatures of the connections $A_{ai}$ and $\Gamma_{ai}$, respectively:
\begin{eqnarray}
	F_{iab} & = & 2\ \partial_{[a}A_{b]i}-\gamma\varepsilon_{ijk}A_a{}^jA_b{}^k, \\
	R_{iab} & = & 2\ \partial_{[a}\Gamma_{b]i}-\varepsilon_{ijk}\Gamma_a{}^j\Gamma_b{}^k.
\end{eqnarray}
The constraints (\ref{constg})-(\ref{consthh}) constitute the Ashtekar-Barbero formulation of general relativity. Hence, we have shown, as promised, that the canonical analysis of the action principle (\ref{ActBFImm}) leads to the phase space of the Ashtekar-Barbero formalism, something analogous to the fact that the canonical analysis of Plebanski's action leads to the Ashtekar variables as shown in section~\ref{Pleb-HamilAna}. We remark that a derivation of the Ashtekar-Barbero variables taking the action (\ref{ActBFImm}) as the starting point is also reported in~\cite{perez2013-16-3}, but there the time gauge was used for solving the constraint (\ref{decompConst}), and so the approach followed by the author breaks the local Lorentz invariance from the very beginning.

\subsubsection{Cosmological constant and matter fields.}
A description of gravity without including its sources would not be complete. In this section we discuss the coupling between gravity and matter in the $BF$ framework, specifically for the cosmological constant (then interpreted as vacuum energy), the scalar field, and the Yang-Mills field. Since the coupling terms we add to (\ref{ActBFImm}) only involve the $B$ field, the matter field, and some auxiliary fields, the equations of motion for the connection and the multiplier $\varphi$ are again (\ref{variatA}) and (\ref{variatphi}), respectively. This implies that the solutions for $B^{IJ}$ are still given by (\ref{solveq1}) and (\ref{solveq2}), and that $A^{IJ}$ is still the spin-connection compatible with the $SO(\sigma)$-frame $e^I$, which determines the spacetime metric $g_{\mu\nu}:=\eta_{IJ}\tensor{e}{_{\mu}^I}e_{\nu}{}^J$, whose inverse is given by $g^{\mu\nu}=\eta^{IJ} \tensor{e}{^{\mu}_I}  \tensor{e}{^{\nu}_J}$ ($e^{\mu}{}_I$ is the inverse of $e_{\mu}{}^I$).

The cosmological constant can be included in the $BF$ formalism by adding to (\ref{ActBFImm}) appropriate volume terms. Let us rename the gravitational action (\ref{ActBFImm}) as $S_{\mathrm{GR}}[A,B,\varphi,\mu]$. Following the approach of~\cite{Smolprd812, MontesVelprd856} (see~\cite{MontesVelprd814} for the coupling of the cosmological constant to the CMPR action), the action reads
\begin{eqnarray}\label{ActCosmoR}
	\fl S[A,B,\varphi,\mu]=S_{\mathrm{GR}}[A,B,\varphi,\mu]+\int_{\mathcal{M}}\left[\mu\lambda +l_1 B_{IJ}\wedge B^{IJ} +l_2 B_{IJ}\wedge *B^{IJ}\right],
\end{eqnarray}
where $\lambda$, $l_1$, and $l_2$ are constants. Notice that now the multiplier $\varphi$ has constant trace
\begin{eqnarray}\label{Newtarce}
	\delta\mu:\hspace{2mm}\varepsilon_{IJKL}\varphi^{IJKL}=\lambda.
\end{eqnarray}
The objective is now to relate $\lambda$, $l_1$, and $l_2$ to the cosmological constant $\Lambda$, for which we use the equations of motion arising from (\ref{ActCosmoR}). The procedure is similar to the one followed in section~\ref{BFEE_derive}, and so we only sketch the results. Consider first the solution (\ref{solveq1}). The second term on the right-hand side of (\ref{ActCosmoR}) now modifies the equation of motion corresponding to the $B$ field, and by combining this equation with the Bianchi identities, the Ricci tensor reads
\begin{eqnarray}\label{Riccilamb1}
	\mathcal{R}_{IJ}=4\kappa_1\left(\frac{1}{8}\varepsilon_{KLMN}\varphi^{KLMN}-\frac{3}{2}\sigma l_2\right)\eta_{IJ}.
\end{eqnarray}
Finally, replacing (\ref{Newtarce}) into (\ref{Riccilamb1}), and comparing the resulting expression with Einstein's equations with a cosmological constant, namely $\mathcal{R}_{IJ}=\Lambda\eta_{IJ}$, the following relation among the constants emerges
\begin{eqnarray}\label{sollamb_1}
	\lambda=\frac{2\Lambda}{\kappa_1}+12\sigma l_2.
\end{eqnarray}
Substituting the expression (\ref{solveq1}) into (\ref{ActCosmoR}), and using (\ref{sollamb_1}), the action collapses to the Holst action with a cosmological constant $\Lambda$:
\begin{eqnarray}
	\fl S[e,A]=\kappa_1\int_{\mathcal{M}} \left\{ \left[*(e^I\wedge e^J)+\frac{\sigma}{\gamma}e^I\wedge e^J\right]\wedge F_{IJ}[A]-\frac{\Lambda}{12}\varepsilon_{IJKL}e^I\wedge e^J\wedge e^K\wedge e^L\right\}. \nonumber\\
\end{eqnarray}

On the other hand, repeating the previous steps using (\ref{solveq2}), we find the following value for $\lambda$
\begin{eqnarray}\label{sollamb_2}
	\lambda=\sigma\left(\frac{2\Lambda}{\kappa_2\gamma}+12 l_2\right).
\end{eqnarray}
Again, the action (\ref{ActCosmoR}) reduces to the Holst action with a cosmological constant, although the Immirzi parameter takes another value:
\begin{eqnarray}
	\fl S[e,A]=\frac{\kappa_2}{\gamma}\int_{\mathcal{M}} \left\{ \left[*(e^I\wedge e^J)+\gamma e^I\wedge e^J\right]\wedge F_{IJ}[A]-\frac{\Lambda}{12}\varepsilon_{IJKL}e^I\wedge e^J\wedge e^K\wedge e^L\right\}. \nonumber\\
\end{eqnarray}

Note that the parameter $l_1$ does not appear in (\ref{sollamb_1}) or (\ref{sollamb_2}). The reason for that can be traced to the term whose coupling constant is $l_1$ in (\ref{ActCosmoR}); the quantity $B_{IJ}\wedge B^{IJ}$ vanishes on shell (that is, when the solutions to the constraint (\ref{variatphi}) are used), and so the term proportional to $l_1$ drops out from the action.

From the Hamiltonian point of view, the terms added to $S_{\mathrm{GR}}[A,B,\varphi,\mu]$ only modify the scalar constraint, which now reads
\begin{eqnarray}\label{scal_lambda}
	\tilde{\tilde{\mathcal{H}}}=\frac{1}{4}\tilde{\eta}^{abc}h_{ad}*{}^{\scriptscriptstyle(\gamma)}\tensor{\tilde{\Pi}}{^{dIJ}} F_{IJbc}-\frac{\sigma\Lambda}{\kappa_1} h\approx 0.
\end{eqnarray} 
The remaining constraints are the same of section~\ref{decomposeact_r}, that is, (\ref{scal_lambda}) together with (\ref{gauss4dr}) and (\ref{vect4dr}) comprise the first-class set, while (\ref{simplipsv}) and (\ref{secund4dr}) make up the second-class set. The second-class constraints can be managed in the same way as in section~\ref{solve2ndclass}. Once we solve them, the resulting constraints are the ones of the previous sections except for the scalar constraint, which gets an additional contribution coming from the cosmological constant term present in (\ref{scal_lambda}). When the gauge is not fixed, the quantity added on the right-hand side of (\ref{scalsolve2nd}) is $-(\sigma\Lambda/\kappa_1)\left(1+\sigma\chi_{k} \chi^{k} \right) \tilde{\tilde{E}}$, where $\tilde{\tilde{E}}:=\det (\tilde{E}^{ai})$ and $h$ are related via $h=\left(1+\sigma\chi_{k} \chi^{k} \right) \tilde{\tilde{E}}$. On the other hand, when the time gauge is imposed, the term $-(\sigma\Lambda/\kappa_1) \tilde{\tilde{E}}$ must be added on the right-hand side of (\ref{consthh}).

The coupling of an scalar field $\phi$ to the action (\ref{ActBFImm}) is done by considering the following action~\cite{MontesVelprd856}
\begin{eqnarray}
	\fl S[A,B,\varphi&,\mu,\pi,\phi]= S_{\mathrm{GR}}[A,B,\varphi,\mu]\nonumber\\
	\fl&\! \! \!+\int_{\mathcal{M}}\left[\left(B_{IJ}\wedge *B^{IJ}\right)(a\pi^{\mu}\partial_{\mu}\phi+bV(\phi))+(\alpha_1\tilde{H}_{\mu\nu}+\alpha_2\tilde{G}_{\mu\nu})\pi^{\mu}\pi^{\nu}d^4 x\right],\label{BFscalarcoupR}
\end{eqnarray}
where $\pi^{\mu}$ are auxiliary fields, $V(\phi)$ is the scalar field potential, and $\tilde{H}_{\mu\nu}$ and $\tilde{G}_{\mu\nu}$ are the Urbantke metrics of weight one defined by
\begin{eqnarray}
	\tilde{H}_{\mu\nu}:=\frac{1}{12}\tilde{\eta}^{\alpha\beta\gamma\delta}\tensor{B}{_{\mu\alpha}^{IJ}}\tensor{B}{_{\beta\gamma}^{KL}}\tensor{B}{_{\delta\nu}^{MN}}\eta_{JK}\eta_{LM}\eta_{NI},\label{Urbmet1}\\
	\tilde{G}_{\mu\nu}:=\frac{1}{24}\tilde{\eta}^{\alpha\beta\gamma\delta}\tensor{B}{_{\mu\alpha}^{IJ}}\tensor{B}{_{\beta\gamma}^{KL}}\tensor{B}{_{\delta\nu}^{MN}}\varepsilon_{JMKL}\eta_{NI}\label{Urbmet2}
\end{eqnarray}
and $a$, $b$, $\alpha_1$, and $\alpha_2$ are constants. Notice that the action (\ref{BFscalarcoupR}) is polynomial in the $B$ field.

The equations of motion arising from (\ref{BFscalarcoupR}) can be shown to imply Einstein's equations with the scalar field as source. Here, we follow an alternative approach that consists in rewriting the action (\ref{BFscalarcoupR}) in terms of the tetrad field. For the solution (\ref{solveq1}), the metric $\tilde{G}_{\mu\nu}$ vanishes, while $\tilde{H}_{\mu\nu}$ yields $\tilde{H}_{\mu\nu}=\kappa_1^3\sigma e g_{\mu\nu}$, with $e:=\det(\tensor{e}{_{\mu}^I})$. After eliminating $\pi^{\mu}$ from (\ref{BFscalarcoupR}), the action reads
\begin{eqnarray}
\fl S[e,A,\phi]=\kappa_1\int_{\mathcal{M}}\left\{\left[*(e^I\wedge e^J)+\frac{\sigma}{\gamma} e^I\wedge e^J\right]\wedge F_{IJ}[A]\right.\nonumber\\
\left. \hspace{8mm}+12\sigma d^4x\sqrt{\sigma g}\left[-\frac{3a^2}{\alpha_1} g^{\mu\nu}\partial_{\mu}\phi\partial_{\nu}\phi+\kappa_1b V(\phi)\right]\right\},
\end{eqnarray}
where $g:=\det(g_{\mu\nu})$. This is the Holst action coupled to a scalar field with potential $V(\phi)$.

Similarly, for the solution (\ref{solveq2}) we have $\tilde{H}_{\mu\nu}=0$ and $\tilde{G}_{\mu\nu}=\kappa_2^3 e g_{\mu\nu}$, and the action (\ref{BFscalarcoupR}) reduces to
\begin{eqnarray}
\fl S[e,A,\phi]=\frac{\kappa_2}{\gamma}\int_{\mathcal{M}}\biggl\{\left[*(e^I\wedge e^J)+\gamma e^I\wedge e^J\right]\wedge F_{IJ}[A]\nonumber\\
\left. \hspace{8mm}+12\gamma d^4x\sqrt{\sigma g}\left[-\frac{3a^2}{\alpha_2} g^{\mu\nu}\partial_{\mu}\phi\partial_{\nu}\phi+\kappa_2b V(\phi)\right]\right\}.
\end{eqnarray}
In this way, we have outlined the equivalence of (\ref{BFscalarcoupR}) with general relativity coupled to a scalar field.

Finally, the coupling of the Yang-Mills field can be implemented through the following action principle~\cite{MontesVelprd856}
\begin{eqnarray}
\fl S[A, B,\varphi,\mu,\mathcal{A},\phi] =&& S_{\mathrm{GR}}[A,B,\varphi,\mu]  + \int_{\mathcal{M}} \mathcal{F}^a[\mathcal{A}]\wedge\left(b\ B^{IJ}+c*B^{IJ}\right)\phi_{aIJ} \nonumber\\
&&  - \frac{1}{2} \int_{\mathcal{M}}  \left(\beta_1B^{IJ}\wedge B^{KL}+\beta_2B^{IJ}\wedge *B^{KL}\right)\tensor{\phi}{^a_{IJ}}\phi_{aKL}  ,\label{Y_Mreal}
\end{eqnarray}
where the internal index $a$ is raised and lowered with the nondegenerate inner product of the Yang-Mills group's Lie algebra, $\mathcal{A}$ is the Yang-Mills connection 1-form whose field strength is given by $\mathcal{F}[\mathcal{A}]:=d\mathcal{A}+\mathcal{A}\wedge\mathcal{A}$, $\tensor{\phi}{^a_{IJ}}$ are Lie algebra-valued auxiliary fields, and $b$, $c$, $\beta_1$, and $\beta_2$ are parameters.

The equivalence of (\ref{Y_Mreal}) with Einstein's theory can be established via equations of motion, but a more economical procedure is seeing how this action principle looks in terms of tetrad field. Using the solution (\ref{solveq1}) for eliminating $\tensor{\phi}{^a_{IJ}}$ from (\ref{Y_Mreal}), the action reads
\begin{eqnarray}
 S[e,A,\mathcal{A}]&=&  \kappa_1\int_{\mathcal{M}} \left [ *(e^I\wedge e^J)+\frac{\sigma}{\gamma} e^I\wedge e^J\right ] \wedge F_{IJ}[A] 
 \nonumber\\
 &&+ \int_{\mathcal{M}}  \left ( d_1\mathcal{F}^a\wedge\mathcal{F}_a+d_2\mathcal{F}^a\wedge*\mathcal{F}_a\right ),\label{Holst-Y-M}
\end{eqnarray}
where $d_1:=[-(\sigma b^2+c^2)\beta_1+2bc\beta_2]/2(\beta_2^2-\sigma\beta_1^2)$ and $d_2:=[(\sigma b^2+c^2)\beta_2-2\sigma bc\beta_1]/2(\beta_2^2-\sigma\beta_1^2)$; this is the Holst action coupled to Yang-Mills theory supplemented with the Pontrjagin term.

Taking now the solution (\ref{solveq2}), an analogous procedure shows that (\ref{Y_Mreal}) reduces to
\begin{eqnarray}
S[e,A,\mathcal{A}]&=& \frac{\kappa_2}{\gamma} \int_{\mathcal{M}} \left [ *(e^I\wedge e^J)+\gamma e^I\wedge e^J \right ] \wedge F_{IJ}[A] \nonumber\\
&&+ \int_{\mathcal{M}} \left ( d_1\mathcal{F}^a\wedge\mathcal{F}_a+d_2\mathcal{F}^a\wedge*\mathcal{F}_a \right ), 
\end{eqnarray}
which is essentially (\ref{Holst-Y-M}). Therefore, we have established that the action (\ref{Y_Mreal}) adequately couples Yang-Mills theory to general relativity in the $BF$ formalism.

An alternative way of coupling Yang-Mills theory to $BF$ gravity consists in enlarging $SO(\sigma)$ to a gauge group $G$ containing $SO(\sigma)$ as subgroup. Then, by a symmetry breaking process, one singles the gravitational sector out, and obtains Yang-Mills theory from the fields in the quotient space $G/SO(\sigma)$~\cite{Smolinprd8012}.

\paragraph{\bf Remarks.} (i) We have not discussed the coupling of fermionic matter to $BF$ gravity. Two obstacles arise for performing the coupling. First, the coupling of fermions to general relativity requires the presence of the tetrad field, but it disappears from the action once the $B$ field is introduced. Second, even in the case of the first-order formalism, there is a variety of different manners in which fermions interact with the gravitational field~\cite{RovelliPerezprd734,Mercuriprd738,Alexandrovcqg2514}. Although spin 1/2 and 3/2 fermions have been successfully coupled to Plebanski's action~\cite{capo1991841}, the coupling of fermionic matter to real general relativity still remains an open issue (see, however,~\cite{Fairbairn2007} for the coupling of Dirac fermions to three-dimensional gravity). (ii) The formulation of general relativity as a constrained $BF$ theory provides a natural way of coupling extended matter to gravity by regarding gauge defects along surfaces~\cite{MPprd.77.104020,Fairbairnjhep09}.

\subsection{MacDowell-Mansouri gravity as a deformed $BF$ theory}\label{MacDowellMansouri}

Four-dimensional general relativity with a nonvanishing cosmological constant admits a formulation in terms of a gauge connection much in the same spirit as Chern-Simons theory does for three-dimensional gravity. This formulation is due to MacDowell and Mansouri~\cite{MacDowell1977} (see~\cite{wisecqg2715} for a geometric interpretation in the framework of Cartan geometry), and considers a larger gauge connection built up from both the $SO(\sigma)$-connection $A_{IJ}$ and the frame\footnote{There is, however, a difference concerning the similarity with Chern-Simons theory. While the Chern-Simons formulation of three-dimensional gravity is invariant under the larger group, the MacDowell-Mansouri action is not, and this is what causes general relativity to emerge.} $e^I$. The larger gauge group itself depends on the signature of the metric as well as on the sign of the cosmological constant, and is given by
\begin{center}
	\begin{tabular}{ l c c }
		& $\Lambda>0$ & $\Lambda<0$ \\
		$\sigma=1$ & $SO(5)$ & $SO(1,4)$ \\
		$\sigma=-1$ & $SO(1,4)$ & $SO(2,3).$ \\
	\end{tabular}
\end{center}
The indefinite groups $SO(1,4)$ and $SO(2,3)$ are known as de Sitter and anti-de Sitter groups, respectively. general relativity is then obtained by breaking the larger group down to $SO(\sigma)$.

Let us call $G$ any of the previous gauge groups. Its Lie algebra $\mathfrak{g}$ admits the Killing-orthogonal and $SO(\sigma)$-invariant vector space splitting given by $\mathfrak{g}\cong\mathfrak{so}(\sigma)\oplus\mathbb{R}^4_{\sigma}$, where $\mathbb{R}^4_{\sigma}$ is $\mathbb{R}^4$ endowed with the inner product defined by $(\eta_{IJ})=\mathrm{diag}(\sigma,1,1,1)$. In the fundamental representation, the 
$\mathfrak{g}$-valued connection $\mathcal{A}$ takes the form
\begin{eqnarray}
\left ( \tensor{\mathcal{A}}{^a_b} \right ) =
\left( {\begin{array}{rc}
	\tensor{A}{^I_J}  & \frac{1}{l}e^I\\      -\frac{\epsilon}{l}e_J & 0     \end{array} } \right ) =\left ( {\begin{array}{cc}
	\tensor{A}{^I_J} & 0\\      0 & 0     \end{array} } \right )+\left ( {\begin{array}{cc}
	0& \frac{1}{l}e^I\\      -\frac{\epsilon}{l}e_J & 0     \end{array} } \right ),\label{MM_gsplit}
\end{eqnarray}
where the indices $a,b,\dots$ run from 0 to 4, $\tensor{A}{^I_J}$ takes values in $\mathfrak{so}(\sigma)$, $e^I$ is a vector in $\mathbb{R}^4_{\sigma}$,  $\epsilon:=\mathrm{sgn}(\Lambda)$, and $l$ is a constant required for dimensional reasons that is related to the cosmological constant via $\Lambda=3\epsilon l^{-2}$. Notice that the group indices are now raised and lowered with the Killing metric
\begin{eqnarray}
(\eta_{ab})=\mathrm{diag}(\sigma,1,1,1,\epsilon)=\left ( {\begin{array}{cc}
	\eta_{IJ}&0\\0&\epsilon\end{array}}\right ).
\end{eqnarray}

Assume that (\ref{MM_gsplit}) defines a connection 1-form valued in $\mathfrak{g}$; then $\tensor{A}{^I_J}$ and $e^I$ are interpreted as an $SO(\sigma)$-connection and an orthonormal frame, respectively. The field strength $\tensor{\mathcal{F}}{^a_b}$ of $\tensor{\mathcal{A}}{^a_b}$ takes the form
\begin{eqnarray}
\tensor{\mathcal{F}}{^I_J}=\tensor{F}{^I_J}-\frac{\Lambda}{3}e^I\wedge e_J,\hspace{5mm} \tensor{\mathcal{F}}{^I_4}=\frac{1}{l} D e^I,\label{M-M_comp}
\end{eqnarray}
with $\tensor{F}{^I_J}$ the curvature of $\tensor{A}{^I_J}$ and $D e^I= d e^I + A^I{}_J \wedge e^J$.

Consider the following action principle of the $BF$-type defined on a $G$-principal bundle over a four-dimensional manifold $\mathcal{M}$~\cite{StarodubtsevArxiv}
\begin{eqnarray}
\fl S[\mathcal{A},\mathcal{B}]=\int_{\mathcal{M}}\left(\mathcal{B}_{ab}\wedge \mathcal{F}^{ab}[\mathcal{A}]-\frac{\alpha}{2}\mathcal{B}^{ab}\wedge \mathcal{B}^{cd}\boldsymbol{\varepsilon}_{abcde}\mathcal{v}^e-\frac{\beta}{2}\mathcal{B}^{ab}\wedge \mathcal{B}_{ab}\right),\label{BfM-M}
\end{eqnarray}
where $\mathcal{B}_{ab}$ are $\mathfrak{g}$-valued 2-forms, $\mathcal{v}$ is a fixed internal vector that breaks the $G$ invariance of the action down to $SO(\sigma)$, $\boldsymbol{\varepsilon}_{abcde}$ is an internal five-dimensional totally antisymmetric symbol ($\boldsymbol{\varepsilon}_{01234}=1$), and $\alpha$ and $\beta$ are constants. Notice that the terms added to the $BF$-action in (\ref{BfM-M}) comprise a true potential term depending only on the $B$-field. The simplest choice for $\mathcal{v}$ is $\mathcal{v}^e=\delta^e_4$, which we assume from now on. In the following, we show that (\ref{BfM-M}) for $\alpha\neq 0$ is an action for general relativity. To start with, we first proceed to integrate out $\mathcal{B}_{ab}$ in (\ref{BfM-M}); the equation of motion corresponding to this field yields
\begin{eqnarray}
\fl \delta \mathcal{B}_{IJ}:\ \mathcal{F}_{IJ}-2\alpha*\mathcal{B}_{IJ}-\beta \mathcal{B}_{IJ}=0,\hspace{5mm} \delta \mathcal{B}_{I4}:\ \mathcal{F}_{I4}-\beta\mathcal{B}_{I4}=0,\label{eqM-M-1}
\end{eqnarray}
where $*$ is the $SO(\sigma)$ Hodge dual operator that has been introduced via the identification $\boldsymbol{\varepsilon}_{IJKL4}=\varepsilon_{IJKL}$. For $\alpha\neq 0$, these equations can be solved for $\mathcal{B}_{ab}$ if $\beta\neq 0$ and $\beta^2-4\sigma\alpha^2\neq 0$, yielding
\begin{eqnarray}
\mathcal{B}_{IJ}=\frac{1}{\beta^2-4\sigma\alpha^2}(\beta\mathcal{F}_{IJ}-2\alpha*\mathcal{F}_{IJ}),\hspace{5mm}\mathcal{B}_{I4}=\frac{1}{\beta}\mathcal{F}_{I4}.
\end{eqnarray}
Substituting this solution back into (\ref{BfM-M}), we obtain
\begin{eqnarray}
\fl S[\mathcal{A}]=\int_{\mathcal{M}}\left[-\frac{\alpha}{\beta^2-4\sigma\alpha^2}\mathcal{F}_{IJ}\wedge *\mathcal{F}^{IJ}+\frac{\beta}{2(\beta^2-4\sigma\alpha^2)}\mathcal{F}_{IJ}\wedge \mathcal{F}^{IJ}+\frac{1}{\beta}\mathcal{F}_{I4}\wedge \mathcal{F}^{I4}\right].\label{M-Mbroken}\nonumber\\
\end{eqnarray}
The first term on the right-hand side of this expression is the MacDowell-Mansouri action, while the other two lead to topological terms. It is worth pointing out that for $\beta=0$, the action (\ref{BfM-M}) collapses to the MacDowell-Mansouri action solely (in this case, the connection is already torsion-free because of the second equation of (\ref{eqM-M-1}) and equation (\ref{M-M_comp})).

Using the expressions in equation (\ref{M-M_comp}), the action (\ref{M-Mbroken}) can be written as
\begin{eqnarray}
\fl S[e,A]=\frac{1}{G_{\mathrm{N}}}\int_{\mathcal{M}}\left\{\left[*(e^I\wedge e^J) -\frac{\sigma}{\gamma}e^I\wedge e^J\right]\wedge F_{IJ}-\frac{\Lambda}{12}\varepsilon_{IJKL}e^I\wedge e^J\wedge e^K\wedge e^L\right.\nonumber\\
\left.\hspace{3mm} -\frac{3}{2\Lambda}F^{IJ}\wedge *F_{IJ}+\frac{3\gamma}{2\Lambda}F^{IJ}\wedge F_{IJ}+\frac{\gamma^2-\sigma}{\gamma}\mathcal{I}_{\mathrm{NY}}\right\},\label{Holstlambtop}
\end{eqnarray}
where $\mathcal{I}_{\mathrm{NY}}:=De^I\wedge De_I-e^I\wedge e^J\wedge F_{IJ}$ is the Nieh-Yan topological invariant~\cite{nieh1982identity,NiehIJMPA}, and $G_{\mathrm{N}}$ and $\gamma$ have the following expressions
\begin{eqnarray}
G_{\mathrm{N}}:=\frac{3(\beta^2-4\sigma\alpha^2)}{2\alpha\Lambda},\hspace{5mm}\gamma:=\frac{\beta}{2\alpha}.
\end{eqnarray}
The first line on the right-hand side of (\ref{Holstlambtop}) is the Holst action with a cosmological constant, where $G_{\mathrm{N}}$ and $\gamma$ are interpreted as the Newton gravitational constant and the Immirzi parameter, respectively, whereas the second line is a combination of the Euler, the Pontrjagin and the Nieh-Yan topological invariants, which can be written as total derivatives and thereby do not contribute to the classical dynamics. In conclusion, when written in terms of $e^I$ and $A_{IJ}$, (\ref{BfM-M}) for $\alpha\neq 0$ and $\beta\neq 0$ (also  $\beta^2-4\sigma\alpha^2\neq 0$) collapses to the Holst action with a cosmological constant up to topological terms, and therefore the action (\ref{BfM-M}) describes general relativity with the Immirzi parameter and a cosmological constant. On the other hand, for $\alpha\neq 0$ and $\beta=0$, the action (\ref{BfM-M}) reduces to Palatini's action with a cosmological constant supplemented with Euler's invariant, as can be seen from (\ref{Holstlambtop}) for vanishing torsion. Note that for $\alpha=0$ the action (\ref{BfM-M}) is topological (it is $G$-invariant $BF$ theory coupled to a volume term), and so it is the deformation introduced through the symmetry breaking term proportional to $\alpha$ the responsible for generating the right propagating degrees of freedom of gravity. 

The canonical analysis of the $BF$-type action (\ref{BfM-M}) is performed in~\cite{Durkacqg2718}, leading to the same constraints arising in the canonical analysis of the Holst action. Boundaries and Noether charges are addressed in~\cite{Durkaprd8312}; in particular, the black hole entropy turns out to be independent of the Immirzi parameter. The coupling of point particles to (\ref{BfM-M}) is explored in~\cite{Freidelprd748}. An extension of (\ref{BfM-M}) for describing $\mathcal{N}=1$ supergravity is reported in~\cite{Durkaprd814}. Finally, the symmetry breaking mechanism outlined in this section can be implemented on the gauge group $SL(5,\mathbb{R})$, with general relativity also emerging~\cite{Mielkeprd834}.

\section{$BF$ formulations of lower-dimensional gravity}\label{BF-lower}

\subsection{Two-dimensional gravity: JT model}\label{jackwteitel}

\subsubsection{Introduction.}

In two dimensions the Einstein tensor vanishes identically, and consequently Einstein's equations are meaningless~\cite{Brown112388}. However, there are some interesting alternatives to Einstein's theory which are based on the Ricci scalar~\cite{Jackiw1984,Teitelboim198341,Teitelboim1984,verlinde1992mgm,callan1992-45.R1005,kim1993-47}. Among these, we will consider the one proposed independently by Jackiw~\cite{Jackiw1984} and Teitelboim~\cite{Teitelboim198341,Teitelboim1984}, since, as we shall see, it is the simplest and closest analog to Einstein's gravity (this model actually is the simplest among a class of  models for two-dimensional gravity called dilaton gravity~\cite{Grumiller2002327}). Let ${\mathcal{M}}$ be a two-dimensional manifold. The action of the JT model is 
\begin{eqnarray}
S[g,\psi]= \int_{\mathcal{M}} d^2x \sqrt{|g|} \psi (\mathcal{R}-2 \Lambda), \label{2D-JTaction}
\end{eqnarray}
where $g$ is the determinant of the two-dimensional metric $g_{\mu\nu}$, $\psi$  is a scalar field, $\mathcal{R}$ is the Ricci scalar, and $\Lambda$ is the cosmological constant. The equations of motion coming from the action are
\begin{eqnarray}
\delta \psi&:&  \mathcal{R}-2 \Lambda=0, \label{2D-JTeq1}\\
\delta g_{\mu\nu}&:&     \nabla_\mu\nabla_\nu \psi + \Lambda g_{\mu\nu} \psi =0     ,  \label{2D-JTeq2}
\end{eqnarray}
where $\nabla_\mu$ is the torsion-free covariant derivative compatible with the metric $g_{\mu\nu}$. The first equation is regarded as the analog of Einstein's equations and entails that the Ricci scalar is constant. This gravitational field equation seems to be natural in certain sense in two dimensions, since  in this case the Ricci scalar contains all the information about the Riemann tensor. The second equation allows us to compute $\psi$ with no additional assumptions on the metric. It is worth mentioning that there is no purely metric action principle leading to (\ref{2D-JTeq1}); that is, without introducing an additional variable. 

\subsubsection{JT model as a $BF$ theory.}

The JT model can be described by a pure $BF$ theory ~\cite{fukuyama1985259,isler1989-63,chamseddine198975} whose  gauge group  $G$ depends on  the signature of the metric  and the sign of the cosmological constant:
\begin{center}
	\begin{tabular}{ l c c }
		& $\Lambda>0$ & $\Lambda<0$ \\
		$\sigma=1$ & $SO(3)$ & $SO(1,2)$ \\
		$\sigma=-1$ & $SO(1,2)$ & $SO(1,2).$ \\
	\end{tabular}
\end{center}
The action principle for the $BF$ description of the model is given by
\begin{eqnarray}
S[A,\phi]=\int_{\mathcal{M}} \phi_i F^i, \label{2D-BFaction}
\end{eqnarray}
where $\phi^i$ ($i=0,1,2$) are $0$-forms taking values in the Lie algebra $\mathfrak{g}$ of $G$, and $A^i$ is the $\mathfrak{g}$-valued  connection 1-form with curvature $F^i=dA^i+(1/2) f^i\,_{jk}A^j\wedge A^k$. The equations of motion resulting from (\ref{2D-BFaction}) are
\begin{eqnarray}
\delta A^i&:& D\phi^i:=d\phi^i+ f^i\,_{jk} A^j\phi^k=0,  \label{2D-BFA}\\
\delta \phi^i&:& F^i=0,\label{2D-BFphi}
\end{eqnarray}
which mean that $\phi^i$ is covariantly constant, and that the connection $A^i$ is flat. The infinitesimal gauge transformation, $\delta A^i = D\varepsilon^i$ and $ \delta \phi^i = - f^i\,_{jk} \varepsilon^j \phi^k$, comes from the identity $- D \phi_i \wedge D \varepsilon^i + F^i \left ( f^j\,_{ik} \varepsilon_j \phi^k  \right ) = d \left ( \varepsilon_i D \phi^i \right )$. The action (\ref{2D-BFaction}) is also diffeomorphism-invariant. However, the diffeomorphism generated by the vector field $v$ 
\begin{eqnarray}
\fl \delta A^i = {\mathcal L}_v A^i = D (v \cdot A^i ) + v \cdot F^i, \quad \delta \phi^i = {\mathcal L}_v \phi^i = - f^i\,_{jk} (v \cdot A^j) \phi^k + v \cdot ( D\phi^i),
\end{eqnarray} 
where $v \cdot A^i$ stands for the contraction of $v$ with $A^i$ and ${\mathcal L}_v$ is the Lie derivative along $v$, is a gauge transformation with (field-dependent) gauge parameter $\varepsilon^i:=v \cdot A^i$ modulo a trivial transformation generated by the equations of motion. 

The JT model (\ref{2D-JTaction}) is obtained from the $BF$-type action (\ref{2D-BFaction}) when the internal gauge symmetry is split as
\begin{eqnarray}
(A^i)=(l^{-1}e^I, \omega), \label{2D-connection}
\end{eqnarray}
where $e^I$ ($I=0,1$) is an orthonormal basis of the tangent space of ${\mathcal M}$, and $l$ is a nonvanishing constant, which is related to the cosmological constant by $\Lambda=\sigma\epsilon l^{-2}$ with $\epsilon:=\mathrm{sgn}(\sigma\Lambda)$. Because the indices $i,j,k, \ldots$ are raised and lowered with the Killing metric
\begin{eqnarray}
(\eta_{ij})=\mathrm{diag}(\sigma,1,\epsilon)=\left ( {\begin{array}{cc}
	\eta_{IJ}&0\\0&\epsilon\end{array}}\right ),
\end{eqnarray}
the structure constants can be written in the form $f^i{}_{jk}=\eta^{il} \varepsilon_{ljk}$ with $\eta^{ij}$ being the inverse of $\eta_{ij}$ and $\varepsilon_{012}=1$.

Using the splitting~(\ref{2D-connection}), equation~(\ref{2D-BFphi}) becomes
\begin{eqnarray}
d e^I + \omega^I\,_J \wedge e^J=0, \label{2deq1}\\
R^I\,_J =\Lambda e^I \wedge e_J\label{2deq2}, 
\end{eqnarray} 
where $R^I{}_J= d \omega^I{}_J + \omega^I{}_K \wedge \omega^K{}_J$ is the curvature of the spacetime connection $\omega^I{}_J$, which has been identified as $\omega^I{}_J := - \varepsilon^I{}_{J} \omega$ with $\varepsilon_{IJ}:=\varepsilon_{IJ2}$. The first equation tells us that $\omega^I{}_J$ is torsion-free (and compatible with the metric $\eta_{IJ}$ because of $d \eta_{IJ} + \omega^K\,_I \eta_{KJ} + \omega^K\,_{J} \eta_{IK}=0$), whereas the second one implies that it has constant curvature. Similarly, using 
\begin{eqnarray}
(\phi^i)=(\phi^I,\psi),\label{splitphi}
\end{eqnarray}
 Equation (\ref{2D-BFA}) becomes
\begin{eqnarray}
d \phi^I + \omega^I{}_J \phi^J +\frac{1}{l} \varepsilon^I{}_{J} e^J \psi =0, \label{splitphi12d1}\\
d \psi + \frac{\epsilon}{l} \varepsilon_{IJ} e^I \phi^J =0. \label{splitphi12d2}
\end{eqnarray}
Handling these equations, we get
\begin{eqnarray}
\nabla_{\mu} \nabla_{\nu} \psi + \Lambda \eta_{IJ} e_{\mu}{}{}^I e_{\nu}{}^J \psi =0,
\end{eqnarray}
where the relation between the spacetime and frame components of the connection $\omega^I{}_J$ has been used. Finally, it is worth mentioning that the action (\ref{2D-BFaction}) can be obtained from a dimensional reduction of a Chern-Simons theory~\cite{Cangemi1992261,SUNGKUKIM1993223}.

\subsubsection{Canonical analysis.}

We assume that the spacetime has a topology $\mathcal{M}\sim\mathbb{R}\times\Omega$,  where $\mathbb{R}$ represents time and $\Omega$ is a spatial 1-manifold without a boundary.  Let $\{t,x\}$ be the spacetime coordinates; the (1+1)-decomposition of the action (\ref{2D-BFaction}) is then
\begin{eqnarray}
S[A_x{},\Pi,A_t]=\int_{\mathbb{R}}dt\int_{\Omega}dx  \left( \Pi_i \dot{A}_x{}^i  + A_t{}^i  D_x \Pi_i  \right) , \label{2D-Hamil}
\end{eqnarray}
where $\Pi_i:=\phi_i$ and $D_x \Pi^i := \partial_x \Pi^i + f^i\,_{jk}A_x{}^j \Pi^k $. Since the variables $A_t{}^i$ appear linearly in the action (\ref{2D-Hamil}), they can be considered as Lagrange multipliers imposing the primary constraints 
\begin{eqnarray}
\mathcal{G}_i:=D_x \Pi_i \approx 0.
\end{eqnarray}
The evolution of $\mathcal{G}_i$ leads to $\dot{\mathcal{G}_i}=f^k\,_{ji} A_t{}^j \mathcal{G}_k\approx 0$, and hence no new constraints emerge. As a result, the constraints $\mathcal{G}_i$ are first class and generate the internal gauge transformations of the theory. There are no constraints associated to diffeomorphisms because, as we have seen, they are built from the internal gauge transformation plus trivial transformations. Since the phase space has three configuration variables $A_x{}^i$, the JT model is topological, as expected.

The quantization of the JT model in the $BF$ formalism has been performed following different approaches; see for instance~\cite{Constantinidis2009} and  references therein.

\subsubsection{Alternative form of the action.}
An alternative form of the action (\ref{2D-BFaction}) can be put forward by eliminating some of the components of the connection $A^i$. Let us analyze this possibility. Because we want to get rid of $A^i$, we are tempted to solve (\ref{2D-BFA}) for $A^i$. This is, however, only partially attainable, since the system of equations defined by (\ref{2D-BFA}) has vanishing determinant. More precisely, the antisymmetric $3\times 3$ matrix $\tensor{M}{^i_j}:=\tensor{f}{^i_{jk}}\phi^k$ is noninvertible, and so the solution of the system of equations is not unique. In fact, this matrix has $\phi^i$ as the only null vector (as long as $\phi_i\phi^i\neq 0$), and then the solution of (\ref{2D-BFA}) involves an arbitrary function multiplying it. Given that $\tensor{M}{^i_j}$ has rank 2, we can solve two of the components of $A^i$ in terms of the remaining fields, leaving one component of the connection unfixed (this component is related to the arbitrary function present in the general solution). Using the splitting (\ref{2D-connection}) and (\ref{splitphi}), one arrives at (\ref{splitphi12d1}) and (\ref{splitphi12d2}). Solving (\ref{splitphi12d1}) for $e^I$ yields to
\begin{eqnarray}
	e^I=\frac{\sigma l}{\psi}\tensor{\varepsilon}{^I_J}D_{\omega}\phi^J,\label{sol2A2d}
\end{eqnarray}
where the covariant derivative is defined by $D_{\omega}\phi^I:=d\phi^I+\tensor{\omega}{^I_J}\phi^J$ (recall that $\tensor{\omega}{^I_J}= - \varepsilon^I{}_{J} \omega$). Bear in mind that (\ref{splitphi12d2}) has not been touched yet, but it implies, after using (\ref{sol2A2d}), that the constraint 
\begin{eqnarray}
	\phi_id\phi^i=\epsilon\psi d\psi+\phi_I d\phi^I=0\label{const2d}
\end{eqnarray}
must be satisfied (this relation implies that locally $\phi^i$ satisfies $\phi_i\phi^i=\mathrm{const}$, which means that these fields live on an internal spherical or de Sitter space, depending on the gauge group). Therefore, the component $\omega$ remains unconstrained in the solution of (\ref{2D-BFA}).

Using once again the splitting (\ref{2D-connection}) and (\ref{splitphi}), the action (\ref{2D-BFaction}) reads
\begin{eqnarray}
	\fl S[\psi,\phi,e,\omega]=\int_{\mathcal{M}}\left[-\frac{\sigma\epsilon}{2}\psi\varepsilon_{IJ}\left(d\omega^{IJ}-\frac{\sigma\epsilon}{l^2}e^I\wedge e^J\right)+\frac{1}{l}\phi_{I}(de^I+\tensor{\omega}{^I_J}\wedge e^J)\right].\label{splitted2d}
\end{eqnarray}
Note that since the second term on the right-hand side of (\ref{splitted2d}) imposes precisely (\ref{2deq1}), $\tensor{\omega}{^I_J}$ becomes the spin-connection compatible with $e^I$, and substituting this back into the action (and using the fact that in two dimensions the curvature takes the form $R^{IJ}=d\omega^{IJ}=(1/2)\mathcal{R}e^I\wedge e^J$, where $\mathcal{R}$ is the curvature scalar), the action collapses (up to a global factor) to the JT action (\ref{2D-JTaction}). On the other hand, replacing (\ref{sol2A2d}) into (\ref{splitted2d}), we obtain
\begin{eqnarray}
\fl	S[\omega,\phi^I,\psi]=\sigma\int_{\mathcal{M}}&\left[-\frac{1}{2\psi}(\epsilon\psi^2+\phi_K\phi^K)\varepsilon_{IJ}d\omega^{IJ}+\frac{1}{2\psi}\varepsilon_{IJ}D_{\omega}\phi^I\wedge D_{\omega}\phi^J\right.\nonumber\\
	&\left.\hspace{2mm}-\frac{1}{\psi^2}\phi^I\varepsilon_{IJ}d\psi\wedge D_{\omega}\phi^J\right].\label{altact2d}
\end{eqnarray}
This action, in which now $\psi$ and $\phi^I$ are dynamical fields, provides an alternative description of the JT model. Notice that we did not use the condition (\ref{const2d}) in the derivation of (\ref{altact2d}). The reason for this is that we only required two of the equations (\ref{2D-BFA}) to eliminate $e^I$ from the action (the ones corresponding to varying (\ref{splitted2d}) with respect to $e^I$), the third one being not necessary for such a purpose. In fact, from (\ref{altact2d}), the equation of motion corresponding to $\omega^{IJ}$ leads precisely to (\ref{const2d}), which shows that our procedure is consistent. 

It is interesting to point out that the spacetime metric tensor, which entered in (\ref{2D-BFaction}) through $e^I$, has completely disappeared in (\ref{altact2d}). Nevertheless, we can recover the metric by identifying it (or the frame) from the equations of motion arising from this action. For $\phi^I$ and $\psi$, they read
\begin{eqnarray}
	\delta \phi^I&:& D_{\omega}\left(\frac{1}{\psi}\varepsilon_{IJ}D_{\omega}\phi^J\right)=-\frac{1}{\psi}\phi_I\varepsilon_{JK}R^{JK}-\frac{2}{\psi^2}\varepsilon_{IJ}d\psi\wedge D_{\omega}\phi^J,\label{eqphi2d}\\
	\delta \psi&:& R^{IJ}=\frac{\epsilon}{\psi^2}D_{\omega}\phi^I\wedge D_{\omega}\phi^J\label{eqpsi2d}.
\end{eqnarray}
Using (\ref{const2d}) in the second term on the right-hand side of (\ref{eqphi2d}), and combining the result with (\ref{eqpsi2d}), equation (\ref{eqphi2d}) reduces to $De^I=0$, where $e^I$ has been identified as in (\ref{sol2A2d}). With this, we recover equation (\ref{splitphi12d1}), while (\ref{const2d}) and (\ref{eqpsi2d}) become (\ref{splitphi12d2}) and (\ref{2deq2}), respectively, showing that all the equations arising from the original action (\ref{2D-BFaction}) can also be obtained from the action (\ref{altact2d}). Notice that the metric tensor emerging from (\ref{altact2d}) is identified as
\begin{equation}
g_{\mu\nu}=\frac{\sigma l^2}{\psi^2}\eta_{IJ}D_{\mu}\phi^I D_{\nu}\phi^J,
\end{equation}
where $D_{\mu}\phi^I:=(D_{\omega}\phi^I)_{\mu}=\partial_{\mu}\phi^I+\tensor{\omega}{_{\mu}^I_J}\phi^J$.

\subsection{Three-dimensional general relativity}

Vacuum Einstein's theory in three dimensions is an $SO(1,2)$ topological $BF$ theory in which the triad plays the role of the $B$ field. As such, it is soluble at both the classical and quantum levels~\cite{WITTEN198846}, and has become an arena for exploring and comparing different approaches to the quantization of the gravitational field~\cite{perez2013-16-3,carlip2003quantum,FreiLivRovcqg208,BonzomLivicqg2519,NouiPerPran2011}. Since we are interested in describing Euclidean and Lorentzian  three-dimensional gravity, let us denote the internal group in this section also as $SO(\sigma)$, where $SO(+1)=SO(3)$ and $SO(-1)=SO(1,2)$.

\subsubsection{Action principle with a cosmological constant.}\label{sec_3dgravity}

Let $\mathcal{M}$ be a three-dimensional manifold. To include the cosmological constant $\Lambda$, we add to the pure $BF$ action a term proportional to the spacetime volume, so that the action reads
\begin{eqnarray}
S[e,A]=-\sigma\int_{\mathcal{M}}\left(e_I\wedge F^I[A]+\frac{\sigma\Lambda}{6}\varepsilon_{IJK}e^I\wedge e^J\wedge e^K\right),\label{3Dgravact}
\end{eqnarray}
where $e^I$ are $\mathfrak{so}(\sigma)$-valued 1-forms, $A^I$ is an $\mathfrak{so}(\sigma)$-valued connection 1-form, and $F^I:=dA^I+(1/2)\tensor{\varepsilon}{^I_{JK}}A^J\wedge A^K$ is its curvature. The spacetime metric is determined by $e^I$ as $g_{\mu\nu}=\eta_{IJ}\tensor{e}{_{\mu}^I}\tensor{e}{_{\nu}^J}$, with $(\eta_{IJ})={\rm diag}(\sigma,1,1)$. The equations of motion arising from (\ref{3Dgravact}) are
\begin{eqnarray}
\delta A^I&:& de^I+\tensor{\varepsilon}{^I_{JK}}A^J\wedge e^K=0,\label{3Dtorfree}\\
\delta e^I&:& F^I=-\frac{\sigma\Lambda}{2}\tensor{\varepsilon}{^I_{JK}} e^J\wedge e^K.\label{3DCurvcte}
\end{eqnarray}
Equation (\ref{3Dtorfree}) says that $A^I$ is the spin-connection compatible with $e^I$, while equation (\ref{3DCurvcte}) establishes that the curvature is constant; they both are the three-dimensional Einstein equations with an arbitrary cosmological constant. The same equations of motion (for a nonvanishing cosmological constant) can also be derived from Witten's exotic action, which endows the phase space with a different symplectic structure~\cite{WITTEN198846}. In addition, three-dimensional general relativity can be formulated as a Chern-Simons theory in which the larger gauge connection is built up from $A^I$ and $e^I$~\cite{WITTEN198846,ACHUCARRO198689,HorneWitten}.

\subsubsection{Pure connection formulation.}

A remarkable feature of the action (\ref{3Dgravact}) is that it allows us to obtain a pure connection formulation for three-dimensional gravity. To be more general, let us consider the action principle (\ref{3Dgravact}) in which the gauge group is enlarged so that it has the form $G_T=SO(\sigma)\times G$, where objects transforming according to $G$ represent matter fields interacting with the gravitational field; this idea was introduced in~\cite{PeldanNPB395239} as a way of unifying gravity and Yang-Mills fields in three dimensions. Now, the indices $I,J,\dots$ run from 0 to $\mathrm{dim}(\mathfrak{g}_T)-1$, and they are raised and lowered with the Killing metric associated to $\mathfrak{g}_T$. We point out that the action (\ref{3Dgravact}) for arbitrary gauge groups is topological~\cite{romano1993,CattaneoJMP3611}, and so, in order to allow for matter fields (and given that the $B$ field is no longer a square matrix), the second term on the right-hand side of (\ref{3Dgravact}) must be replaced with the determinant of the spacetime metric, which in this case also takes the form $g_{\mu\nu}=\tensor{e}{_{\mu}^I}e_{\nu I}$; the generalized action reads 
\begin{eqnarray}
S[e,A]=-\sigma\int_{\mathcal{M}}d^3 x\left(e_{\mu I}\tilde{F}^{\mu I}+\sigma\Lambda\sqrt{\sigma g}\right),\label{3Dgravactlarg}
\end{eqnarray}
where $\tilde{F}^{\mu I}:=(1/2)\tilde{\eta}^{\mu\nu\lambda}\tensor{F}{^I_{\nu\lambda}}$ and\footnote{The actions (\ref{3Dgravact}) and (\ref{3Dgravactlarg}) have the general form $S[A,B]=\int_{\mathcal{M}}\left[\mathrm{Tr}(B\wedge F)+V(B)\right]$, where $V(B)$ is a potential term. The form of this potential could kill some of the gauge degrees of freedom of $BF$ theory, turning them into physical ones. This is exactly what the potential term of (\ref{3Dgravactlarg}) does, opening space for Yang-Mills fields to propagate (in a diffeomorphism-invariant way, since, for instance, the potential term of (\ref{BFYM_1}) generates the dynamics of Yang-Mills fields, but breaks diffeomorphism invariance), whereas the potential term of (\ref{3Dgravact}) does not spoil the topological character of the theory (even for arbitrary gauge group). The same is also true in four dimensions, where the potential term in (\ref{PleLike-act2}) or in (\ref{BfM-M}) breaks the topological symmetry of the $BF$ term and generates the dynamics of general relativity.} $g:=\det(g_{\mu\nu})$. We now have to integrate out the $B$ field from the action. The equation of motion corresponding to $e^I$ yields
\begin{eqnarray}
\tilde{F}^{\mu I}=-\sigma\Lambda\sqrt{\sigma g}e^{\mu I},\label{3dmote}
\end{eqnarray}
with $e^{\mu I}:=g^{\mu\nu}\tensor{e}{_{\nu}^I}$. This equation can be solved for $e^I$ only if $\Lambda\neq 0$, which we assume. The solution is
\begin{eqnarray}
e^{\mu I}=-\sigma\mathrm{sgn}(\Lambda)\left|\frac{\sigma}{\Lambda^2}\det(\tilde{F}^{\nu J}\tensor{\tilde{F}}{^{\lambda}_J})\right|^{-1/4}\tilde{F}^{\mu I}.\label{3dsolveeinF}
\end{eqnarray}
Notice that the inverse of the spacetime metric then takes the form
\begin{eqnarray}
g^{\mu\nu}=\left|\frac{\sigma}{\Lambda^2}\det(\tilde{F}^{\lambda J}\tensor{\tilde{F}}{^{\rho}_J})\right|^{-1/2}\tilde{F}^{\mu I}\tensor{\tilde{F}}{^{\nu}_I},\label{metric3dconnec}
\end{eqnarray}
which can be considered as a three-dimensional version of the Urbantke metric in the sense that it is constructed from the 2-forms (in this case, from the curvature $F^I$). Plugging (\ref{3dsolveeinF}) back into (\ref{3Dgravactlarg}), we obtain~\cite{PeldanNPB395239}
\begin{eqnarray}
S[A]=2\mathrm{sgn}(\Lambda)\int_{\mathcal{M}}d^3 x\left|\frac{\sigma}{\Lambda^2}\det(\tilde{F}^{\mu I}\tensor{\tilde{F}}{^{\nu}_I})\right|^{1/4}.\label{pure3dYM}
\end{eqnarray}
The action (\ref{pure3dYM}) embodies the pure connection formulation of a model that unifies both gravity and Yang-Mills theory. When $G_T=SO(\sigma)$, this action reduces to
\begin{eqnarray}
S[A]=2\mathrm{sgn}(\Lambda)\int_{\mathcal{M}}d^3 x\left|\frac{1}{\Lambda}\det(\tilde{F}^{\mu I})\right|^{1/2}.\label{pure3d}
\end{eqnarray} 
which defines a diffeomorphism-invariant gauge theory and constitutes the pure connection formulation of three-dimensional general relativity with a nonvanishing cosmological constant. In this sense, the action (\ref{pure3d}) can be considered as the three-dimensional version of the pure connection action (\ref{ActionKrasnovprl}). The equation of motion arising from (\ref{pure3dYM}) is
\begin{eqnarray}
D_{\lambda}\left(\sqrt{\sigma g}g^{\lambda\rho}g^{\mu\nu}\tensor{F}{^I_{\rho\nu}}\right)=0\ \Longleftrightarrow\ D*F^I=0,\label{motpure3d}
\end{eqnarray}
where $D$ is the $G_T$-covariant derivative, and the Hodge dual operator $*$ is computed with respect to the metric (\ref{metric3dconnec}) (this metric emerges dynamically, so that we are compelled to identify it with the spacetime metric). This equation is also valid for $G_T=SO(\sigma)$; hence, we conclude that three-dimensional general relativity with a nonvanishing cosmological constant is a diffeomorphism-invariant $SO(\sigma)$ Yang-Mills theory. For $G_T=SO(\sigma)\times G$, the fields in $G$ satisfy Yang-Mills equations, and thus the action (\ref{3Dgravactlarg}) or (\ref{pure3dYM}) provides a unification of gravity and Yang-Mills theory; it is worth pointing out that this unification coincides with the usual coupling of Yang-Mills fields to gravity only in the case of weak Yang-Mills fields~\cite{PeldanNPB395239}.

\subsubsection{Canonical analysis of the pure connection formulation.}

We now perform the canonical analysis of the action (\ref{pure3dYM}) to fill out a gap in the literature. We assume that the spacetime has the topology $\mathcal{M}\sim\mathbb{R}\times\Omega$, with $\Omega$ a spatial 2-manifold without boundary. Let us define the quantity $\tilde{\tilde{h}}^{\mu\nu}:=\tilde{F}^{\mu I}\tensor{\tilde{F}}{^{\nu}_I}$. Using this definition, we can show that
\begin{eqnarray}
2\mathrm{sgn}(\Lambda)\left|\frac{\sigma}{\Lambda^2}\det(\tilde{\tilde{h}}^{\mu\nu})\right|^{1/4}=\tilde{\Pi}^{aI}\dot{A}_{aI}+A_{tI}D_{a}\tilde{\Pi}^{aI}-\partial_{a}(\tilde{\Pi}^{aI}A_{aI}),\label{acthamilpure}
\end{eqnarray}
where $\tilde{\Pi}$ has the following expression
\begin{eqnarray}
\tilde{\Pi}^{aI}:=\tilde{\eta}^{ab}\mathrm{sgn}(\Lambda)\left|\frac{\sigma}{\Lambda^2}\det(\tilde{\tilde{h}}^{\mu\nu})\right|^{1/4}(\tilde{\tilde{h}}^{-1})_{b\mu}\tilde{F}^{\mu I},\label{can3dmom}
\end{eqnarray}
with $\tilde{\eta}^{ab}:=\tilde{\eta}^{0ab}$ the two-dimensional Levi-Civita tensor density. According to (\ref{acthamilpure}), the pairs $(A_{aI},\tilde{\Pi}^{aI})$ parametrize the phase space of the theory. On the other hand, the definition (\ref{can3dmom}) implies the following constraints
\begin{eqnarray}
\tilde{\mathcal{V}}_a&:=&\tilde{\Pi}^{bI}F_{Iba}=0,\label{vec3d}\\
\tilde{\tilde{\mathcal{H}}}&:=&\frac{1}{4}\tilde{\eta}^{ab}\tilde{\eta}^{cd}F^I{}_{ab} F_{Icd}-\sigma\Lambda^2\det(\tilde{\Pi}^{aI}\tilde{\Pi}^b{}_I)=0.\label{hamil3d}
\end{eqnarray}
Taking (\ref{acthamilpure}), (\ref{vec3d}), and (\ref{hamil3d}) into account, the extended action corresponding to (\ref{pure3dYM}) reads
\begin{eqnarray}
\fl S[A_a,\tilde{\Pi}^a,A_t,\underaccent{\tilde}{N},N^a]=\int_{\mathbb{R}}dt\int_{\Omega}d^2 x\left(\tilde{\Pi}^{aI}\dot{A}_{aI}+A_{tI}\tilde{\mathcal{G}}^I+\underaccent{\tilde}{N}\tilde{\tilde{\mathcal{H}}}+N^a\tilde{\mathcal{V}}_a\right),
\end{eqnarray}
where $\tilde{\mathcal{G}}^I:=D_{a}\tilde{\Pi}^{aI}$. The Lagrange multipliers $A_t$, $\underaccent{\tilde}{N}$, and $N^a$ impose the constraints $\tilde{\mathcal{G}}^I\approx0$, $\tilde{\tilde{\mathcal{H}}}\approx0$, and $\tilde{\mathcal{V}}_a\approx0$, respectively. These constraints are first class and therefore generate the gauge symmetries of the theory: $\tilde{\mathcal{G}}^I$ generates local $G_T$-transformations, whereas $\tilde{\tilde{\mathcal{H}}}$ and $\tilde{\mathcal{V}}_a$ generate spacetime diffeomorphisms. The count of degrees of freedom is straightforward: we have $2\times \mathrm{dim}(\mathfrak{g}_T)$ canonical pairs $(A_{aI},\tilde{\Pi}^{aI})$ subject to $\mathrm{dim}(\mathfrak{g}_T)+3$ first-class constraints, which leave a reduced phase space with $\mathrm{dim}(\mathfrak{g}_T)-3$ local physical degrees of freedom. For $G_T=SO(\sigma)$ the theory is topological, as expected, and while the constraints $\tilde{\mathcal{G}}^I$ and $\tilde{\mathcal{V}}_a$ are the same as the Gauss and vector constraints of the three-dimensional Ashtekar formalism~\cite{BengtIJMPA4}, respectively, the constraint $\tilde{\tilde{\mathcal{H}}}$ is related to the three-dimensional Ashtekar scalar constraint $\tilde{\tilde{\mathcal{H}}}_{\mathrm{Ash}}$ by
\begin{eqnarray}
	\hspace{-3mm}\tilde{\tilde{\mathcal{H}}}=\frac{\sigma}{\det(\tilde{\Pi}^{cL}\tilde{\Pi}^d{}_L)}\left[\frac{1}{2}\varepsilon_{IJK}\tilde{\Pi}^{aI}\tilde{\Pi}^{bJ}\tensor{F}{^K_{ab}}+\Lambda \det(\tilde{\Pi}^{aI}\tilde{\Pi}^b{}_I)\right]\tilde{\tilde{\mathcal{H}}}_{\mathrm{Ash}},
\end{eqnarray}
where 
\begin{eqnarray}
\hspace{-3mm}	\tilde{\tilde{\mathcal{H}}}_{\mathrm{Ash}}:=\frac{1}{2}\varepsilon_{IJK}\tilde{\Pi}^{aI}\tilde{\Pi}^{bJ}\tensor{F}{^K_{ab}}-\Lambda \det(\tilde{\Pi}^{aI}\tilde{\Pi}^b{}_I),
\end{eqnarray}
and a term quadratic in $\tilde{\mathcal{V}}_a$ has been neglected.

\subsubsection{Plebanski-like action.}\label{3dconst}

In this section we follow the approach of~\cite{GeiNouicqg2913}, where it is shown that Euclidean three-dimensional general relativity can be formulated as an $SO(4)$-constrained $BF$ theory. Consider an $SO(4)$-principal bundle over a three-dimensional manifold $\mathcal{M}$; the action principle is given by 
\begin{eqnarray}
S[A,B,\tilde{\phi}]=\int_{\mathcal{M}}B^{IJ}\wedge F_{IJ}[A]+\int_{\mathcal{M}}d^3 x\tilde{\phi}^{\mu\nu}B_{\mu IJ}*B_{\nu}{}^{IJ},\label{BFC3d}
\end{eqnarray}
where $B^{IJ}$ are $\mathfrak{so}(4)$-valued 1-forms and $\tilde{\phi}^{\mu\nu}$ are symmetric Lagrange multipliers (we follow the conventions of section~\ref{BF4_act}). $\tilde{\phi}^{\mu\nu}$ imposes the following simplicity constraints
\begin{eqnarray}
\varepsilon_{IJKL}B_{\mu}{}^{IJ}B_{\nu}{}^{KL}=0.\label{const3dbf}
\end{eqnarray}
As in section \ref{BFEE_derive}, this constraint has two nondegenerate solutions~\cite{GeiNouicqg2913}
\begin{eqnarray}
B^{IJ}=\tensor{\varepsilon}{^{IJ}_{KL}}\chi^K e^L,\label{3dsol1}\\
B^{IJ}=(\chi^I e^J-\chi^J e^I),\label{3dsol2}
\end{eqnarray}
where $\chi^I$ is a scalar taking values in $\mathbb{R}^4$ and $e^I$ is a set of four 1-forms. Note that the number of constraints in equation (\ref{const3dbf}) is 6, which implies that the 18 components $\tensor{B}{_{\mu}^{IJ}}$ should be parametrized in terms of 12 independent variables. However, $\chi^I$ and $\tensor{e}{_{\mu}^I}$ amount to 16 variables, meaning that all them are not independent; indeed, there are two symmetries which allows us to fix four of them: (i)  we have the rescaling symmetry $\chi^I\rightarrow\alpha\chi^I$, $e^I\rightarrow\alpha^{-1}e^I$, $\alpha\neq 0$, which fixes one component of $\chi^I$ (for instance, we may choose $\chi^I$ with unit norm), and (ii) there is a translational symmetry $\tensor{e}{_{\mu}^I}\rightarrow\tensor{e}{_{\mu}^I}+\beta_{\mu}\chi^I$, with $\beta_{\mu}$ a 1-form, whose immediate consequence is that we can make $\tensor{e}{_{\mu}^0}=0$, reducing in three the number of variables (this choice breaks the $SO(4)$ symmetry, though).

Substituting (\ref{3dsol1}) into (\ref{BFC3d}) we obtain
\begin{eqnarray}
S[e,\chi,A]=\int_{\mathcal{M}}\varepsilon_{IJKL}\chi^Ie^J\wedge F^{KL}[A].\label{Ac3dsol1}
\end{eqnarray}
Using the time gauge $\chi^i=0$ ($i=1,2,3$) combined with the rescaling symmetry, we can set $\chi^I=\delta^I_0$, and after eliminating some of the components of the connection, the action (\ref{Ac3dsol1}) reads
\begin{eqnarray}
S[e,\omega]=\int_{\mathcal{M}}\varepsilon_{ijk}e^i\wedge F^{jk}[\omega],
\end{eqnarray}
where $\omega_{ij}:=A_{ij}$. This is Palatini's action in three dimensions.

Let us see how the action (\ref{BFC3d}) looks for the solutions (\ref{3dsol1}) and (\ref{3dsol2}) in the second-order formalism. Recall that the Lie algebra $\mathfrak{so}(4)$ separates into the self-dual and anti-self-dual subalgebras as $\mathfrak{so}(4)={}^{(+)}\mathfrak{so}(3)\oplus{}^{(-)}\mathfrak{so}(3)$. The generators of these subalgebras take the form ${}^{(\pm)}\tensor{J}{_i^{IJ}}=(1/2)[-\tensor{\varepsilon}{_{0i}^{IJ}}\pm(\delta_i^I\delta_0^J-\delta_0^I\delta_i^J)]$ and satisfy $[{}^{(\pm)}J_i,{}^{(\pm)}J_j]=\tensor{\varepsilon}{_{ij}^k}{}^{(\pm)}J_k$, $[{}^{(+)}J_i,{}^{(-)}J_j]=0$, ${}^{(\pm)}\tensor{J}{_i^{IJ}}{}^{(\pm)}J{_{jIJ}}=\delta_{ij}$, ${}^{(+)}\tensor{J}{_i^{IJ}}{}^{(-)}J{_{jIJ}}=0$, and $*{}^{(\pm)}\tensor{J}{_i^{IJ}}=\pm {}^{(\pm)}\tensor{J}{_i^{IJ}}$.

Separating $B$ into its self-dual and anti-self-dual parts, namely $B={}^{(+)}B^i{}^{(+)}J_i+{}^{(-)}B^i{}^{(-)}J_i$ with ${}^{(\pm)}B^i=B^{IJ}{}^{(\pm)}J^i{}_{IJ}$, the action (\ref{BFC3d}) reads (we omit the second term on the right-hand side, since we assume that we use the solutions for the $B$ field)
\begin{eqnarray}
\fl S[{}^{(+)}B,{}^{(-)}B,{}^{(+)}A,{}^{(-)}A]=\int_{\mathcal{M}}\left[{}^{(+)}B^i\wedge F_i[{}^{(+)}A]+{}^{(-)}B^i\wedge F_i[{}^{(-)}A]\right],\label{actself3d}
\end{eqnarray}
where $F^i[A]:=dA^i+(1/2)\tensor{\varepsilon}{^i_{jk}}A^j\wedge A^k$. The equation of motion corresponding to ${}^{(\pm)}A^i$ then implies that, for $\det ({}^{(\pm)}\tensor{B}{_{\mu}^i})\neq 0$, ${}^{(\pm)}A^i$ is the spin-connection in each sector.

Now, introduce the metric ${}^{(\pm)}g_{\mu\nu}:={}^{(\pm)}\tensor{B}{_{\mu}^i}{}^{(\pm)}\tensor{B}{_{\nu}^j}\delta_{ij}$. Bearing in mind that ${}^{(\pm)}A^i$ is the spin-connection compatible with ${}^{(\pm)}B^i$, the action (\ref{actself3d}) reduces to
\begin{eqnarray}
\fl S[{}^{(+)}g,{}^{(-)}g]=-\frac{\epsilon^{+}}{2}\int_{\mathcal{M}}d^3x\sqrt{{}^{(+)}g}\mathcal{R}[{}^{(+)}g_{\mu\nu}]-\frac{\epsilon^{-}}{2}\int_{\mathcal{M}}d^3x\sqrt{{}^{(-)}g}\mathcal{R}[{}^{(-)}g_{\mu\nu}],\label{act2ndself}
\end{eqnarray}
where $\epsilon^{\pm}:=\mathrm{sgn}[\det ({}^{(\pm)}\tensor{B}{_{\mu}^i})]$, ${}^{(\pm)}g:=\det({}^{(\pm)}g_{\mu\nu})$, and $\mathcal{R}[g_{\mu\nu}]$ is the curvature scalar (we use the convention ${}^{(\pm)}\tensor{A}{^i_j}=\tensor{\varepsilon}{^i_{kj}}{}^{(\pm)}A^k$).

Using the solution (\ref{3dsol1}), we have ${}^{(\pm)}B^i=\mp\tensor{\varepsilon}{^i_{jk}}\chi^je^k+\chi^ie^0-\chi^0e^i$, which implies $\epsilon^{+}=\epsilon^{-}=\epsilon$, and the metric, being the same in both sectors, takes the form
\begin{eqnarray}
\fl {}^{(\pm)}g_{\mu\nu}=g_{\mu\nu}=(\chi^i\chi_i+\chi_0^2)\tensor{e}{_{\mu}^j}e_{\nu j}+\chi^i\chi_i\tensor{e}{_{\mu}^0}\tensor{e}{_{\nu}^0}-\chi_0\chi_i(\tensor{e}{_{\mu}^0}\tensor{e}{_{\nu}^i}+\tensor{e}{_{\mu}^i}\tensor{e}{_{\nu}^0})-\chi_i\chi_j\tensor{e}{_{\mu}^i}\tensor{e}{_{\nu}^j}.\label{metric3d}\nonumber\\
\end{eqnarray}
Putting these results together, (\ref{act2ndself}) reads
\begin{eqnarray}
S[g]=-\epsilon\int_{\mathcal{M}}d^3x\sqrt{g}\mathcal{R}[g_{\mu\nu}],\label{act2nd3di}
\end{eqnarray}
which is the Einstein-Hilbert action. The solution (\ref{3dsol1}) deservedly has been named gravitational sector.

On the hand, for the solution (\ref{3dsol2}) we have ${}^{(\pm)}B^i=-\tensor{\varepsilon}{^i_{jk}}\chi^je^k\pm(\chi^ie^0-\chi^0e^i)$ (notice that it is $\pm$ times the one corresponding to the other solution), which implies $\epsilon^{+}=-\epsilon^{-}=\epsilon$, while the metric has the same expression (\ref{metric3d}) for both sectors. Taking these results into account, (\ref{act2ndself}) vanishes. Therefore, since the action vanishes on-shell, the solution (\ref{3dsol2}) is referred to as topological sector.

The action (\ref{BFC3d}) admits the following Holst-like generalization
\begin{eqnarray}
\fl S[A,B,\tilde{\phi}]=\int_{\mathcal{M}}\left(B^{IJ}+\frac{1}{\gamma}*B^{IJ}\right)\wedge F_{IJ}[A]+\int_{\mathcal{M}}d^3 x\tilde{\phi}^{\mu\nu}B_{\mu IJ}*B_{\nu}{}^{IJ},\label{BFC3dHolst}
\end{eqnarray}
where $\gamma$ is a three-dimensional Immirzi-like parameter. Using the gravitational solution (\ref{3dsol1}), (\ref{BFC3dHolst}) collapses to
\begin{eqnarray}
S[e,\chi,A]=\int_{\mathcal{M}}\left(\epsilon_{IJKL}+\frac{2}{\gamma}\delta_{IJKL}\right)\chi^Ie^J\wedge F^{KL}[A],\label{Holst3d}
\end{eqnarray}
where $\delta_{IJKL}:=(1/2)(\delta_{IK}\delta_{JL}-\delta_{IL}\delta_{JK})$. As we previously showed, the second term on the right-hand side of this action vanishes on-shell, and so the parameter $\gamma$ drops out from the action, hence playing the same role as the Immirzi parameter. On the other side, for the solution (\ref{3dsol2}), the action (\ref{BFC3dHolst}) reduces to (\ref{Holst3d}) with a different value of the Immirzi parameter.

The action (\ref{Holst3d}), which can be obtained from a symmetry reduction of the four-dimensional Holst action~\cite{Geillergrg459}, has been used to investigate the role of the Immirzi parameter in three-dimensional loop quantum gravity~\cite{AchNouiprd9110} (see~\cite{BonzomLivicqg2519} for an alternative way of introducing an Immirzi-like parameter in three-dimensional gravity). The canonical analysis and the spinfoam quantization of the action (\ref{BFC3d}) are studied in~\cite{GeiNouicqg2913}.

Extending the previous result, the action (\ref{BFC3dHolst}) can similarly be obtained from a symmetry reduction of a $BF$ formulation for four-dimensional general relativity with internal group $SO(4)$. Consider the following action defined on a 4-manifold $\mathcal{M}_4$ (compare with~\cite{BuffHennetcqg2122})
\begin{eqnarray}
\fl S[A,B,\tilde{\varphi},\mu]=\int_{\mathcal{M}_4}\left(B^{IJ}+\frac{1}{\gamma}*B^{IJ}\right)\wedge F_{IJ}[A]+\frac{1}{4}\int_{\mathcal{M}_4}d^4x\tilde{\varphi}^{\mu\nu\lambda\sigma}B_{\mu\nu IJ}*\tensor{B}{_{\lambda\sigma}^{ IJ}}\nonumber\\
+\int_{\mathcal{M}_4}\mu\underaccent{\tilde}{\eta}_{\mu\nu\lambda\sigma}\tilde{\varphi}^{\mu\nu\lambda\sigma},\label{BFalt4D}
\end{eqnarray}
where the multiplier $\tilde{\varphi}^{\mu\nu\lambda\sigma}$ has the same symmetries as the multiplier $\varphi^{IJKL}$ in the classically equivalent action (\ref{ActBFImm}). We now perform a spacetime reduction. Assume that $\mathcal{M}_4$ has the topology $\mathcal{M}_4=\mathcal{M}\times \mathbb{I}$, where $\mathcal{M}$ is a three-dimensional manifold and $\mathbb{I}$ is a real spatial interval with coordinate $x^3$. Furthermore, assume that the fields entering in  (\ref{BFalt4D}) do not depend on $x^3$ and that $B_{\mu\nu}{}^{IJ}=0$ for $\mu,\nu<3$. The action (\ref{BFalt4D}) takes the form
\begin{eqnarray}
\fl S[A,B,\tilde{\varphi}]=-\int_{\mathbb{I}}dx^3\int_{\mathcal{M}}dv\biggl[ &\frac{1}{2}\tilde{\eta}^{\mu\nu\lambda3}\left(\tensor{B}{_{\mu 3}^{IJ}}+\frac{1}{\gamma}*\tensor{B}{_{\mu3}^{IJ}}\right)F_{IJ\nu\lambda}\nonumber\\
&+\tilde{\varphi}^{\mu 3\nu 3}B_{\mu 3 IJ}*\tensor{B}{_{\nu 3}^{ IJ}}\biggr],\label{reduced4d} 
\end{eqnarray}
where $dv=dx^0 \wedge dx^1 \wedge dx^2$, the Greek indices run from 0 to 2, and we have used the equation of motion corresponding to $\mu$ to eliminate the third term on the right-hand side of (\ref{BFalt4D}). Notice that the components $\tensor{A}{_3^{IJ}}$ have disappeared from the action, and thus it only depends on the components $\tensor{A}{_{\mu}^{IJ}}$ ($\mu<3$) through $\tensor{F}{^{IJ}_{\mu\nu}}$. By defining $\tilde{\eta}^{\mu\nu\lambda}:=\tilde{\eta}^{\mu\nu\lambda3}$, $\tensor{B}{_{\mu}^{IJ}}:=\tensor{B}{_{\mu 3}^{IJ}}$, and $\tilde{\phi}^{\mu\nu}:=\tilde{\varphi}^{\mu 3\nu 3}$, the action (\ref{reduced4d}) reads
\begin{eqnarray}
\fl S[A,B,\tilde{\varphi}]=-\int_{\mathbb{I}}dx^3\int_{\mathcal{M}}dv\left[\frac{1}{2}\tilde{\eta}^{\mu\nu\lambda}\left(\tensor{B}{_{\mu}^{IJ}}+\frac{1}{\gamma}*\tensor{B}{_{\mu}^{IJ}}\right)F_{IJ\nu\lambda}+\tilde{\phi}^{\mu\nu}B_{\mu IJ}*\tensor{B}{_{\nu}^{ IJ}}\right],\nonumber\\
\end{eqnarray}
which is, up to a global factor, the three-dimensional action (\ref{BFC3dHolst}).

\section{$BF$ formulation of higher-dimensional general relativity}\label{BF-higher}

Palatini's action can be extended to a $D$-dimensional manifold $\mathcal{M}$ ($D>4$) by appropriately adding the number of frames necessary for the integrand to become an adequate $D$-form. For this purpose, we take as the internal group either SO($D$) for the  Euclidean case or SO(1,$D-1$) for the Lorentzian case; the internal indices $I,J,\dots$ now run from 1 to $D$, while the internal metric has the form  $(\eta_{IJ})=\textrm{diag}(\sigma,1,\dots,1)$. The action principle for general relativity in $D$ dimensions then takes the form
\begin{eqnarray}\label{PalaD}
S[e,A]=\int_{\mathcal{M}}*(e^I\wedge e^J)\wedge F_{IJ}[A],
\end{eqnarray}
where $A$ is a connection 1-form taking values in the Lie algebra of the corresponding gauge group, $F$ is the curvature of this connection, and
\begin{eqnarray}\label{staree}
*(e^I\wedge e^J):=\frac{1}{(D-2)!}\tensor{\varepsilon}{^{IJ}_{K_1\dots K_{D-2}}}e^{K_1}\wedge\dots\wedge e^{K_{D-2}}.
\end{eqnarray}
Notice that $*$ is nothing but the Hodge dual which in this case takes 2-forms to ($D-2$)-forms, and thus makes the integrand well-defined. With this notation, the action (\ref{PalaD}) is even valid for $D$ equal to 3 and 4, and, as in those cases, the equations of motion arising from it lead to the Einstein field equations in vacuum.

The action (\ref{PalaD}) can be equivalently written as

\begin{eqnarray}\label{PalaDc}
S[e,A]=\int_{\mathcal{M}} d^D x \, e\ \tensor{e}{^{\mu}_I}\tensor{e}{^{\nu}_J}\tensor{F}{^{IJ}_{\mu\nu}},
\end{eqnarray}
where $e:=\det(\tensor{e}{_{\mu}^I})$ and $\tensor{e}{^{\mu}_I}$ is the inverse of $\tensor{e}{_{\mu}^I}$. This form of the action is more convenient for passing to its corresponding $BF$ formulation. In section \ref{BF4_act} we saw that Palatini's action in four dimensions could be written as a constrained $BF$ theory with quadratic constraints on the $B$ field. It turns out that the action (\ref{PalaD}) can also be written in the same way, although the constraint on the $B$ field is more complicated. In particular, the quadratic constraints imposed on the $B$ field are no longer independent in this case, and so, one has to be careful when accounting for the degrees of freedom in such a formulation. 

For passing from the action (\ref{PalaD}) to its $BF$ formulation, we need to impose a constraint on the $B$ field so that it takes the form $B^{IJ}=*(e^I\wedge e^J)$ when such a constraint is solved, as (\ref{PalaD}) indicates. Notice that now the $B$ field is not a 2-form as in section~\ref{BF4_act}, but a ($D-2$)-form. One can, however, use a bivector (an antisymmetric tensor of rank 2) to build the $BF$ formulation, and that is what we do below. From the ($D-2$)-form $B^{IJ}$ we construct the densitized bivector defined by
\begin{eqnarray}
\tensor{\tilde{B}}{^{\mu\nu}_{IJ}}:=\frac{1}{2(D-2)!}\tilde{\eta}^{\mu\nu\lambda_1\dots\lambda_{D-2}}B_{\lambda_1\dots\lambda_{D-2}IJ}.
\end{eqnarray}
The action that describes higher-dimensional general relativity as a constrained $BF$ theory then reads 
\begin{eqnarray}
\fl S[A,\tilde{B},\Phi,\tilde{\mu}]=\int_{\mathcal{M}} d^Dx&\left[\tensor{\tilde{B}}{^{\mu\nu}_{IJ}}\tensor{F}{^{IJ}_{\mu\nu}}+\Phi^{[\alpha]IJKL}\underaccent{\tilde}{\eta}_{[\alpha]\mu\nu\lambda\rho}\tensor{\tilde{B}}{^{\mu\nu}_{IJ}}\tensor{\tilde{B}}{^{\lambda\rho}_{KL}}\right.\nonumber\\
&\left.\ +\tilde{\mu}^{[M]}_{[\alpha]}\varepsilon_{[M]IJKL}\Phi^{[\alpha]IJKL}\right].\label{BFactD}
\end{eqnarray}
Here $[\alpha]$ and $[M]$ are sets of completely antisymmetric spacetime and internal indices of length $D-4$, respectively, and both $\Phi$ and $\tilde{\mu}$ are Lagrange multipliers. While the former has the index symmetries $\Phi^{[\alpha]IJKL}=\Phi^{[\alpha][IJ][KL]}=\Phi^{[\alpha]KLIJ}$ and imposes the desired constraint on the $B$ field, the latter imposes a constraint on $\Phi$ itself that can be considered as a higher-dimensional version of the constraint $G_2$ of section~\ref{BF4_act} (see equation (\ref{G_func})). We point out that the action (\ref{BFactD}) differs from the one reported in~\cite{Freid_Puz} in that the former considers the multiplier $\Phi$ whose antisymmetrization in the internal indices vanishes (as enforced by $\tilde{\mu}$), whereas the latter treats the case where the antisymmetrization in the Lagrange multiplier's spacetime indices vanishes; nevertheless, both action principles are classically equivalent, since in both cases the constraint on the $B$ field is enough to ensure that it comes from a frame. The approach of~\cite{Freid_Puz} turns out to be more suitable for exploring the quantum theory, though.

The variation of the action with respect to $\Phi$ yields
\begin{eqnarray}\label{ConstD}
\underaccent{\tilde}{\eta}_{[\alpha]\mu\nu\lambda\rho}\tensor{\tilde{B}}{^{\mu\nu}_{IJ}}\tensor{\tilde{B}}{^{\lambda\rho}_{KL}}+\tilde{\mu}^{[M]}_{[\alpha]}\varepsilon_{[M]IJKL}=0,
\end{eqnarray}
which looks similar to (\ref{variatphi}) with $a_1=0$, and is, in fact, its generalization to higher dimensions. These constraints are also referred to as {\it simplicity constraints}. It turns out that for $D>4$ the relations (\ref{ConstD}) are no longer independent among themselves, but it can be shown that the solution of (\ref{ConstD}) is~\cite{Freid_Puz}
\begin{eqnarray}
\tensor{\tilde{B}}{^{\mu\nu}_{IJ}}=e\ \tensor{e}{^{\mu}_I}\tensor{e}{^{\nu}_J},
\end{eqnarray}
for some (invertible) frame $\tensor{e}{_{\mu}^I}$. By substituting this solution back into the action (\ref{BFactD}) we recover Palatini's action (\ref{PalaDc}).

We can convince ourselves that the constraints (\ref{ConstD}) are not independent by simply counting the number of equations implied by them. This number equals the number of Lagrange multipliers $\Phi$, which is given by ${D\choose 4}{D\choose 2}\left[{D\choose 2}+1\right]/2$. On the other hand, taking into account both $\tilde{B}$ and $\tilde{\mu}$, the number of independent variables present in (\ref{ConstD}) is ${D\choose 2}^2+{D\choose 4}^2$. One can easily verify that for $D>4$ the number of equations exceeds the number of independent variables, and since the simplicity constraints must produce the $D^2$ degrees of freedom entailed by the frame $e^I$, then there must be nontrivial relations among them. Nevertheless, it is possible to find a subset of independent constraints, which amounts to gauge fixing a gauge symmetry occurring on the multiplier $\Phi$; see details in~\cite{Freid_Puz}. The case $D=4$ is exactly the same studied in section~\ref{sect_Real_F} for the constraint $G_2$.

The canonical analysis of the action reported in~\cite{Freid_Puz} is performed in~\cite{BodenThiecqg304}, the latter being part of a series of papers generalizing loop quantum gravity to higher dimensions. As in section~\ref{decomposeact_r}, the simplicity constraint generates second-class constraints that this time are reducible. By carefully addressing this issue, the number of independent second-class constraints is finally identified and the resulting propagating degrees of freedom match those of general relativity in the same dimension, namely $D(D-3)/2$.

\section{Conclusions}

We reviewed classical general relativity in all dimensions as a deformation of $BF$ theory, focusing on the Lagrangian and the Hamiltonian aspects of these $BF$-type  formulations. We also discussed other models of gravity (different from general relativity) which can be cast in the $BF$ formalism; they are the Husain-Kucha\v{r} model (see section~\ref{HusainKuchar}), the non-metric theories for four-dimensional gravity (see section~\ref{nonmetric}), and the JT model for two-dimensional gravity (see section~\ref{jackwteitel}). Since $BF$ theories define topological field theories, it is actually outstanding that one may add degrees of freedom into the game by introducing either constraints on the $B$ field or potential terms built up from it. The role of these terms is to kill some of the gauge degrees of freedom of the $BF$ theory, which breaks its topological character and gives rise to the physical degrees of freedom of the theory one wants to describe within this formalism. In particular, general relativity can be formulated as a background independent  constrained $BF$ theory (see sections \ref{sec:plebanskiformulation}, \ref{sec_Plebanski-like}, \ref{sect_Real_F}, \ref{3dconst}, and \ref{BF-higher}) or as a $BF$ theory plus a potential term (see sections \ref{sec_Plebanski-like}, \ref{MacDowellMansouri}, and \ref{sec_3dgravity}). In the former approach, the (quadratic) constraint on the $B$ field is introduced so that the orthonormal frame be brought back into the formalism, leading to Palatini-like actions for general relativity. In the latter, the potential term (polynomial or not in the $B$ field) destroys a lot of gauge invariance of the original $BF$ action and generates the gravitational degrees of freedom.

The $BF$-type formulations of general relativity then stand at the same level of other formulations of general relativity such as the first-order formulation, and can be used as starting points for investigating many other issues related to gravity (see, for instance,~\cite{kras201143}). In particular, given that these formulations are neighbors of $BF$ theories, they have opened new ways of tackling the problem of quantizing gravity. Since the quantization of the topological $BF$ theory has been wholly achieved, the constraints or potential terms added to the $BF$ action can be integrated into the quantum theory by properly imposing the constraints on the discretized theory (see~\cite{perez2013-16-3} and references therein) or by treating the potential terms in a perturbative way~\cite{StarodubtsevArxiv,Rovelligrg2006} (see~\cite{FreidKrassATMP} for a general spinfoam treatment of these potential terms). One of the most promising nonperturbative approaches to solve the quantum gravity puzzle is the spinfoam approach~\cite{perez2013-16-3}, which tries to achieve a fully covariant path integral quantization of gravity taking as the starting  point an action for general relativity (in any dimension) formulated as a $BF$ theory. In the four-dimensional setting, a spinfoam corresponds to a transition amplitude between 3-geometry states (or spin-networks) at different times, and so this approach could provide a novel viewpoint for understanding the quantum dynamics of gravity, something that is lacking in the loop framework.

\addcontentsline{toc}{section}{Acknowledments}
\ack 

We are grateful to Riccardo Capovilla, Jorge Romero and Jos\'e A. Zapata for
their careful reading of the manuscript. On the occasion of Carlo Rovelli's 60th birthday,
Merced Montesinos would like to express his deep gratitude to Carlo for
his support over the years. Furthermore, Merced Montesinos thanks Emel\'i Cortina for inspiring him. This work was supported in part by Consejo Nacional de Ciencia y Tecnolog\'{\i}a, M\'exico, Grant Numbers 167477-F and 237004-F.

\section*{References}\addcontentsline{toc}{section}{References}
\bibliography{references}

\end{document}